\documentclass[floatfix, superscriptaddress,reprint,amsmath,amssymb,
pra,showkeys]{revtex4-2}

\usepackage{graphics}
\usepackage{dcolumn}
\usepackage{bm}

\usepackage{array}
\usepackage{booktabs}
\usepackage{multirow}
\usepackage{esint}
\usepackage{color}
\usepackage{graphicx}
\usepackage{epstopdf}
\usepackage[version=4]{mhchem}
\usepackage{verbatim}
\usepackage{float}
\usepackage{subfigure}
\usepackage{fancyvrb}
\usepackage[export]{adjustbox}
\usepackage{svg} 
\usepackage{comment} 

\usepackage{float}
\makeatletter
\let\newfloat\newfloat@ltx
\makeatother
\usepackage{algorithm} 
\usepackage{algpseudocode} 

\usepackage{tikz}
\usepackage{pgfplots}
\usepackage{pgfplotstable}
\pgfplotsset{compat = newest}
\usetikzlibrary{shapes.geometric, arrows}

\usepackage{xr-hyper}

\usepackage{soul} 

\newcommand{\RNum}[1]{\uppercase\expandafter{\romannumeral #1\relax}}
\usepackage{textcomp, gensymb}
\DeclareUnicodeCharacter{0308}{HERE!HERE!}

\usepackage{scalerel}
\usepackage{tikz}
\usetikzlibrary{svg.path}
\definecolor{orcidlogocol}{HTML}{A6CE39}
\tikzset{
  orcidlogo/.pic={
    \fill[orcidlogocol] svg{M256,128c0,70.7-57.3,128-128,128C57.3,256,0,198.7,0,128C0,57.3,57.3,0,128,0C198.7,0,256,57.3,256,128z};
    \fill[white] svg{M86.3,186.2H70.9V79.1h15.4v48.4V186.2z}
                 svg{M108.9,79.1h41.6c39.6,0,57,28.3,57,53.6c0,27.5-21.5,53.6-56.8,53.6h-41.8V79.1z M124.3,172.4h24.5c34.9,0,42.9-26.5,42.9-39.7c0-21.5-13.7-39.7-43.7-39.7h-23.7V172.4z}
                 svg{M88.7,56.8c0,5.5-4.5,10.1-10.1,10.1c-5.6,0-10.1-4.6-10.1-10.1c0-5.6,4.5-10.1,10.1-10.1C84.2,46.7,88.7,51.3,88.7,56.8z};
  }
}
\newcommand\orcidicon[1]{\href{https://orcid.org/#1}{\mbox{\scalerel*{
\begin{tikzpicture}[yscale=-1,transform shape]
\pic{orcidlogo};
\end{tikzpicture}
}{|}}}} 

\begin{document}
%

\newcommand{\rz}{\mathbb{R}}
\newcommand{\gz}{\mathbb{Z}}
\newcommand{\cz}{\mathbb{C}}
\newcommand{\qz}{\mathbb{Q}}
\newcommand{\nz}{\mathbb{N}}

\newcommand{\bfa}{{\bf a}}
\newcommand{\bfb}{{\bf b}}
\newcommand{\bfc}{{\bf c}}
\newcommand{\bfd}{{\bf d}}
\newcommand{\bfe}{{\bf e}}
\newcommand{\bff}{{\bf f}}
\newcommand{\bfg}{{\bf g}}
\newcommand{\bfh}{{\bf h}}
\newcommand{\bfi}{{\bf i}}
\newcommand{\bfj}{{\bf j}}
\newcommand{\bfk}{{\bf k}}
\newcommand{\bfl}{{\bf l}}
\newcommand{\bfm}{{\bf m}}
\newcommand{\bfn}{{\bf n}}
\newcommand{\bfo}{{\bf o}}
\newcommand{\bfp}{{\bf p}}
\newcommand{\bfq}{{\bf q}}
\newcommand{\bfr}{{\bf r}}
\newcommand{\bfs}{{\bf s}}
\newcommand{\bft}{{\bf t}}
\newcommand{\bfu}{{\bf u}}
\newcommand{\bfv}{{\bf v}}
\newcommand{\bfw}{{\bf w}}
\newcommand{\bfx}{{\bf x}}
\newcommand{\bfy}{{\bf y}}
\newcommand{\tbfy}{{\tilde{\bfy}}}
\newcommand{\bfz}{{\bf z}}
\newcommand{\bfA}{{\bf A}}
\newcommand{\bfB}{{\bf B}}
\newcommand{\bfC}{{\bf C}}
\newcommand{\bfD}{{\bf D}}
\newcommand{\bfE}{{\bf E}}
\newcommand{\bfF}{{\bf F}}
\newcommand{\bfG}{{\bf G}}
\newcommand{\bfH}{{\bf H}}
\newcommand{\bfI}{{\bf I}}
\newcommand{\bfJ}{{\bf J}}
\newcommand{\bfK}{{\bf K}}
\newcommand{\bfL}{{\bf L}}
\newcommand{\bfM}{{\bf M}}
\newcommand{\bfN}{{\bf N}}
\newcommand{\bfO}{{\bf O}}
\newcommand{\bfP}{{\bf P}}
\newcommand{\bfQ}{{\bf Q}}
\newcommand{\bfR}{{\bf R}}
\newcommand{\bfS}{{\bf S}}
\newcommand{\bfT}{{\bf T}}
\newcommand{\bfU}{{\bf U}}
\newcommand{\bfV}{{\bf V}}
\newcommand{\bfW}{{\bf W}}
\newcommand{\bfX}{{\bf X}}
\newcommand{\bfY}{{\bf Y}}
\newcommand{\bfZ}{{\bf Z}}
\newcommand{\vphi}{{\varphi}}
\newcommand{\eps}{{\varepsilon}}
\newcommand{\Nhat}{\hat{\mbox{\tiny {\bf N}}}}
\newcommand{\ehat}{\hat{\bf e}}
\newcommand{\nhat}{\hat{\bf n}}
\newcommand{\uhat}{\hat{\bf u}}
\newcommand{\phihat}{\hat{\varphi}}
\newcommand{\xihat}{\hat{\xi}}
\newcommand{\fhat}{\hat{f}}
\newcommand{\Vhat}{\hat{V}}
\newcommand{\adj}{{\mbox{adj }}}
\newcommand{\beq}{\begin{equation}}
\newcommand{\eeq}{\end{equation}}
\newcommand{\beqs}{\begin{eqnarray}}
\newcommand{\eeqs}{\end{eqnarray}}
\newcommand{\beql}{\begin{equation} \label}

\newcommand{\normp}[1]{\| #1 \|}
\newcommand{\brho}{\boldsymbol{\rho}}
\newcommand{\expchar}[2]{\textrm{e}^{\frac{2\pi \textrm{i} #1}{#2}}}
\newcommand{\expcharconj}[2]{\textrm{e}^{-\frac{2\pi \textrm{i} #1}{#2}}}

\newcommand{\half}{\frac{1}{2}}
\newcommand{\calA}{{\cal A}}
\newcommand{\calB}{{\cal B}}
\newcommand{\calC}{{\cal C}}
\newcommand{\calD}{{\cal D}}
\newcommand{\calE}{{\cal E}}
\newcommand{\calF}{{\cal F}}
\newcommand{\calG}{{\cal G}}
\newcommand{\calH}{{\cal H}}
\newcommand{\calI}{{\cal I}}
\newcommand{\calJ}{{\cal J}}
\newcommand{\calK}{{\cal K}}
\newcommand{\calL}{{\cal L}}
\newcommand{\calM}{{\cal M}}
\newcommand{\calN}{{\cal N}}
\newcommand{\calO}{{\cal O}}
\newcommand{\calP}{{\cal P}}
\newcommand{\calQ}{{\cal Q}}
\newcommand{\calR}{{\cal R}}
\newcommand{\calS}{{\cal S}}
\newcommand{\calT}{{\cal T}}
\newcommand{\calU}{{\cal U}}
\newcommand{\calV}{{\cal V}}
\newcommand{\calW}{{\cal W}}
\newcommand{\calX}{{\cal X}}
\newcommand{\calY}{{\cal Y}}
\newcommand{\calZ}{{\cal Z}}

\newcommand{\hilb}{\mathsf{H}}
\newcommand{\thilb}{\tilde{\mathsf{H}}}
\newcommand{\hamil}{\mathfrak{H}}
\newcommand{\thrbyto}{\frac{3}{2}}
\newcommand{\Fhat}{\hat{F}}
\newcommand{\tlambda}{\tilde{\lambda}}
\newcommand{\fivbyto}{\frac{5}{2}}

%
%
%
%


\newenvironment{myproof}[1][Proof:]{\begin{trivlist}
\item[\hskip \labelsep {\bfseries #1}]}{\qed\end{trivlist}}

\newenvironment{myexample}[1][Example:]{\begin{trivlist}
\item[\hskip \labelsep {\bfseries #1}]}{\end{trivlist}}

\newenvironment{myremarks}[1][Remarks:]{\begin{trivlist}
\item[\hskip \labelsep {\bfseries #1}]}{\end{trivlist}}

\newenvironment{myremark}[1][Remark:]{\begin{trivlist}
\item[\hskip \labelsep {\bfseries #1}]}{\end{trivlist}}


\newcommand{\abs}[1]{\lvert#1\rvert}
\newcommand{\norm}[2]{\lVert#1\rVert_{#2}}
\newcommand{\innprod}[3]{\langle#1,#2\rangle_{#3}}
\newcommand{\biginnprod}[3]{\mathbf{\Big\langle}#1,#2\mathbf{\Big\rangle}_{#3}}
\newcommand{\esssup}[2]{\displaystyle\text{ess sup}_{#1}#2}
\newcommand{\Lpspc}[3]{\textsf{L}^{#1}_{#2}(#3)}
\newcommand{\infm}[2]{\displaystyle\inf_{#1}{#2}}
\newcommand{\supm}[2]{\displaystyle\sup_{#1}{#2}}
\newcommand{\argmin}[2]{\displaystyle\textrm{argmin}_{#1}{#2}}
\newcommand{\argmax}[2]{\displaystyle\textrm{argmax}_{#1}{#2}}
\newcommand{\mspan}[1]{\text{span}\{#1\}}
\newcommand{\pd}[2]{\frac{\partial#1}{\partial#2}}
\newcommand{\hpd}[3]{\frac{\partial^{#3}#1}{\partial#2^{#3}}}
\newcommand{\od}[2]{\frac{d#1}{d#2}}
\newcommand{\hod}[3]{\frac{d^{#3}#1}{d#2^{#3}}}
\newcommand{\isom}[2]{(#1|#2)}
\newcommand{\orb}[2]{\text{Orb}(#1,#2)}
\newcommand{\stab}[2]{\text{Stab}(#1,#2)}
\newcommand{\ball}[2]{\calB_{#1}(#2)}
\newcommand{\ccf}[1]{\mathsf{C}_{\mathsf{c}}(#1)}
\newcommand{\spt}[1]{\text{spt}.(#1)}
\newcommand{\mattr}[1]{\text{Tr.}(#1)}

\let\oldFootnote\footnote
\newcommand\nextToken\relax

\renewcommand\footnote[1]{%
    \oldFootnote{#1}\futurelet\nextToken\isFootnote}

\newcommand\isFootnote{%
    \ifx\footnote\nextToken\textsuperscript{,}\fi}
    
\newcommand{\dbltilde}[1]{\accentset{\approx}{#1}}


\title{Electronic Structure Prediction of Multi-million Atom Systems Through Uncertainty Quantification Enabled Transfer Learning}

\author{Shashank Pathrudkar \orcidicon{0000-0001-8546-8056}}
\affiliation{Department of Mechanical Engineering--Engineering Mechanics, Michigan Technological University}
\author{Ponkrshnan Thiagarajan \orcidicon{0000-0003-3946-3902}}
\affiliation{Department of Mechanical Engineering--Engineering Mechanics, Michigan Technological University}
\author{Shivang Agarwal \orcidicon{0000-0001-9231-5461}}
\affiliation{Department of Electrical and Computer Engineering, University of California, Los Angeles, CA 90095, USA}
\author{Amartya S. Banerjee \orcidicon{0000-0001-5916-9167}}
\email{asbanerjee@ucla.edu}
\affiliation{Department of Materials Science and Engineering, University of California, Los Angeles, CA 90095, USA}
\author{Susanta Ghosh \orcidicon{0000-0002-6262-4121}}
\email{susantag@mtu.edu}
\affiliation{Department of Mechanical Engineering--Engineering Mechanics, Michigan Technological University}
\affiliation{Faculty member of the Center for Data Sciences, Michigan Technological University}
\date{\today}

\begin{abstract}
The ground state electron density --- obtainable using Kohn-Sham Density Functional Theory (KS-DFT) simulations --- contains a wealth of material information, making its prediction via machine learning (ML) models attractive. However, the computational expense of KS-DFT scales cubically with system size which tends to stymie training data generation, making it difficult to develop quantifiably accurate ML models that are applicable across many scales and system configurations. Here, we address this fundamental challenge by employing transfer learning to leverage the multi-scale nature of the training data, while comprehensively sampling system configurations using thermalization.  Our ML models are less reliant on heuristics, and being based on Bayesian neural networks, enable uncertainty quantification. We show that our models incur significantly lower data generation costs while allowing confident --- and when verifiable, accurate --- predictions for a wide variety of bulk systems well beyond training, including systems with defects, different alloy compositions, and at unprecedented, multi-million-atom scales. Moreover, such predictions can be carried out using only modest computational resources.
\end{abstract}
\maketitle

\begin{figure*}[htbp]
    \centering
    \includegraphics[width=0.95\linewidth]{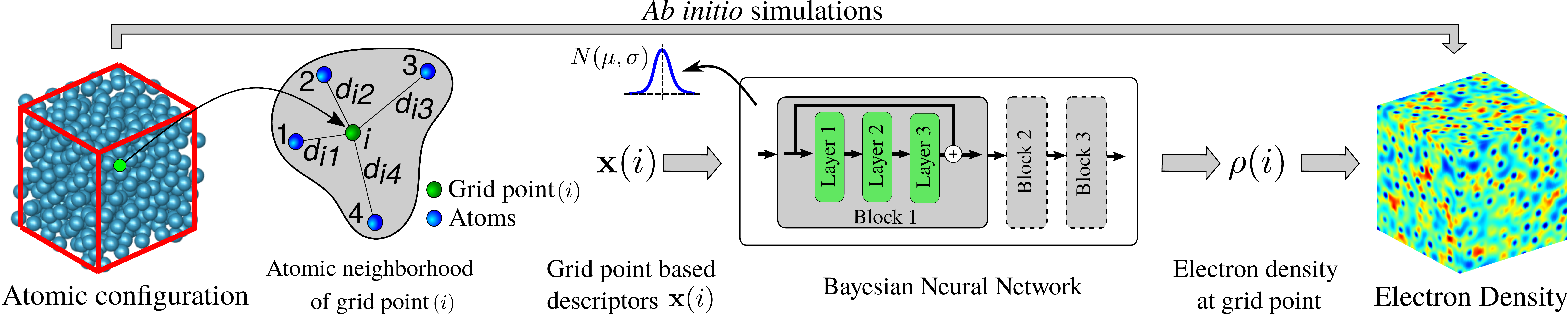}
    \caption{Overview of the present Machine Learning (ML) model. The first step is the training data generation via \textit{ab initio} simulations shown by the arrow at the top. The second step is to generate atomic neighborhood descriptors $\mathbf{x}(i)$ for each grid point, $i$, in the training configurations. The third step is to create a probabilistic map (Bayesian Neural Network with DenseNet like blocks consisting of skip connections) from atomic neighborhood descriptors $\mathbf{x}(i)$ to the charge density at the corresponding grid point $\rho(i)$. The trained model is then used for inference which includes (i) descriptor generation for all grid points in the query configuration, (ii) forward propagation through the Bayesian Neural Network, and (iii) aggregation of the point-wise charge density $\rho(i)$ to obtain the charge density field $\rho$.}
    \label{fig:model_schematic}
\end{figure*} 

\section{Introduction}
Over the past several decades, Density Functional Theory (DFT) calculations based on the Kohn-Sham formulation \citep{kohn1965self, hohenberg1964inhomogeneous} have emerged as a fundamental tool in the prediction of electronic structure. Today, they stand as the de facto workhorse of computational materials simulations \cite{van2014density, makkar2021review,  hafner2006toward, mattsson2004designing}, offering broad applicability and versatility. Although formulated in terms of orbitals, the fundamental unknown in Kohn Sham Density Functional Theory (KS-DFT) is the \emph{electron density}, from which many ground state material properties --- including structural parameters, elastic constants,  magnetic properties, phonons/vibrational spectra, etc., may be inferred. The ground state electron density is also the starting point for calculations of excited state phenomena, including those related to optical and transport properties \citep{martin2016interacting, datta2005quantum}. 

In spite of their popularity, conventional KS-DFT calculations scale in a cubic manner with respect to the number of atoms within the simulation cell, making calculations of large and complex systems computationally burdensome. To address this challenge, a number of different approaches, which vary in their computational expense and their range of applicability, have been proposed over the years. Such techniques generally avoid explicit diagonalization of the Kohn-Sham Hamiltonian in favor of computing the single particle density matrix \citep{goedecker1999linear}. Many of these methods are able to scale linearly with respect to the system size when bulk insulators or metals at high temperatures are considered \citep{goedecker1999linear, bowler2002recent, artacho1999linear,  skylaris2005introducing, pratapa2016spectral}, while others exhibit sub-quadratic scaling when used for calculations of low-dimensional materials (i.e., nanostructures)\citep{JPCM_25_295501_2013_PEXSI, lin2011selinv}. Contrary to these specialized approaches, there are only a handful of first-principles electronic structure calculation techniques that operate universally across bulk metallic, insulating, and semiconducting systems,  while performing more favorably than traditional cubic scaling methods (especially, close to room temperature). However, existing techniques in this category, e.g. \citep{motamarri2014subquadratic, CMS2009}, tend to face convergence issues due to aggressive use of density matrix truncation, and in any case, have only been demonstrated for systems containing at most a few thousand atoms, due to their overall computational cost. Keeping these developments in mind, a separate thread of research has also explored reducing computational wall times by lowering the prefactor associated with the cubic cost of Hamiltonian diagonalization, while ensuring good parallel scalability of the methods on large scale high-performance computing platforms \citep{banerjee2016chebyshev, banerjee2018two, marek2014elpa, gavini2023roadmap}. In spite of demonstrations of these and related methods to study a few large example problems (e.g. \citep{hu20222, hu2021high, dogan2023real}), their routine application to complex condensed matter systems, using modest, everyday computing resources appears infeasible.

The importance of being able to routinely predict the electronic structure of generic bulk materials, especially, metallic and semiconducting systems with a large number of representative atoms within the simulation cell, cannot be overemphasized. Computational techniques that can perform such calculations accurately and efficiently have the potential to unlock insights into a variety of material phenomena and can lead to the guided design of new materials with optimized properties. Examples of materials problems where such computational techniques can push the state-of-the-art include elucidating the core structure of defects at realistic concentrations, the electronic and magnetic properties of disordered alloys and quasicrystals \citep{wei1990electronic, jaros1985electronic, fischer2002critical, de2005electronic}, and the mechanical strength and failure characteristics of modern, compositionally complex refractory materials \citep{senkov2011mechanical, senkov2010refractory}. Moreover, such techniques are also likely to carry over to the study of low dimensional matter and help unlock the complex electronic features of emergent materials such as van-der-Waals heterostructures \citep{geim2013van} and moir\'{e} superlattices \citep{carr2020electronic}. {Notably, a separate direction of work has also explored improving Density Functional Theory predictions themselves, by trying to learn the Hohenberg-Kohn functional or exchange correlation potentials\cite{snyder2012finding, brockherde2017bypassing, kanungo2019exact}. This direction of work will not have much bearing on the discussion that follows below.}

An attractive alternative path to overcoming the cubic scaling bottleneck of KS-DFT --- one that has found much attention in recent years --- is the use of Machine Learning (ML) models as surrogates \cite{schleder2019dft, kulik2022roadmap}. Indeed, a  significant amount of research has already been devoted to the development of ML models that predict the energies and forces of atomic configurations matching with KS-DFT calculations, thus spawning ML-based \emph{interatomic potentials} that can be used for molecular dynamics calculations with \emph{ab initio} accuracy \cite{csanyi2004learn, behler2007generalized, seko2014sparse, wang2018deepmd, chen2022universal, freitas2022machine}. Parallelly, researchers have also explored direct prediction of the ground state electron density via ML models trained on the self-consistent electron density obtained from KS-DFT simulations \cite{lewis2021learning, jorgensen2022equivariant, zepeda2021deep, chandrasekaran2019solving, fiedler2023predicting, brockherde2017bypassing}. This latter approach is particularly appealing, since, in principle, the ground state density is rich in information that goes well beyond energies and atomic forces, and such details can often be extracted through simple post-processing steps. {Development of ML models of the electron density can also lead to electronic-structure-aware potentials, which are likely to overcome limitations of existing Machine Learning Interatomic Potentials, particularly in the context of reactive systems \cite{deng2023chgnet, ko2023recent}}. Having access to the electron density as an intermediate verifiable quantity is generally found to also increase the quality of ML predictions of various material properties \cite{lewis2021learning, pathrudkar2022machine}, and can allow training of additional ML models. Such models can use the density as a descriptor to predict specific quantities, such as defect properties of complex alloys \citep{arora2022charge, medasani2016predicting} and bonding information \citep{alred2018machine}. Two distinct approaches have been explored in prior studies to predict electron density via Machine Learning, differing in how they represent the density -- the output of the machine learning model.  One strategy involves representing the density by expanding it as a sum of atom-centered basis functions \cite{grisafi2018transferable, fabrizio2019electron}. The other involves predicting the electron density at each grid point in a simulation cell. Both strategies aim to predict the electron density using only the atomic coordinates as inputs. While the former strategy allows for a compact representation of the electron density, it requires the determination of an optimized basis set that is tuned to specific chemical species. It has been shown in \cite{grisafi2018transferable} that the error in the density decomposition through this strategy can be reduced to as low as 1\%. In contrast, the latter strategy does not require such optimization but poses a challenge in terms of inference - where the prediction for a single simulation cell requires inference on thousands of grid points (even at the grid points in a vacuum region). The former strategy has shown good results for molecules \cite{grisafi2018transferable} while the latter has shown great promise in density models for bulk materials especially metals \cite{zepeda2021deep, ellis2021accelerating, fiedler2023predicting}. In this work, we use the latter approach.

{For physical reasons, the predicted electron density is expected to obey transformations consistent with overall rotation and translation of the material system. Moreover, it should remain invariant under permutation of atomic indices. 
To ensure such properties, several authors have employed equivariant-neural networks \cite{thomas2018tensor, jorgensen2022equivariant, koker2023higher, nigam2022equivariant, unke2021se}. An alternative to such approaches, which is sufficient for scalar valued quantities such as electron density, is to employ invariant descriptors  \cite{koker2023higher,thomas2018tensor,chandrasekaran2019solving,zepeda2021deep}. We adopt this latter approach in this work and show through numerical examples that using invariant features and predicting electron density as a scalar valued variable indeed preserves the desired transformation properties. }

A key challenge in building surrogate models of the ground state electron density from KS-DFT calculations is the process of data generation itself, which can incur significant offline cost \citep{teh2021machine}. In recent work \citep{pathrudkar2022machine}, we have demonstrated how this issue can be addressed for chiral nanomaterials \citep{aiello2022chirality}. For such forms of matter, the presence of underlying structural symmetries allows for significant dimensionality reduction of the predicted fields, and the use of specialized algorithms for ground state KS-DFT calculations \citep{banerjee2021ab, yu2022density, agarwal2022solution}.  However, such strategies cannot be adopted for bulk materials with complex unit cells, as considered here.  For generic bulk systems, due to the confining effects of periodic boundary conditions, small unit-cell simulations alone cannot represent a wide variety of configurations. To obtain ML models that can work equally well across scales and for a variety of configurations (e.g. defects \cite{woodward2002flexible, gavini2007vacancy}), data from large systems is also essential. However, due to the aforementioned cubic scaling of KS-DFT calculations, it is relatively inexpensive to generate a lot of training data using small sized systems (say, a few tens of atoms), while larger systems (a few hundred atoms) are far more burdensome, stymieing the data generation process. Previous work on electron density prediction \cite{chandrasekaran2019solving, fiedler2023predicting} has been made possible by using data from large systems exclusively. However, this strategy is likely to fail when complex systems such as multi-principal element alloys are dealt with, due to the large computational cells required for such systems. This is especially true while studying compositional variations in such systems since such calculations are expected to increase the overall computational expense of the process significantly.

In this work, we propose a machine-learning model that accurately predicts the ground state electron density of bulk materials at any scale, while quantifying the associated uncertainties. Once trained, our model significantly outperforms conventional KS-DFT-based computations in terms of speed. To address the high cost of training data generation associated with KS-DFT simulations of larger systems --- a key challenge in developing effective ML surrogates of KS-DFT --- we adopt a transfer learning (TL) approach \citep{zhuang2020comprehensive}. Thus, our model is first trained using a large quantity of cheaply generated data from simulations of small systems, following which, a part of the model is retrained using a small amount of data from simulations of a few large systems. This strategy significantly lowers the training cost of the ML model, without compromising its accuracy. Along with the predicted electron density fields, our model also produces a detailed spatial map of the uncertainty, that enables us to assess the confidence in our predictions for very large scale systems (thousands of atoms and beyond), for which direct validation via comparison against KS-DFT simulations data is not possible. 
The uncertainty quantification (UQ) properties of our models are achieved through the use of Bayesian Neural Networks (BNNs), which systematically obtain the variance in prediction through their stochastic parameters,  and tend to regularize better than alternative approaches \cite{blundell2015weight,thiagarajan2021explanation, thiagarajan2023jensen}. They allow us to systematically judge the generalizability of our ML model, and open the door to Active Learning approaches \citep{settles2009active} that can be used to further reduce the work of data generation in the future.

To predict the electron density at a given point, the ML model encodes the local atomic neighborhood information in the form of descriptors, that are then fed as inputs to the BNN.  Our neighborhood descriptors are rather simple: they include distance and angle information from nearby atoms in the form of scalar products and avoid {choosing the basis set}  and ``handcrafted'' descriptors adopted by other workers \cite{huo2022unified, bartok2013representing, rupp2012fast, behler2007generalized, musil2021physics}. Additionally, we have carried out a systematic algorithmic procedure to select the optimal set of descriptors, thus effectively addressing the challenge associated with the high dimensionality of the descriptor-space. We explain this feature selection process in section \ref{subsec:Methods,optimaldescriptors}. To sample this descriptor space effectively, we have employed thermalization, i.e., ab initio molecular dynamics (AIMD) simulations at various temperatures, which has allowed us to carry out accurate predictions for systems far from training. Overall, our ML model reduces the use of heuristics adopted by previous workers in notable ways, making the process of ML based prediction of electronic structure much more systematic. Notably, the point-wise prediction of the electronic fields via the trained ML model, make this calculation scale linearly with respect to the system size, enabling a wide variety of calculations across scales. 

In the following sections, we demonstrate the effectiveness of our model by predicting the ground state electron density for bulk metallic and semiconducting alloy systems. In particular, we present: (i) Predictions and error estimates for systems well beyond the training data, including systems with defects and varying alloy compositions; (ii) Demonstration of the effectiveness of the transfer learning approach; (iii) Uncertainty quantification capabilities of the model, and the decomposition of the uncertainty into epistemic and aleatoric parts; and (iv) Computational advantage of the model over conventional KS-DFT calculations, and the use of the model to predict the electron density of systems containing millions of atoms.

\section{Results}


\begin{figure*}[htbp]
    \centering
    \includegraphics[width=0.8\linewidth]{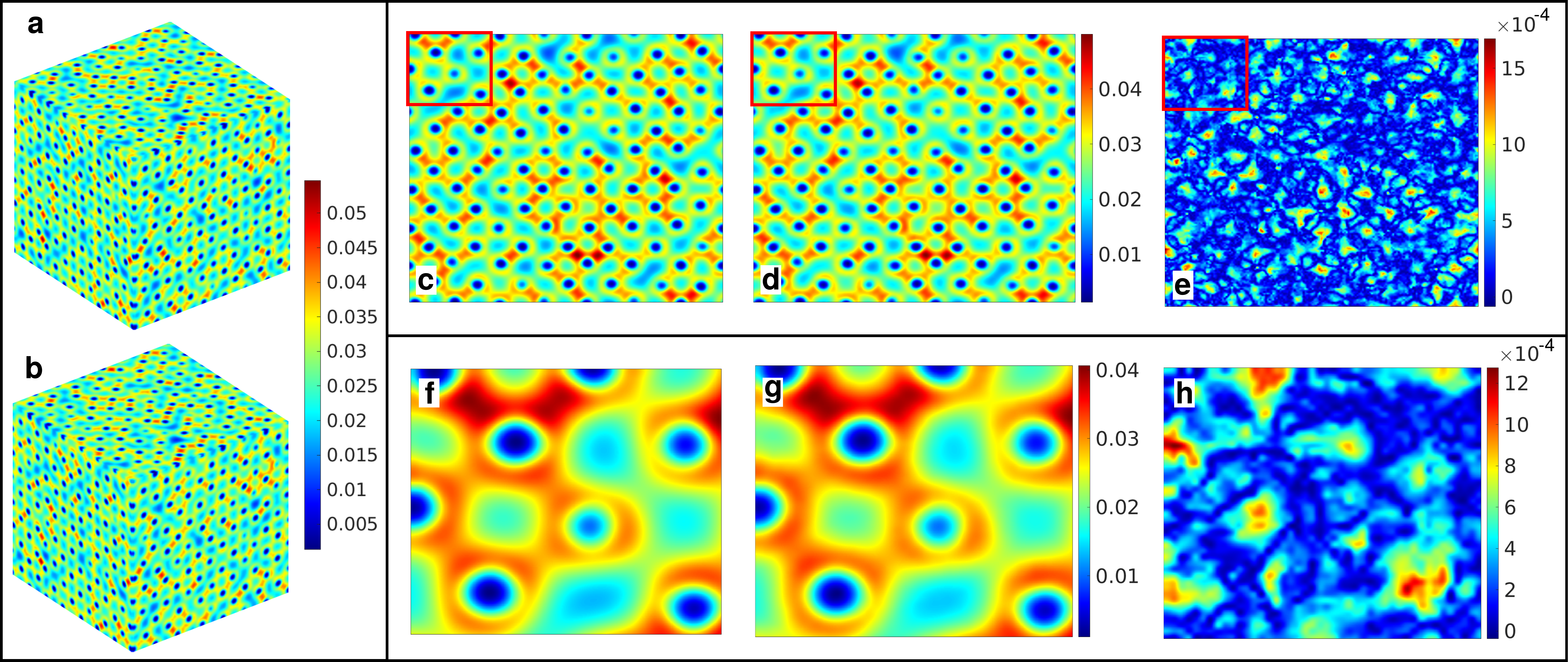}\label{fig:1372_pred}
    \caption{$1372$ atom aluminum simulation cell at $631$ K. Electron densities (a) calculated by  DFT and (b) predicted by  ML. The two-dimensional slice of (b) that has the highest mean squared error, as calculated by (c) DFT and predicted by (d) ML.  (e) Corresponding absolute error in ML with respect to DFT. (f) - (h) Magnified view of the  rectangular areas in (c) - (e) respectively. The unit for electron density is $\text{e}\, \text{Bohr}^{-3}$, where $\text{e}$ denotes the electronic charge.}
\end{figure*} 

\begin{figure*}[htbp]
    \centering
    \includegraphics[width=0.8\linewidth]{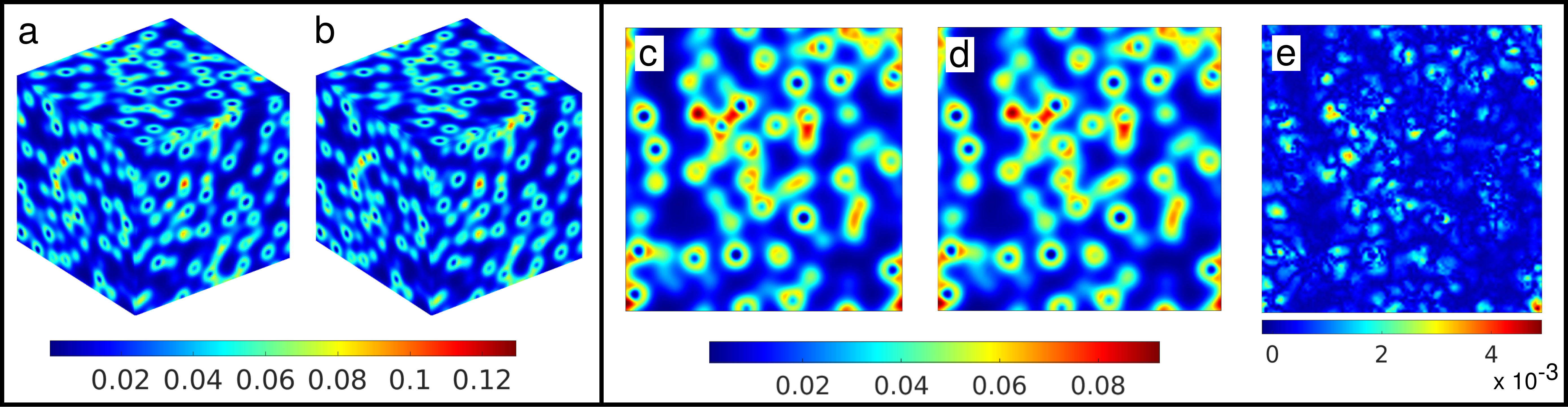}\label{fig:sige}
    \caption{512 atoms Si$_{0.5}$Ge$_{0.5}$ simulation cell at $2300$ K. Electron densities (a) calculated by  DFT and (b) predicted by  ML. The two-dimensional slice of (b) that has the highest mean squared error, as calculated by (c) DFT and predicted by (d) ML.  (e) Corresponding absolute error in ML with respect to DFT.The unit for electron density is $\text{e}\, \text{Bohr}^{-3}$, where $\text{e}$ denotes the electronic charge.}
\end{figure*} 

\begin{figure*}[htbp]
    \centering
    \includegraphics[width=0.8\linewidth]{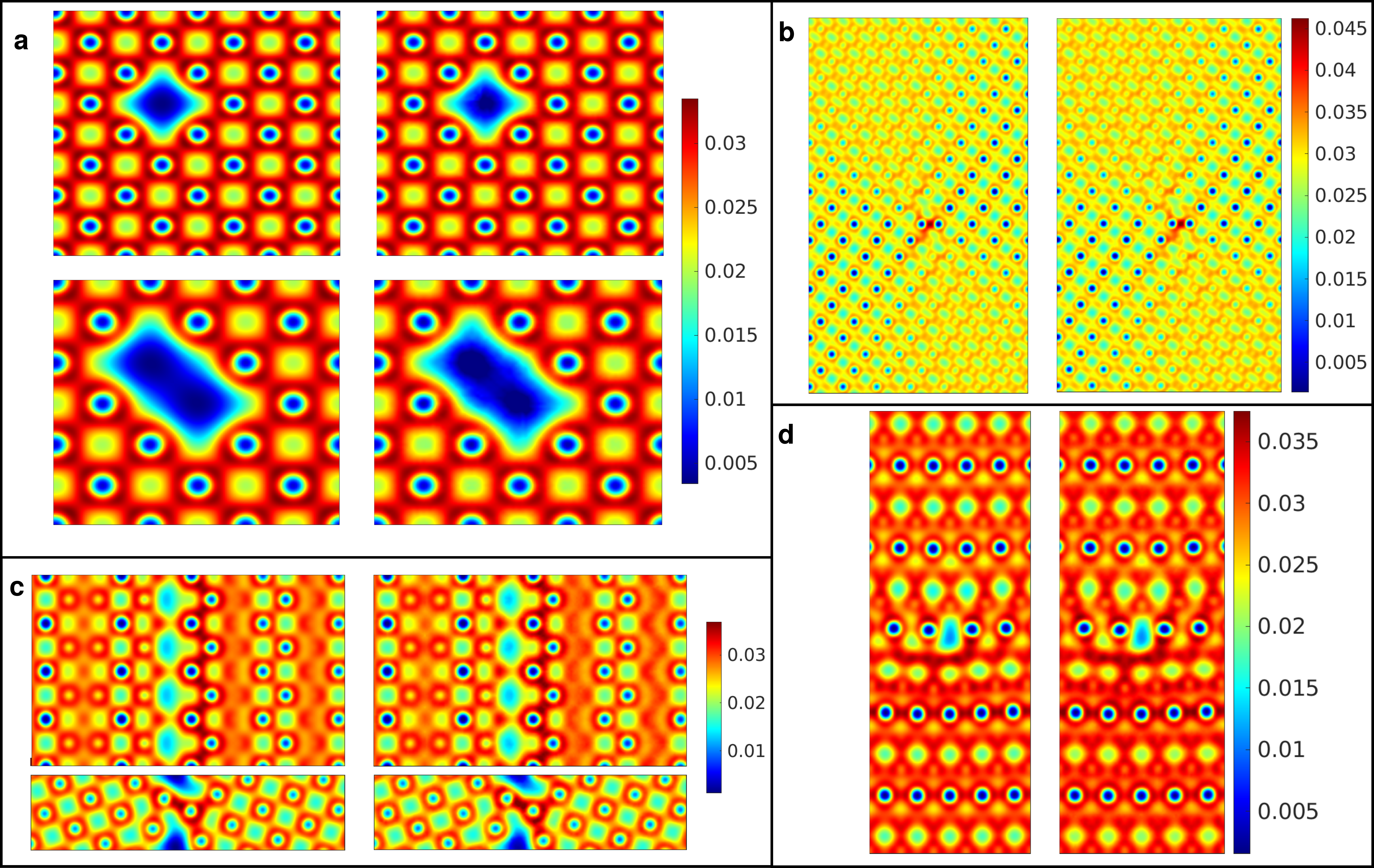}\label{fig:defect}
    \caption{Electron density contours for aluminum systems with localized and extended defects --- Left: calculated by DFT, Right: predicted by ML. (a) (Top) Mono-vacancy in 256 atom aluminum system, (Bottom) Di-Vacancy in 108 atom aluminum system, (b) (1 1 0) plane of a perfect screw dislocation in aluminum with Burgers vector $\frac{a_0}{2}[110]$, and line direction along $[110]$. The coordinate system was aligned along $[1\bar{1}2]$--$[\bar{1}11]$--$[110]$, (c) (Top) (0 1 0) plane, (bottom) (0 0 1) plane of a $[001]$ symmetric tilt grain boundary ($0\degree$ inclination angle) in aluminum, (d) Edge dislocation in aluminum with Burgers vector $\frac{a_0}{2}[110]$. The coordinate system was aligned along $[110]$--$[\bar{1}11]$--$[1\bar{1}2]$ and the dislocation was created by removing a half-plane of atoms below the glide plane. The unit for electron density is $\text{e}\, \text{Bohr}^{-3}$, where $\text{e}$ denotes the electronic charge.}
\end{figure*} 

\begin{figure*}[htbp]
    \centering
    \includegraphics[width=0.8\linewidth]{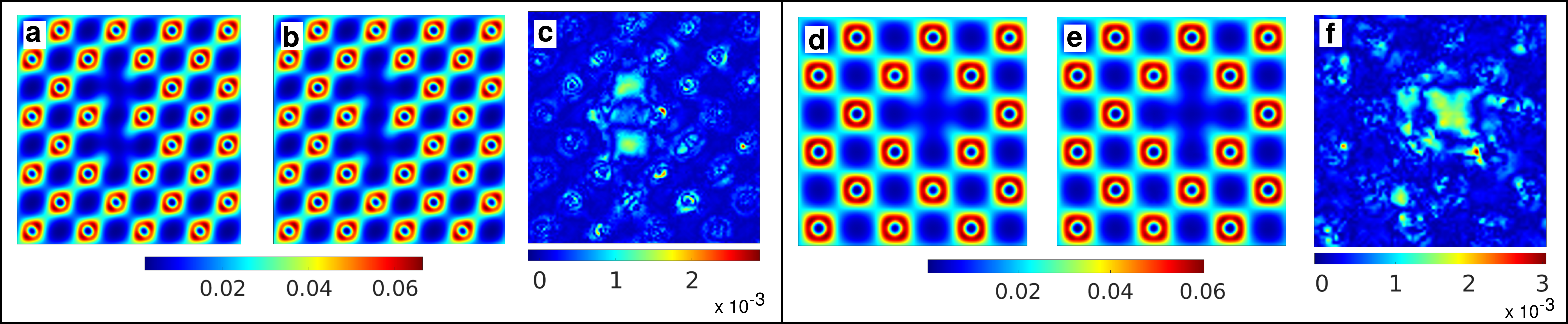}\label{fig:sige_defect}
    \caption{Electron density contours and absolute error in ML for SiGe systems with (a-c) Si double vacancy defect in 512 atom system (d-f) Ge single vacancy defect in 216 atom system. Densities (a,d) calculated by DFT, (b,e) predicted by ML, and (c,f) error in ML predictions. Note that the training data for the above systems did not include any defects. The unit for electron density is $\text{e}\, \text{Bohr}^{-3}$, where $\text{e}$ denotes the electronic charge.}
\end{figure*} 

\begin{figure*}[htbp]
    \centering    \includegraphics[width=0.75\linewidth]{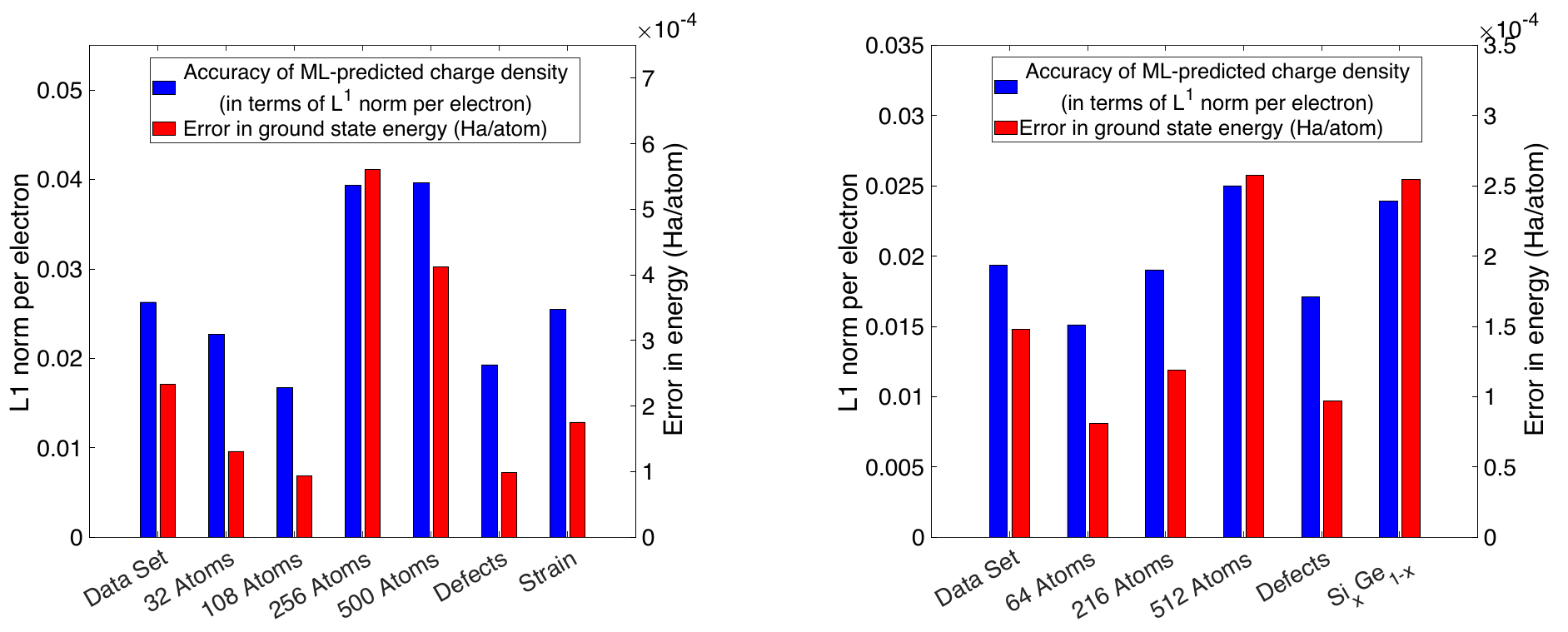}
    \caption{A comparison of the accuracy in the prediction of the charge density (in terms of the L$^1$ norm per electron between $\rho^{\text{DFT}}$ and $\rho^{\text{scaled}}$), and the error (in Ha/atom) in the ground state total energy computed using $\rho^{\text{DFT}}$ and $\rho^{\text{scaled}}$, for \ce{Al} (left), and \ce{SiGe} (right) systems. $\rho^{\text{scaled}}$ is the scaled ML predicted electron density as given in Eq. \ref{eq:scaled_density}. We observe that the errors are far better than chemical accuracy, i.e., errors below 1 kcal mol$^{-1}$ or $1.6$ milli-Hartree atom$^{-1}$, for both systems, even while considering various types of defects and compositional variations. Note that for \ce{Si_{x}Ge_{1-x}}, we chose $x=0.4,0.45,0.55,0.6$.} 
    \label{fig:Al_SiGe_bar_plot_error}
\end{figure*} 

\begin{figure*}[htbp]
    \centering
   \includegraphics[width=0.72\linewidth]{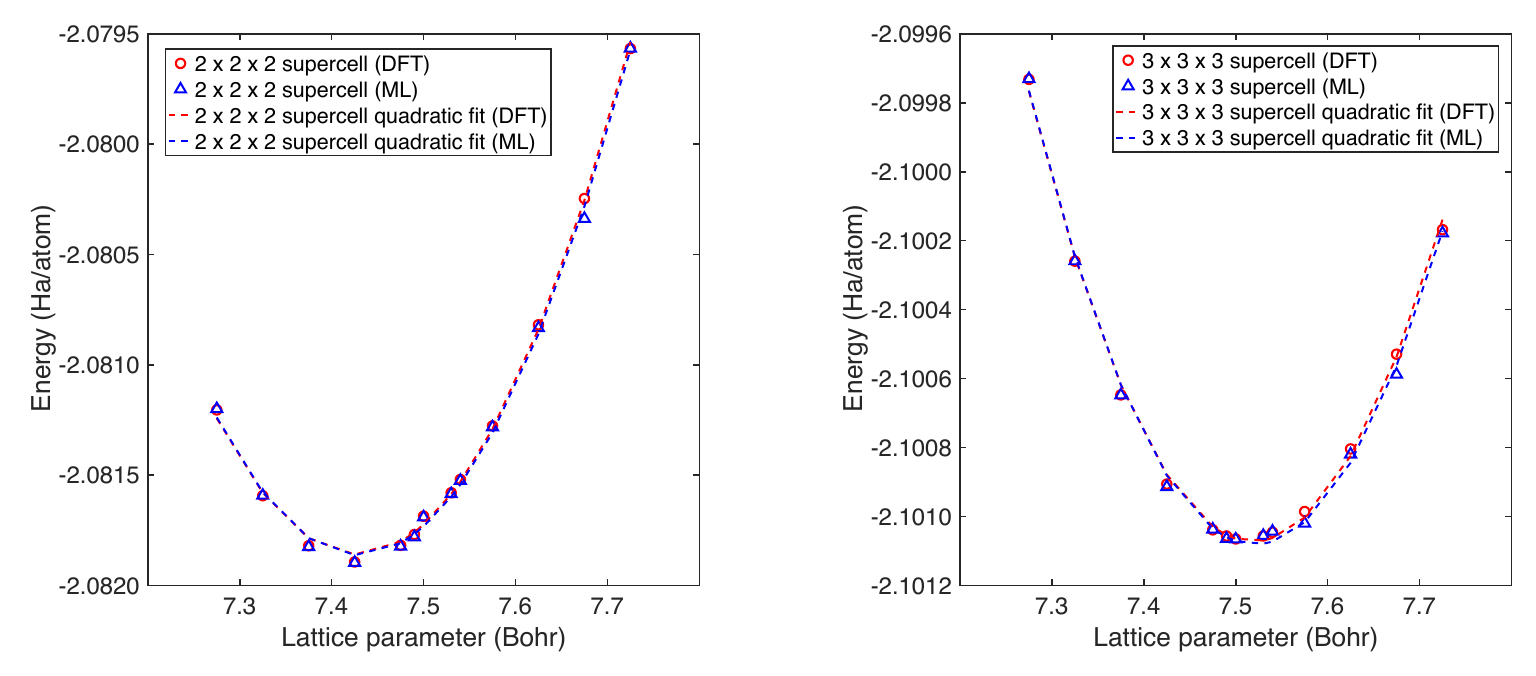}
    \caption{The energy curve with respect to different lattice parameters for a $2\times2\times2$ (left) and $3\times3\times3$ (right) supercell of aluminum atoms. Overall, we see excellent agreement in the energies (well within chemical accuracy). The lattice parameter (related to the first derivative of the energy plot) calculated in each case agrees with the DFT-calculated lattice parameter to  $ \mathcal{O} (10^{-2}) $ Bohr or better (i.e., it is accurate to a fraction of a percent). The bulk modulus calculated (related to the second derivative of the energy plot) from DFT data and ML predictions agree to within $1\%$. For the $3\times3\times3$ supercell, the bulk modulus calculated via DFT calculations is $76.39$ GPa, close to the experimental value of about $76$ GPa \cite{raju2002high}. The value calculated from ML predictions is $75.80$ GPa.}
    \label{fig:bulk_modulus}
\end{figure*} 


\begin{figure*} [htbp]
    \centering
    \includegraphics[width=0.8\linewidth]{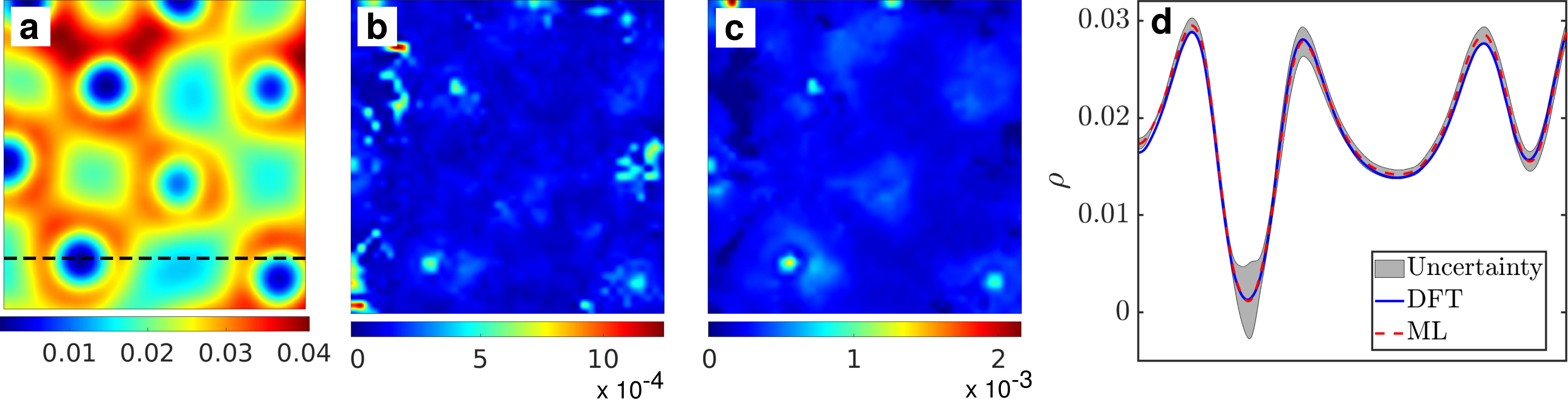}\label{fig:1372uq}\\
    \caption{Uncertainty quantification for 1372 atom aluminum system. (a) ML prediction of the electron density, (b) Epistemic Uncertainty (c) Aleatoric Uncertainty (d) Total Uncertainty shown along the dotted line from the ML prediction slice. The uncertainty represents the bound $\pm 3\sigma_{total}$, where, $\sigma_{total}$ is the total uncertainty. {The unit for electron density is $\text{e}\, \text{Bohr}^{-3}$, where $\text{e}$ denotes the electronic charge.}}
\end{figure*} 

\begin{figure*} [htbp]
    \centering
    \includegraphics[width=0.8\linewidth]{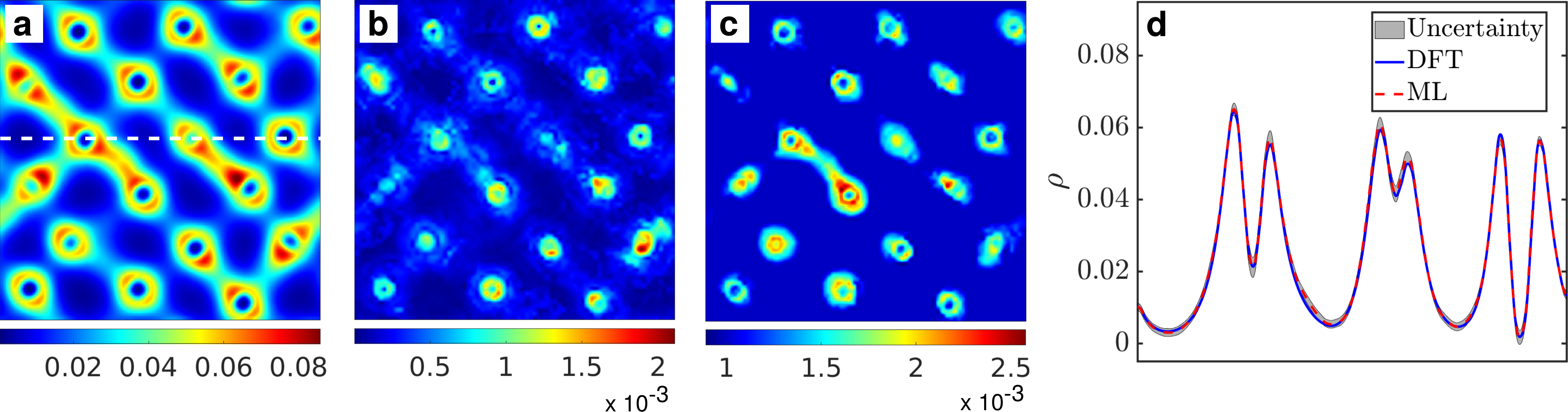}
    \label{fig:uq_SiGe_4060} 
    \caption{Uncertainty quantification for Si$_{0.4}$Ge$_{0.6}$  system. (a) ML prediction of the electron density, (b) Epistemic Uncertainty (c) Aleatoric Uncertainty (d) Total Uncertainty shown along the dotted line from the ML prediction slice. The uncertainty represents the bound $\pm 3\sigma_{total}$, where, $\sigma_{total}$ is the total uncertainty. {The unit for electron density is $\text{e}\, \text{Bohr}^{-3}$, where $\text{e}$ denotes the electronic charge.}}
\end{figure*}

\begin{figure} [htbp]
    \centering    \includegraphics[width=0.99\linewidth]{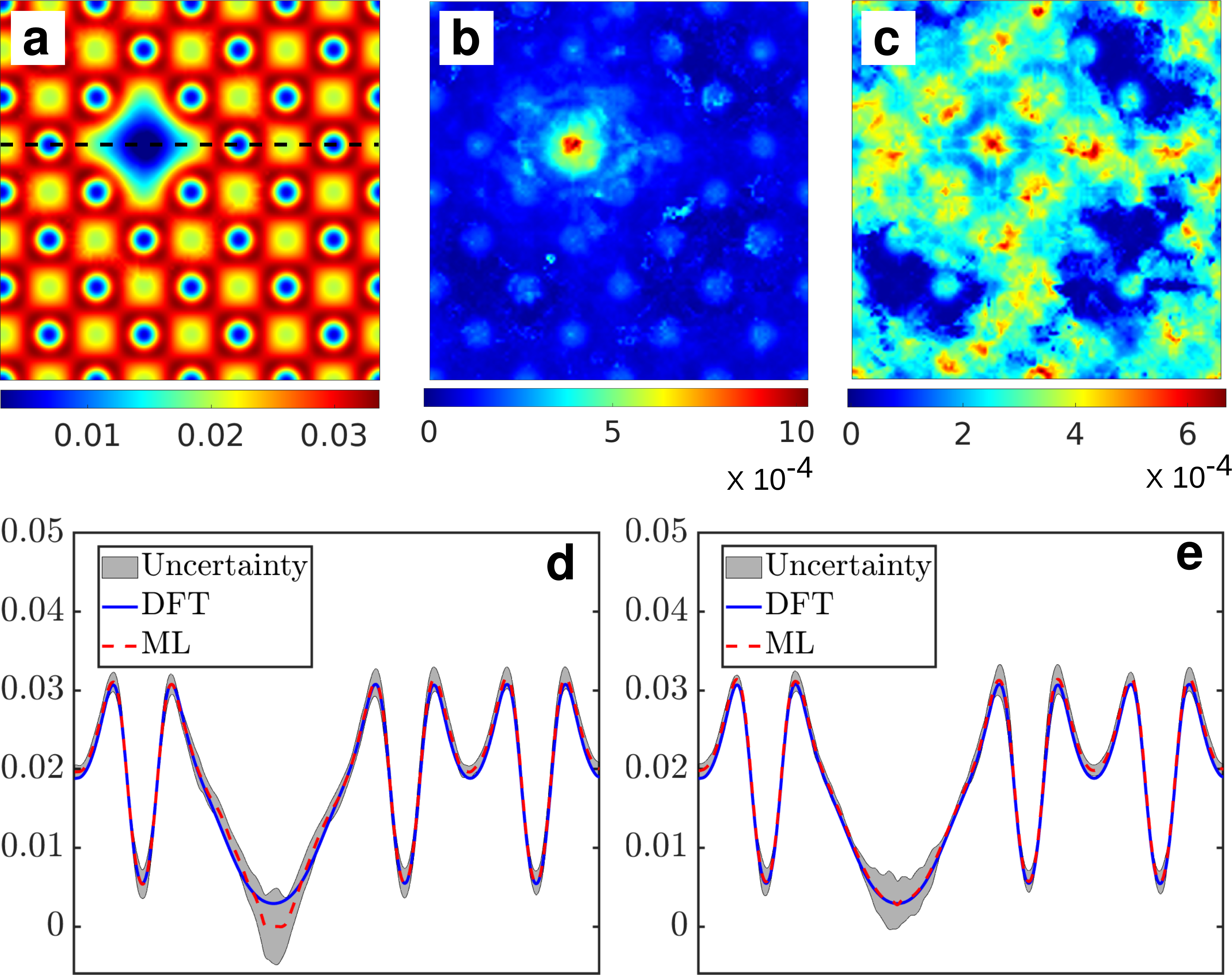}
    \caption{Uncertainty quantification for a 256 atom aluminum system with a mono vacancy defect. (a) ML prediction of the electron density shown on the defect plane, (b)) Epistemic uncertainty (c) Aleatoric uncertainty d) Uncertainty shown along the black dotted line from the ML prediction slice. The uncertainty represents the bound $\pm 3\sigma_{total}$, where, $\sigma_{total}$ is the total uncertainty. Note that the model used to make the predictions in (a-d) is not trained on the defect data, as opposed to the model used for (e), where defect data from the 108 atom aluminum system was used to train the model. The uncertainty and error at the location of the defect reduce with the addition of defect data in the training, as evident from (d) and (e). {The unit for electron density is $\text{e}\, \text{Bohr}^{-3}$, where $\text{e}$ denotes the electronic charge.}}
    \label{fig:uq_defect_train} 
\end{figure}

In this section, we present electron density predictions by the proposed machine learning (ML) model for two types of bulk materials --- pure aluminum and alloys of silicon-germanium. These serve as prototypical examples of metallic and covalently bonded semiconducting systems, respectively. These materials were chosen for their technological importance and because the nature of their electronic fields is quite distinct (see Fig.~\ref{fig:histogram_rho} in the supplemental material), thus presenting distinct challenges to the ML model. Additionally, being metallic, the aluminum systems do not show simple localized electronic features often observed in insulators \citep{resta1999electron, prodan2005nearsightedness}, further complicating electron density prediction. 

The overview of the present ML model is given in Fig. \ref{fig:model_schematic}. The models are trained using a transfer learning approach, with thermalization used to sample a variety of system configurations. 
In the case of aluminum (Al), the model is trained initially on a 32-atom and subsequently on a 108-atom system. Corresponding system sizes for silicon germanium (SiGe) are 64 and 216 atoms respectively. Details of the ML model are provided in  section~\ref{sec:Methods}.

We evaluate the performance of the ML models for a wide variety of test systems, which are by choice, well beyond the training data. This is ensured by choosing system sizes far beyond training, strained systems, systems containing defects, or alloy compositions not included in the training. We assess the  accuracy of the ML models by comparing predicted  electron density fields and ground state energies against DFT simulations.  In addition, we quantify the uncertainty in the model's predictions. 
We decompose the total uncertainty into two parts: ``aleatoric'' and ``epistemic''. 
The first is a result of inherent variability in the data, while the second is a result of insufficient knowledge about the model parameters due to limited training data. The inherent variability in the data might arise due to approximations and round-off errors incurred in the DFT simulations and calculation of the ML model descriptors. On the other hand, the modeling uncertainty arises due to the lack of or incompleteness in the data. This lack of data is inevitable since it is impossible to exhaustively sample all possible atomic configurations during the data generation process. Decomposing the total uncertainty into these two parts helps distinguish the contributions of inherent randomness and incompleteness in the data to the total uncertainty. In the present work, a  ``heteroscedastic'' noise model is used to compute the aleatoric uncertainty, which captures the spatial variation of the noise/variance in the data. 

\subsection{Error Estimation}

To evaluate the accuracy of the model, we calculated the Root Mean Squared Error (RMSE) for the entire {test} dataset, including systems of the same size as the training data {as well as sizes bigger than training data}. For aluminum, the RMSE was determined to be $4.1\times 10^{-4}$, while for SiGe, it was $7.1\times 10^{-4}$, which shows an improvement over RMSE values for Al available in \cite{chandrasekaran2019solving}. {The L$^1$ norm per electron for Aluminuum is $2.63 \times 10^{-2}$ and for SiGe it is $1.94 \times 10^{-2}$ for the test dataset.} Additionally, the normalized RMSE is obtained by dividing the RMSE value by the range of respective $\rho$ values for aluminum and SiGe. The normalized RMSE for aluminum and SiGe {test dataset} was found to be $7.9\times 10^{-3}$ for both materials. {Details of training and test dataset are presented in SM section \ref{sec:DetailsBNN}}. To assess the generalizability of the model, we evaluate the accuracy of the ML model using systems much larger than those used in training, but accessible to DFT. We consider two prototypical systems, an Aluminium system having 1372 atoms (Fig.~\ref{fig:1372_pred}) and a Silicon Germanium (Si$_{0.5}\,$Ge$_{0.5}$) system having 512 atoms  (Fig.~\ref{fig:sige}). The model shows remarkable accuracy for both of these large systems. The RMSE is $3.8\times 10^{-4} $ and $7.1 \times 10^{-4}$ for aluminum and SiGe respectively, which confirms the high accuracy of the model for system sizes beyond those used in training. 

We now evaluate the performance of the ML model for systems containing extended and localized defects, although such systems were not used in training. We consider the following defects: mono-vacancies, di-vacancies, grain boundaries, edge, and screw dislocations for \ce{Al}, and mono-vacancies and di-vacancies  for \ce{SiGe}. The electron density fields predicted by the ML models match with the DFT calculations extremely well, as shown in Figs.~\ref{fig:defect} and \ref{fig:sige_defect}. The error magnitudes (measured as the $L^{1}$ norm of the difference in electron density fields, per electron) are about $2\times10^{-2}$ (see Fig.~\ref{fig:Al_SiGe_bar_plot_error}). {The corresponding NRMSE is $7.14 \times 10^{-3}$ }.  We show in Section \ref{subsec:UQ}, that the model errors and uncertainty can be both brought down significantly, by including a single snapshot with defects, during training.   

Another stringent test of the generalizability of the ML models is performed by investigating Si$_{x}\,$Ge$_{1-x}$ alloys, for $x \neq 0.5$. Although only equi-atomic alloy compositions (i.e., $x=0.5$) were used for training, the error in prediction (measured as the $L^{1}$ norm of the difference in electron density fields, per electron) is lower than $3\times10^{-2}$ (see Fig.~\ref{fig:Al_SiGe_bar_plot_error}).  {The corresponding RMSE is $8.04 \times 10^{-4}$ and NRMSE is $7.32 \times 10^{-3}$ }. {We would like to make a note that we observed good accuracy in the immediate neighborhood  ($x = 0.4 \;\text{to}\; 0.6$) of the training data ($x = 0.5$). Prediction for $x = 0.4$ is shown in Fig. 6(ii). The prediction accuracy however decreases as we move far away from the training data composition. This generalization performance far away from the training data is expected.}  We have also carried out tests with aluminum systems subjected to volumetric strains, for which the results were similarly good.

Our electron density errors are somewhat lower than compared to the earlier works \citep{zepeda2021deep, chandrasekaran2019solving}, At the same time, thanks to the sampling and transfer learning techniques adopted by us, the amount of time spent on DFT calculations used for producing the training data is also smaller. To further put into context the errors in the electron density, we evaluate the ground state energies from the charge densities predicted by the ML model through a postprocessing step and compare these with the true ground state energies computed via DFT. Details on the methodology for postprocessing can be found in the `Methods' section, and a summary of our postprocessing results can be seen in Fig.~\ref{fig:Al_SiGe_bar_plot_error}, and in Tables \ref{tab:Al_error} and  \ref{tab:SiGe_error}, in the supplemental material. On average, the errors are well within chemical accuracy for all {test} systems considered and are generally $\mathcal{O}(10^{-4})$ Ha atom$^{-1}$, as seen in Fig. \ref{fig:Al_SiGe_bar_plot_error}. Furthermore, not only are the energies accurate, but the derivatives of the energies, e.g., with respect to the supercell lattice parameter, are found to be quite accurate as well (see Fig. \ref{fig:bulk_modulus}). This enables us to utilize the ML model to predict the optimum lattice parameter --- which is related to the first derivative of the energy curve, and the bulk modulus --- which is related to the second derivative of the energy curve, accurately. We observe that the lattice parameter is predicted accurately to a fraction of a percent, and the bulk modulus is predicted to within $1\%$ of the DFT value (which itself is close to experimental values \citep{raju2002high}). Further details can be found in the supplemental material. This demonstrates the utility of the ML models to predict not only the electron density but also other relevant physical properties.

Overall, the generalizability of our models is strongly suggestive that our use of thermalization to sample the space of atomic configurations, and the use of transfer-learning to limit training data generation of large systems are both very effective. We discuss uncertainties arising from the use of these strategies and due to the neural network model, in addition to the noise in the data, in the following sections.


\begin{figure*}[htbp]
    \centering
    \includegraphics[width=0.995\linewidth]{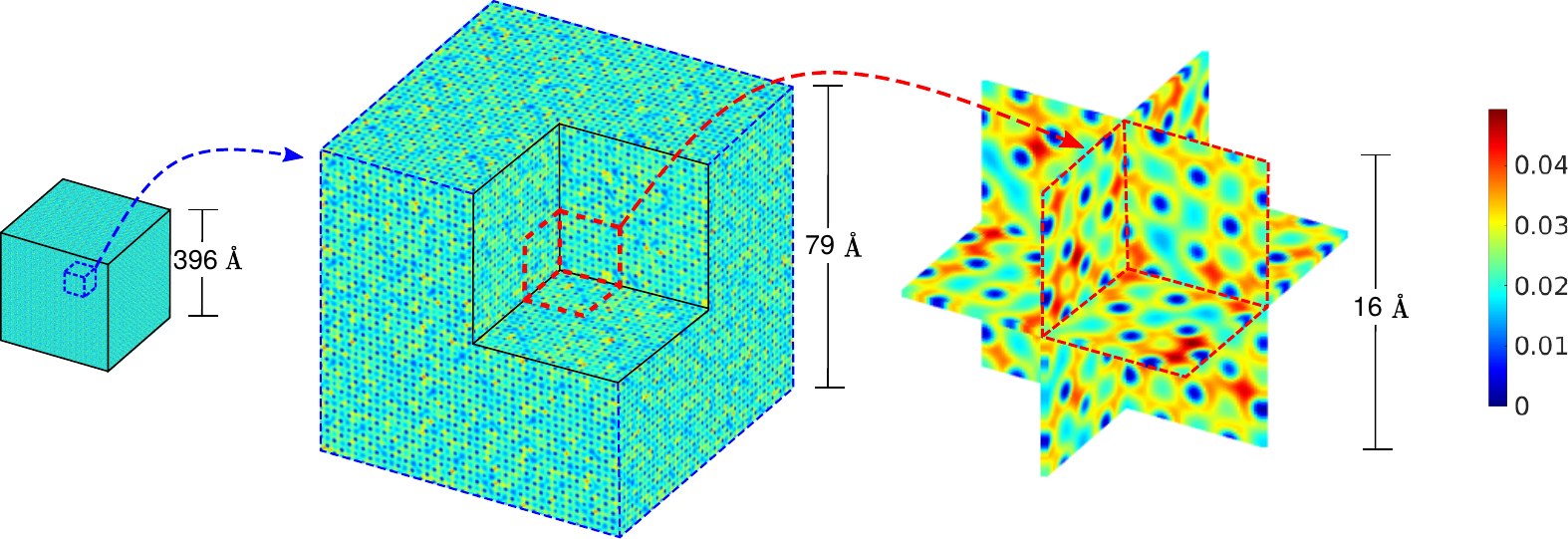}\label{fig:4M}
    \caption{Prediction of electronic structure for aluminum system containing $\approx$ $4.1$ million atoms. {The unit for electron density is $\text{e}\, \text{Bohr}^{-3}$, where $\text{e}$ denotes the electronic charge.}}
\end{figure*} 

\begin{figure*}[htbp]
    \centering
    \includegraphics[width=0.995\linewidth]{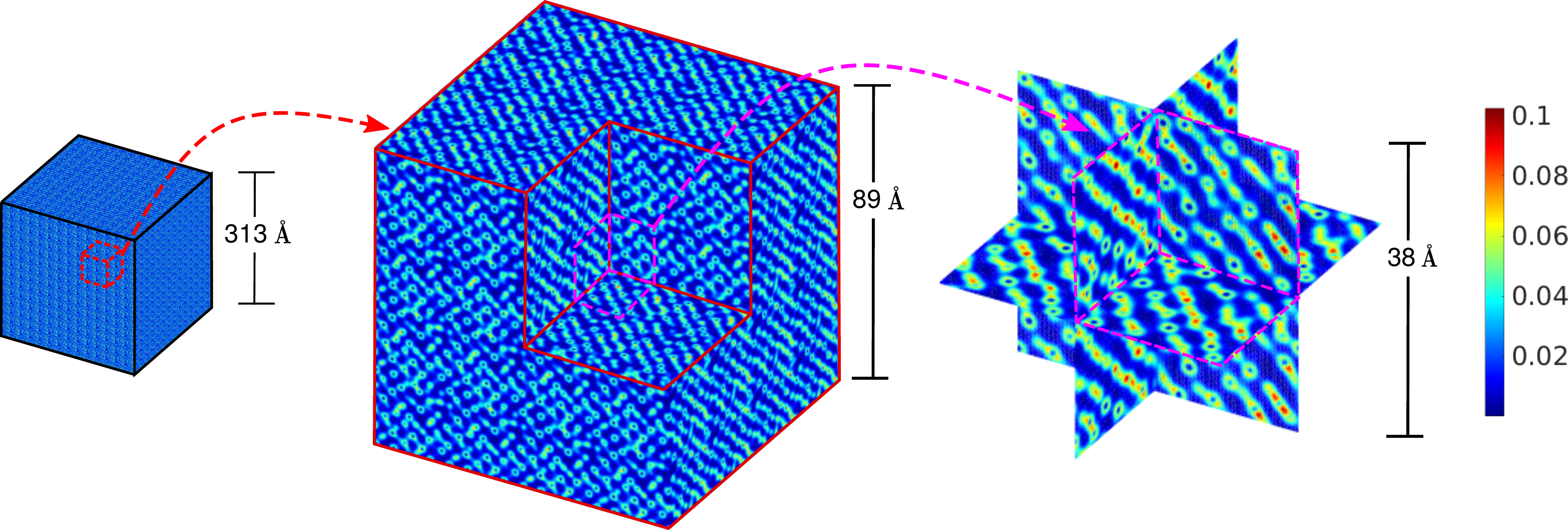}\label{fig:1M_SiGe}
    \caption{Prediction of electronic structure for $\text{Si}_{0.5}\text{Ge}_{0.5}$ system containing $\approx$ $1.4$ million atoms. {The unit for electron density is $\text{e}\, \text{Bohr}^{-3}$, where $\text{e}$ denotes the electronic charge.}}
\end{figure*}


\subsection{Uncertainty quantification} \label{subsec:UQ}
The present work uses a Bayesian Neural Network (BNN) which provides a systematic route to uncertainty quantification (UQ) through its stochastic parameters as opposed to other methods for UQ, for instance ensemble averaging \cite{fowler2019managing}. 
Estimates of epistemic and aleatoric uncertainties for the following  systems are shown: a  defect-free \ce{Al} system with $1372$ atoms (Fig.~\ref{fig:1372uq}), a $256$-atom \ce{Al} system with a mono-vacancy (Fig.~\ref{fig:uq_defect_train}(a-d)), and a \ce{Si$_{0.4}$Ge$_{0.6}$} alloy (Fig.~\ref{fig:uq_SiGe_4060}). Note, for the results in Fig.~\ref{fig:uq_defect_train}(a-d)   the training data does not contain any systems having defects, and for the results in Fig.~\ref{fig:uq_SiGe_4060} the training data contains only $50-50$ composition.  

In these systems, the aleatoric uncertainty has the same order of magnitude as the epistemic uncertainty. This implies that the uncertainty due to the inherent randomness in the data is of a similar order as the modeling uncertainty. 
The aleatoric uncertainty is significantly higher near the nuclei (Fig.~\ref{fig:1372uq} and Fig.~\ref{fig:uq_SiGe_4060}) and also higher near the vacancy (Fig~\ref{fig:uq_defect_train}). This indicates that the training data has high variability at those locations. 
The epistemic uncertainty is  high near the nucleus (Fig.~\ref{fig:1372uq} and Fig.~\ref{fig:uq_SiGe_4060}) since only a small fraction of grid points are adjacent to nuclei, resulting in the scarcity of training data for such points. The paucity of data near a nucleus is shown through the distribution of electron density in Fig.~\ref{fig:histogram_rho} of the supplemental material. 
For the system with vacancy, the aleatoric uncertainty is higher in most regions, as shown in Fig.~\ref{fig:uq_defect_train}(c). However, the epistemic uncertainty is significantly higher only at the vacancy (Fig.~\ref{fig:uq_defect_train}(b)), which might be attributed to the complete absence of data from systems with defects in the training. 

To investigate the effect of adding data from systems with defects in the training, we added a single snapshot of $108$ atom aluminum simulation with mono vacancy defect to the training data. This reduces the error at the defect site significantly and also reduces the  uncertainty (Fig.~\ref{fig:uq_defect_train}(e)). However, the uncertainty is still quite higher at the defect site because the data is biased against the defect site. That is, the amount of training data available at the defect site is much less than the data away from it. Thus, this analysis distinguishes uncertainty from inaccuracy. 

To investigate the effect of adding data from larger systems in training, we compare two models. The first model is trained with data from the $32$-atom system. The second model uses a transfer learning approach where it is initially trained using the data from the $32$-atom system and then a part of the model is retrained using data from the $108$-atom system. We observe a significant reduction in the error and in the epistemic uncertainty for the transfer learned model as compared to the one without transfer learning. The RMSE on the test system ($256$ atom) decreases by $50$\% when the model is transfer learned using $108$ atom data. The addition of the $108$-atom system's data to the training data decreases epistemic uncertainty as well since the $108$-atom system is less restricted by periodic boundary conditions than the $32$-atom system. Further, it is also statistically more similar to the larger systems used for testing as shown in Fig.~\ref{fig:histogram_rho_all} of the supplemental material. 
These findings demonstrate the effectiveness of the Bayesian Neural Network in pinpointing atomic arrangements or physical sites where more data is essential for enhancing the ML model's performance. Additionally, they highlight its ability to measure biases in the training dataset.
The total uncertainty in the predictions provides a confidence interval for the ML prediction.  This analysis provides an upper bound of uncertainty arising out of  two key heuristic strategies adopted in our ML model: data generation through thermalization of the systems and transfer learning. 

\subsection{Computational efficiency gains and confident prediction for very large system sizes} \label{sec:results_computational_advantage}
{Conventional KS-DFT calculations scale as $\mathcal{O}({N_\text{a}}^3)$ with respect to the number of atoms $N_\text{a}$, whereas, our ML model scales linearly (i.e., $\mathcal{O}(N_\text{a})$), as shown in Fig.~\ref{fig:time_comp_Al_SiGe}. This provides computational advantage for ML model over KS-DFT with increasing number of atoms.} For example, even with 500 atoms, the calculation wall times for ML model is 2 orders of magnitude lower than KS-DFT. {The linear scaling behavior of the ML model with respect to the number of atoms can be understood as follows. As the number of atoms within the simulation domain increases, so does the total simulation domain size, leading to a linear increase in the total number of grid points (keeping the mesh size constant, to maintain calculation accuracy). Since the machine learning inference is performed for each grid point, while using information from a fixed number of atoms in the local neighborhood of the grid point, the inference time is constant for each grid point. Thus the total ML prediction time scales linearly with the total number of grid points, and hence the number of atoms in the system.}

Taking advantage of this trend, the ML model can be used to predict the electronic structure for system sizes far beyond the reach of conventional calculation techniques, including systems  containing millions of atoms, as demonstrated next. We anticipate that with suitable parallel programming strategies (the ML prediction process is embarrassingly parallel) and computational infrastructure, the present strategy can be used to predict the electronic structure of systems with hundreds of millions or even billions of atoms. {Recently, there have been attempts at electronic structure predictions at million atom scales}. In \cite{zhou2023device}, a machine learning based potential is developed for germanium–antimony–tellurium alloys, effectively working for device scale systems containing over half a million atoms. Another contribution comes from Fiedler et al. \cite{fiedler2023predicting}, where they present a model predicting electronic structure for systems containing over 100,000 atoms.

We show the electron densities, as calculated by our ML model, for a four million  atom system of \ce{Al} and a one million atom system of \ce{SiGe}, in Figs.~\ref{fig:4M} and \ref{fig:1M_SiGe} respectively. In addition to predicting electron densities, we also quantify uncertainties for these systems. We found that the ML model predicts larger systems with equally high certainty as smaller systems (see Fig.~\ref{fig:4Muq} of supplemental material). 
The confidence interval obtained by the total uncertainty provides a route to assessing the reliability of predictions for these million atom systems for which KS-DFT calculations are simply not feasible. {A direct comparison of ML obtained electron density with DFT for large systems is not done till date, mainly because simulating such systems with DFT is impractical. However, recent advancements in DFT techniques hold promise for simulating large-scale systems \cite{SURYANARAYANA2018288, das2022dft, gavini2023roadmap}. In future, it will be worthwhile to compare ML predicted electron density for large systems and the electron density obtained through DFT, utilizing these recently introduced DFT techniques.}

\begin{figure} [htbp] 
    \centering
\includegraphics[width=0.8\linewidth]
{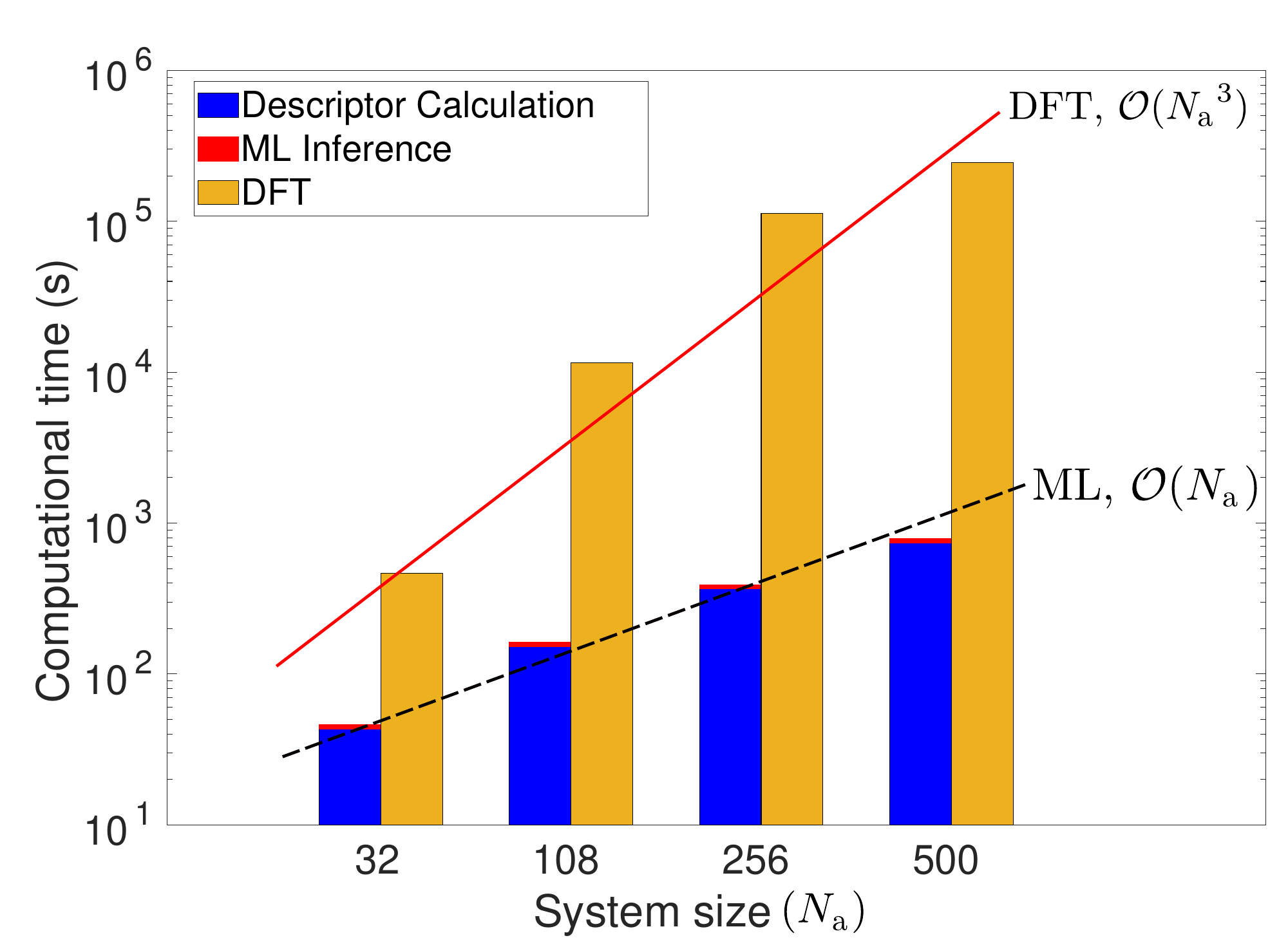}
\includegraphics[width=0.8\linewidth]
{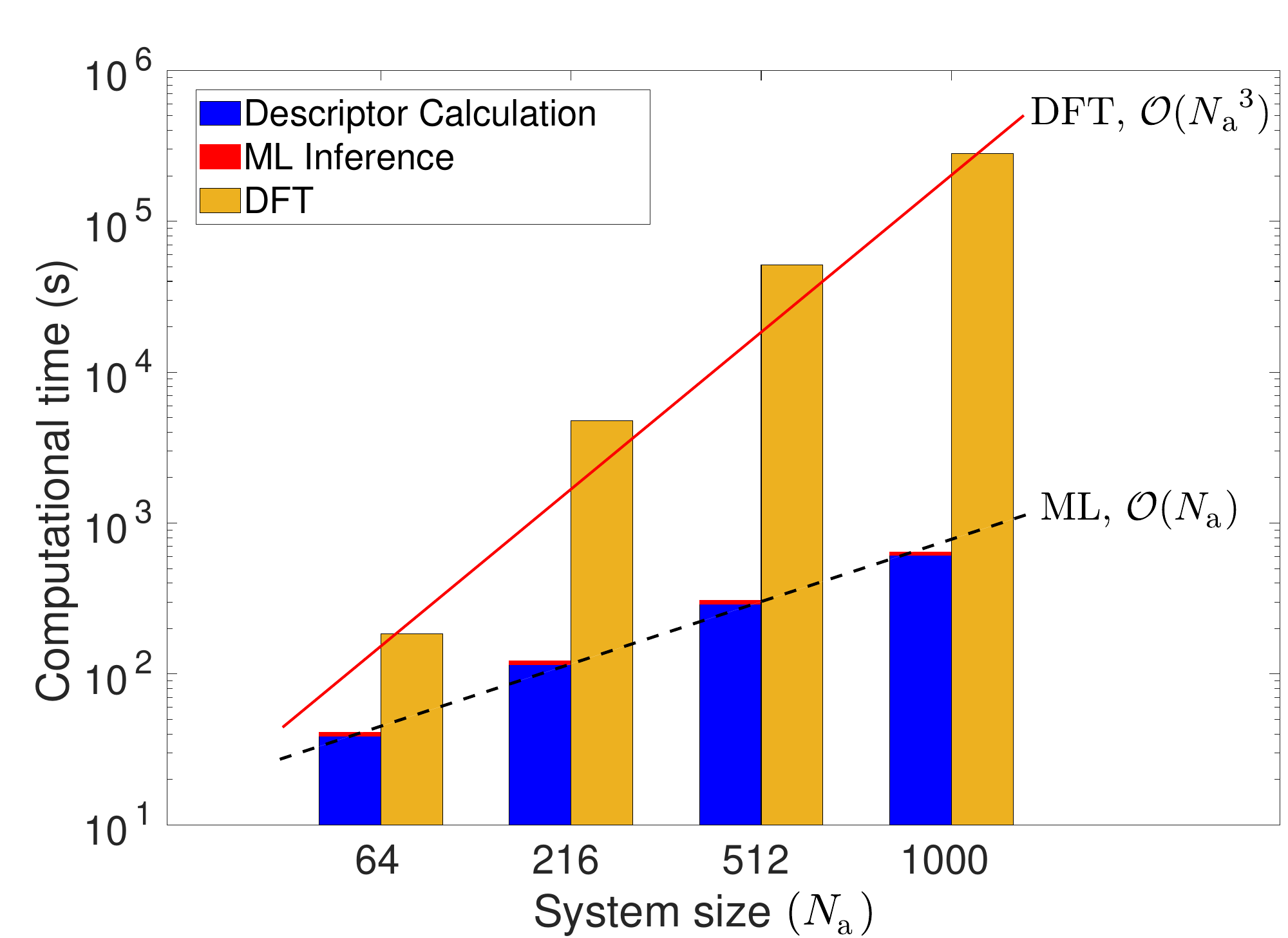}
\caption{{Computational time comparison between DFT calculations and prediction via trained ML model. (Top) Aluminum, (Bottom) SiGe. The DFT calculations scale  $\mathcal{O}({N_\text{a}}^3)$ with respect to the system size (number of atoms $N_\text{a}$), whereas, the present ML model scales linearly (i.e., $\mathcal{O}(N_\text{a})$). The time calculations were performed using the same number of CPU cores and on the same system (Perlmutter CPU). }}
    \label{fig:time_comp_Al_SiGe}
\end{figure}

\subsection{Reduction of training data generation cost via transfer learning}
One of the key challenges in developing an accurate ML model for electronic structure prediction is the high computational cost associated with the generation of the training data through KS-DFT, especially for predicting the electron density for systems across length-scales. A straightforward approach would involve data generation using  sufficiently large systems  wherein the electron density obtained from DFT is unaffected by the boundary constraints. However, simulations of larger bulk systems are significantly more expensive than smaller systems. {To address the computational burden of simulating large systems, strategies such as ``fragmentation" have been used in electronic structure calculations \cite{wang2008linearly, yang1995density}. Further, certain recent studies on Machine Learning Interatomic Potentials suggest utilizing portions of a larger system for training the models \cite{herbold2023machine, herbold2022hessian}. To the best of our knowledge, there is no corresponding  work that utilizes fragmentation in ML modeling of the electron density.} In this work, to address the issue, we employed a transfer learning (TL) approach. We first trained the ML model on smaller systems and subsequently trained a part of the neural network using data from larger systems. This strategy allows us to obtain an efficient ML model that requires fewer simulations of expensive large-scale systems compared to what would have been otherwise required without the TL approach. The effectiveness of the TL approach stems from its ability to retain information from a large quantity of cheaper, smaller scale simulation data. {We would like to note however, that the transfer learning approach is inherently bound by the practical constraints associated with simulating the largest feasible system size.}

As an illustration of the above principles, we show in Fig. \ref{fig:transfer_learning_main}, the RMSE obtained on $256$ atom data (system larger than what was used in the training data) using the TL model and the non-TL model. We also show the time required to generate the training data for both models.  For the \ce{Al} systems, we trained the TL model with $32$-atom data first and then $108$-atom data. In contrast, the non-TL model was trained only on the $108$-atom data.

The non-TL model requires significantly more $108$-atom data than the TL model to achieve a comparable RMSE on the 256-atom dataset. Moreover, the TL model's training data generation time is approximately $55\%$ less than that of the non-TL model. This represents a substantial computational saving in developing the ML model for electronic structure prediction, making the transfer learning approach a valuable tool to expedite such model development. Similar savings in training data generation time were observed for SiGe as shown in Fig. \ref{fig:transfer_learning_main}. In the case of SiGe, the TL model was first trained using 64 atom data and then transfer learned using 216 atom data. 

\begin{figure} [htbp]
    \centering
    \subfigure[Aluminum]{\includegraphics[width=0.49\linewidth]{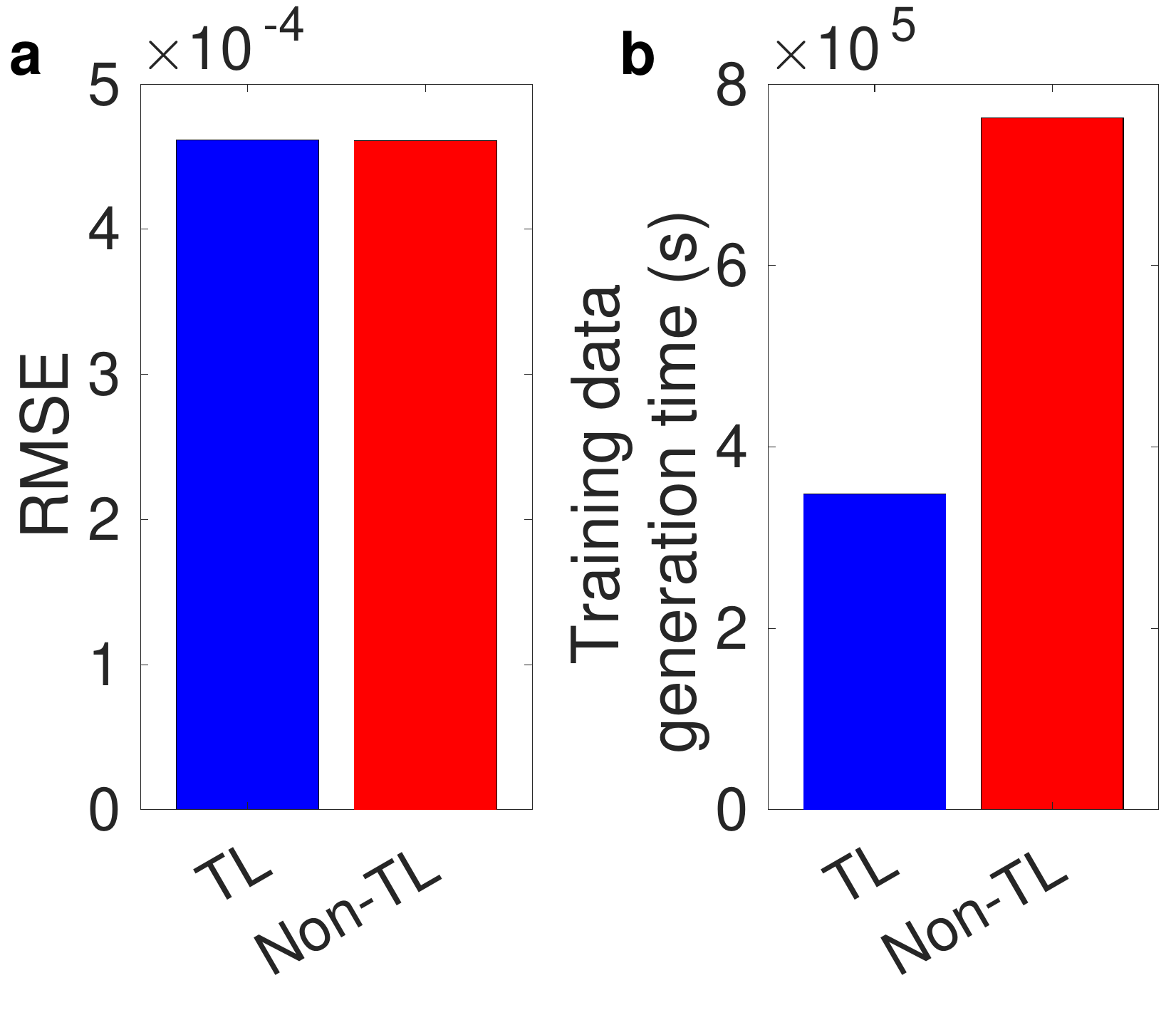}}
    \subfigure[SiGe]{\includegraphics[width=0.49\linewidth]{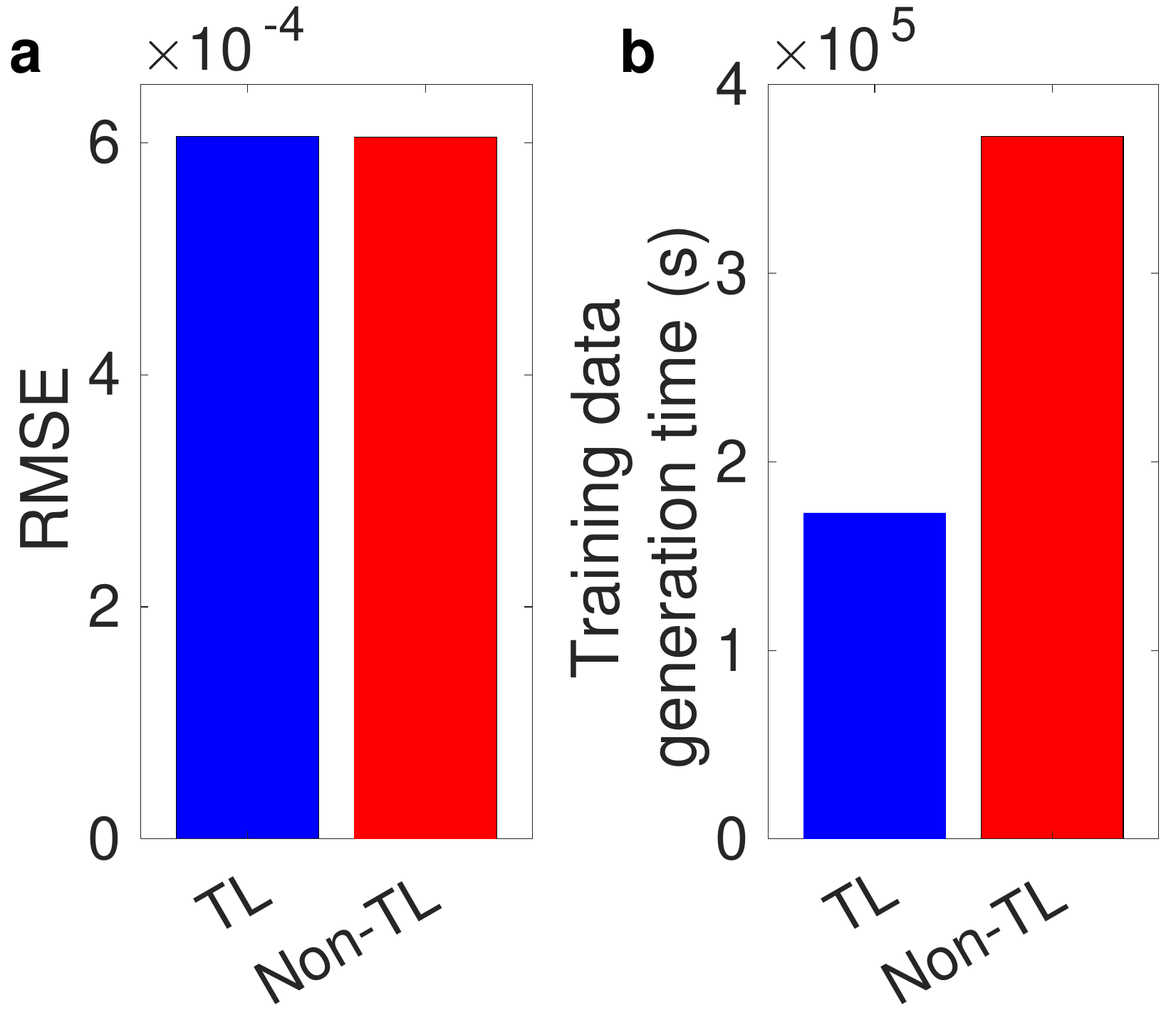}}
    \caption{Models with Transfer Learning (TL) and without Transfer Learning (Non-TL): (a) Root mean square error (RMSE) on the test dataset and (b) Computational time to generate the training data. In the case of aluminum, the TL model is trained using 32 and 108 atom data. For SiGe, the TL model was trained using 64 and 216 atom data. In the case of aluminum, the non-TL model is trained using 108 atom data. Whereas, in the case of SiGe, the non-TL model is trained using 216 atom data.}
    \label{fig:transfer_learning_main}
\end{figure} 

\section{Discussions}

We have developed an uncertainty quantification (UQ) enabled machine learning (ML) model that creates a map from the descriptors of atomic configurations to the electron densities.
We use simple scalar product-based descriptors to represent the atomic neighborhood of a point in space. These descriptors, while being easy to compute, satisfy translational, rotational, and permutational invariances. In addition, they avoid any handcrafting. We systematically identify the optimal set of descriptors for a given dataset.
Once trained, our model enables predictions across multiple length scales and supports embarrassingly parallel implementation. As far as we can tell, our work is the first attempt to systematically quantify uncertainties in ML predicted electron densities across different scales relevant to materials physics. 
To alleviate the high cost of training data generation via KS-DFT, we propose a two-pronged strategy: i) we use thermalization to comprehensively sample system configurations, leading to a highly transferable ML model; and ii) we employ transfer learning to train the model using a large amount of inexpensively generated data from small systems while retraining a part of the model using a small amount of data from more expensive  calculations of larger systems. 
The transfer learning procedure is systematically guided by the probability distributions of the data. This approach enables us to determine the maximum size of the training system, reducing dependence on heuristic selection.
As a result of these strategies, the cost of training data generation is reduced by more than $50$\%, while the models continue to be highly transferable across a large variety of material configurations. Our use of Bayesian Neural Networks (BNNs) allows the uncertainty associated with these aforementioned strategies to be accurately assessed, thus enabling confident predictions in scenarios involving millions of atoms, for which ground-truth data from conventional KS-DFT calculations is infeasible to obtain. Overall, our ML model significantly decreases the reliance on heuristics used by prior researchers, streamlining the process of ML-based electronic structure prediction and making it more systematic.

We demonstrate the versatility of the proposed machine learning models by accurately predicting electron densities for multiple materials and configurations. We focus on bulk aluminum and Silicon-Germanium alloy systems. The ML model shows remarkable accuracy when compared with DFT calculations, even for systems containing thousands of atoms.  {In the future, a similar model can be developed to test the applicability of the present descriptors and ML framework for molecules across structural and chemical space \cite{de2016comparing, bartok2017machine, deringer2021gaussian, grisafi2018symmetry}}. As mentioned above, the ML model also has excellent generalization capabilities, as it can predict electron densities for systems with localized and extended defects, and varying alloy compositions, even when the data from such systems were not included in the training. It is likely that the ensemble averaging over model parameters in the BNNs, along with comprehensive sampling of the  descriptor space via system  thermalization together contribute to the model generalization capabilities. Our findings also show a strong agreement between physical parameters calculated from the DFT and ML electron densities (e.g. lattice constants and bulk moduli). 

To rigorously quantify uncertainties in the predicted electron density, we adopt a Bayesian approach. Uncertainty quantification by a Bayesian neural network (BNN) is mathematically well-founded and offers a more reliable measure of uncertainty in comparison to non-Bayesian approaches such as the method of ensemble averaging.  
Further, we can decompose the total uncertainty into aleatoric and epistemic parts. This decomposition allows us to distinguish and analyze the contributions to the uncertainty arising from (i) inherent noise in the training data (i.e. aleatoric uncertainty)  and (ii) insufficient knowledge about the model parameters due to the lack of information in the training data (i.e. epistemic uncertainty). The aleatoric uncertainty or the noise in the data is considered irreducible, whereas the epistemic uncertainty can be reduced by collecting more training data. As mentioned earlier, the UQ capability of the model allows us to establish an upper bound on the uncertainty caused by two key heuristic strategies present in our ML model, namely, data generation via the thermalization of systems and transfer learning. 

The reliability of the ML models is apparent from the low uncertainty of its prediction for systems across various length-scales and configurations. 
Furthermore, the magnitude of uncertainty for the million-atom systems is similar to that of smaller systems for which the accuracy of the ML model has been established. This allows us to have confidence in the ML predictions of systems involving multi-million atoms, which are far beyond the reach of conventional DFT calculations. 

The ML model can achieve a remarkable speed-up of more than two orders of magnitude over DFT calculations, even for systems involving a few hundred atoms.  As shown here, these computational efficiency gains by the ML model can be further pushed to regimes involving multi-million atoms, not accessible via conventional KS-DFT calculations.

In the future, we intend to leverage the uncertainty quantification aspects of this model to implement an active learning framework. This framework will enable us to selectively generate training data, reducing the necessity of extensive datasets and significantly lowering the computational cost associated with data generation. Moreover, we anticipate that the computational efficiencies offered via the transfer learning approach, are likely to be even more dramatic while considering more complex materials systems, e.g. compositionally complex alloys \citep{ikeda2019ab, george2019high}.

\section{Methods}\label{sec:Methods}
\subsection{\textit{Ab Initio} Molecular Dynamics}
To generate training data for the model, \textit{Ab Initio} Molecular Dynamics (AIMD) simulations were performed using the finite-difference based SPARC code \cite{xu2021sparc,xu2020m, ghosh2017sparc}. We used the GGA PBE exchange-correlation functional \cite{perdew1996generalized} and ONCV pseudopotentials \cite{hamann2013optimized}. For aluminum, a mesh spacing of $0.25$ Bohrs was used while for \ce{SiGe}, a mesh spacing of $0.4$ Bohrs was used. These parameters are more than sufficient to produce accurate energies and forces for the pseudopotentials chosen, as was determined through convergence tests. A tolerance of $10^{-6}$ was used for self-consistent field (SCF) convergence and the Periodic-Pulay \citep{banerjee2016periodic} scheme was deployed for  convergence acceleration. These parameters and pseudopotential choices were seen to produce the correct lattice parameters and bulk modulus values for the systems considered here, giving us confidence that the DFT data being produced is well rooted in the materials physics.

For AIMD runs, a standard NVT-Nos\'e Hoover thermostat \cite{evans1985nose} was used,  and Fermi-Dirac smearing at an electronic temperature of $631.554$ K was applied. The time step between successive AIMD steps was $1$ femtosecond. The atomic configuration and the electron density of the system were captured at regular intervals, with sufficient temporal spacing between snapshots to avoid the collection of data from correlated atomic arrangements. To sample a larger subspace of realistic atomic configurations, we performed AIMD simulations at temperatures ranging from $315$ K to about twice the melting point of the system, i.e.\ $1866$ K for \ce{Al} and $2600$ K for \ce{SiGe}. Bulk disordered \ce{SiGe} alloy systems were generated by assigning atoms randomly to each species, consistent with the composition. 

We also generate DFT data for systems with defects and systems under strain, in order to demonstrate the ability of our ML model to predict unseen configurations. 
To this end, we tested the ML model on monovacancies and divacancies, edge and screw dislocations, and grain boundaries.  For vacancy defects, we generated monovacancies by removing an atom from a random location, and divacancies by removing two random neighboring atoms before running AIMD simulations. Edge and screw dislocations for aluminum systems were generated using Atomsk \cite{hirel2015atomsk}. Further details can be found in Fig.~\ref{fig:defect}. Grain boundary configurations were obtained based on geometric considerations of the tilt angle --- so that an overall periodic supercell could be obtained, and by removing extra atoms at the interface. For aluminum, we also tested an isotropic lattice compression and expansion of up to $5$\%; these systems were generated by scaling the lattice vectors accordingly (while holding the fractional atomic coordinates fixed). 

\subsection{Machine learning map for charge density prediction}

Our ML model maps the coordinates $\{\textbf{R}_I\}_{ I = 1}^{N_\text{a}}$ and species (with atomic numbers $\{Z_I\}_{ I = 1}^{N_\text{a}}$) of the atoms, and a set of grid points $\{\textbf{r}_i\}_{i = 1}^{N_{\text{grid}}}$ in a computational domain, to the electron density values at those grid points. Here, $N_\text{a}$ and $N_{\text{grid}}$ refer to the number of atoms and the number of grid points, within the computational domain, respectively. We compute the aforementioned map in two steps. \emph{First}, given the atomic coordinates and species information,  we calculate atomic neighborhood descriptors for each  grid point. \emph{Second}, a neural network is used to map the descriptors to the electron density at each grid point. These two steps are discussed in more detail subsequently. 

\subsection{Atomic neighborhood descriptors}  In this work, we use a set of scalar product-based descriptors to encode the local atomic environment. The scalar product-based descriptors for the grid point at $\textbf{r}_i$ consist of distance between the grid point and the atoms at $\textbf{R}_I$; and the cosine of angle at the grid point $\textbf{r}_i$ made by the pair of atoms at $\textbf{R}_I$ and $\textbf{R}_J$.  Here $i = 1, \ldots, {N_{\text{grid}}}$ and $I,J = 1, \ldots, N_\text{a}$.
We refer to the collections of distances i.e.,  $\vert\vert \textbf{r}_i - \textbf{R}_I \vert\vert$ as set \RNum{1} descriptors, and the collections of the cosines of the angles i.e., $\frac{(\textbf{r}_i - \textbf{R}_I) \cdot (\textbf{r}i - \textbf{R}_{J})}{||\textbf{r}_i - \textbf{R}_I|| \, || \textbf{r}i - \textbf{R}_{J} ||}$ are referred to as set \RNum{2} descriptors.

Higher order scalar products such as the scalar triple product, and the scalar quadruple product which involve more than two atoms at a time can also be considered. However, these additional scalar products are not included in the descriptor set in this work since they do not appear to increase the accuracy of predictions.

{Since the predicted electron density is a scalar valued variable,  invariance of the input features is sufficient to ensure equivariance of the predicted electron density under rotation, translation, and permutation of atomic indices (as mentioned in \cite{koker2023higher,thomas2018tensor}).  Since the features of our ML model are scalar products and are sorted, they are invariant with respect to rotation, translation, and permutation of atomic indices. In section I of the supplemental material we show through a numerical example that our model is indeed equivariant. Further details of the descriptor calculation are also presented in that section.}

\subsection{Selection of optimal set of descriptors}\label{subsec:Methods,optimaldescriptors}

As has been pointed out by previous work on ML prediction of electronic structure \cite{zepeda2021deep, chandrasekaran2019solving}, the nearsightedness principle \cite{kohn1996density, prodan2005nearsightedness} and screening effects \citep{ashcroft2022solid} indicate that the electron density at a grid point has little influence from atoms sufficiently far away. This suggests that only descriptors arising from atoms close enough to a grid point need to be considered in the ML model, a fact which is commensurate with our findings in Fig.~\ref{fig:feature_convergence}. 

{Using an excessive number of descriptors can increase the time required for descriptor-calculation, training, and inference, is susceptible to curse of dimensionality, and affect prediction performance \cite{Hamer_Dupont2021JMRL,Guyon_Elisseeff2003JMRL}.}
On the other hand, utilizing an insufficient number of descriptors can result in an inadequate representation of the atomic environments and lead to an inaccurate ML model.

Based on this rationale, we propose a procedure to select an optimal set of descriptors for a given atomic system. We select a set of $M$ ($M \leq N_\text{a}$) nearest atoms from the grid point to compute the descriptors and perform a convergence analysis to strike a balance between the aforementioned conditions to determine the optimal value of $M$. {It is noteworthy that the selection of optimal descriptors has been explored in previous works, in connection with Behler-Parinello symmetry functions such as \cite{gastegger2018wacsf} and \cite{imbalzano2018automatic}. These systematic procedures for descriptor selection eliminate trial-and-error operations typically involved in finalizing a descriptor set. In \cite{imbalzano2018automatic}, the authors have demonstrated for Behler-Parinello symmetry functions that using an optimal set of descriptors enhances the efficiency of machine learning models.}

For $M$ nearest atoms, we will have $N_\text{set\,\RNum{1}}$ distance descriptors, and $N_\text{set\,\RNum{2}}$  angle descriptors, with $N_\text{set\,\RNum{1}} = M$ and $N_\text{set\,\RNum{2}} \leq\, ^{M}C_{2}$ .

The total number of descriptors is $N_{\text{desc}} = N_\text{set \RNum{1}} + N_\text{set\,\RNum{2}}$. 
To optimize $N_{\text{desc}}$, we first optimize $N_\text{set \RNum{1}}$, till the error converges as shown in Fig. \ref{fig:feature_convergence}. Subsequently, we optimize $N_\text{set\,\RNum{2}}$. To do this, we consider a nearer subset of atoms of size $M_a \leq M$, and for each of these $M_a$ atoms, we consider the angle subtended at the grid point, by the atoms and their $k$ nearest neighbors. This results in $N_\text{set\,\RNum{2}} = M_a \times k$, angle based descriptors, with $M_a$ and $k$ varied to yield the best results, as shown in Fig. \ref{fig:feature_convergence}. The pseudo-code for this process can be found in Algorithms \ref{algo:1} and  \ref{algo:2} in the supplemental material. Further details on feature convergence analysis are provided in the supplemental material. 

\begin{figure}[htbp]
    \centering
   \includegraphics[width=0.9\linewidth]{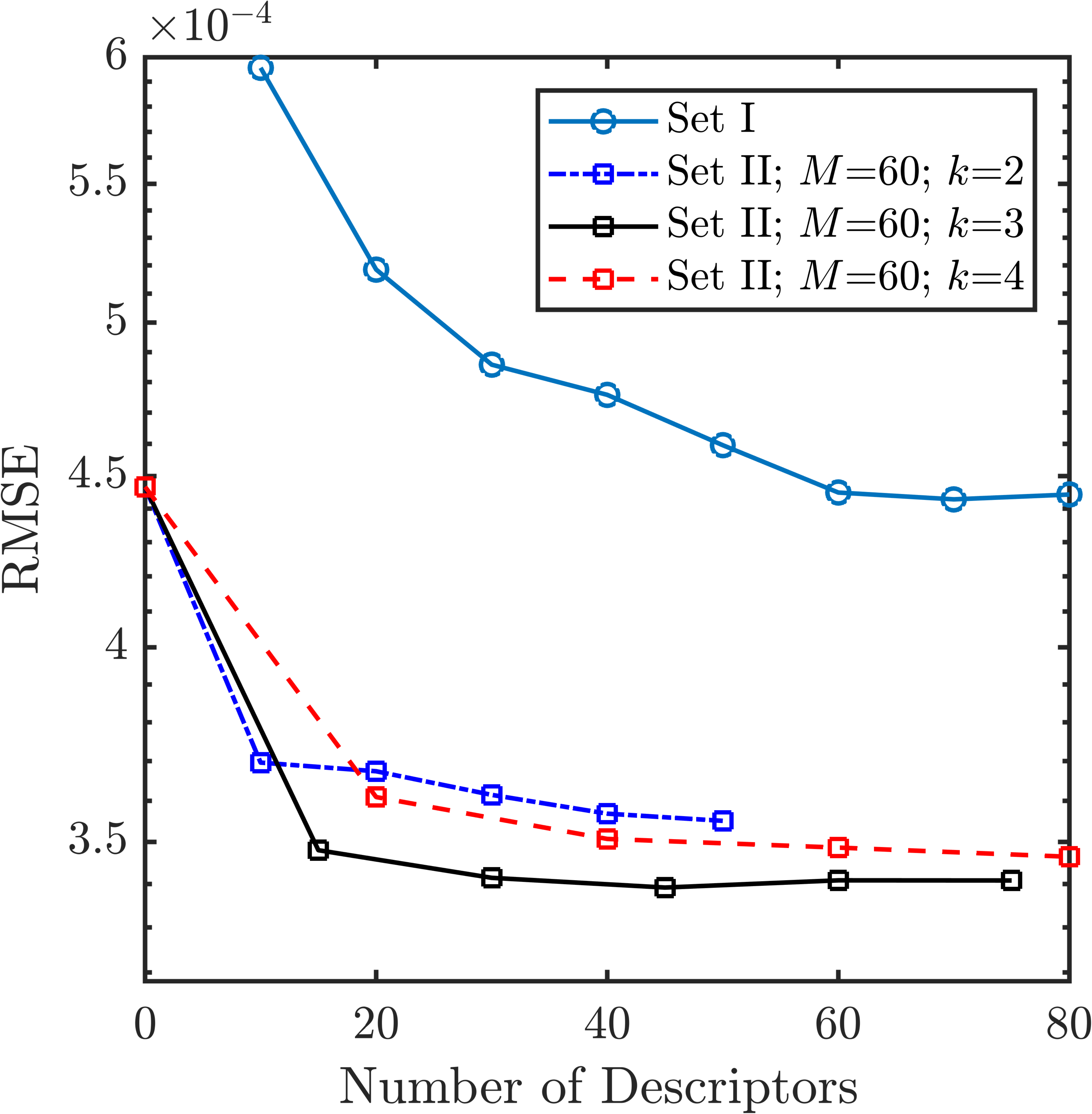}
    \caption{Convergence of error with respect to the number of descriptors, shown for aluminum. The blue line shows the convergence with respect to $N_\text{set\,\RNum{1}}$, while the other three lines show convergence with respect to $N_\text{set\,\RNum{2}}$. The optimal $N_\text{set\,\RNum{1}}$ and $N_\text{set\,\RNum{2}}$ are obtained where their test RMSE values converge. }
    \label{fig:feature_convergence}
\end{figure} 

\subsection{Bayesian Neural Network}
Bayesian Neural Networks (BNNs) have stochastic parameters in contrast to deterministic parameters used in conventional neural networks. BNNs provide a mathematically rigorous and efficient way to quantify uncertainties in their prediction.

We use a Bayesian neural network to estimate the probability $P(\rho|\mathbf{x},\mathcal{D})$ of the output electron density $\rho$ for a given input descriptor $\mathbf{x} \in \mathbb{R}^{N_\text{desc}}$ and training data set $\mathcal{D} = \{\mathbf{x}_i,\rho_i\}_{i=1}^{N_d}$. The probability is evaluated as: 
\begin{equation}\label{eq:marg_rho}
    P(\rho|\mathbf{x},\mathcal{D}) = \int_{\Omega_{w}} P(\rho|\mathbf{x},\mathbf{w}) P(\mathbf{w}|\mathcal{D}) d\mathbf{w}\,.
\end{equation}
Here $\mathbf{w} \in \Omega_{w}$ is the set of  parameters of the network and $N_d$ is the size of the training data set. Through this marginalization over parameters, a BNN provides a route to overcome modeling biases via averaging over an ensemble of networks. Given a prior distribution $P(\mathbf{w})$ on the parameters, the posterior distribution of the parameters $P(\mathbf{w}|\mathcal{D})$ are learned via the Bayes' rule as $P(\mathbf{w}|\mathcal{D}) = P(\mathcal{D}|\mathbf{w})P(\mathbf{w})/P(\mathcal{D})$, where $ P(\mathcal{D}|\mathbf{w})$ is the likelihood of the data. 

This posterior distribution of parameters $P(\mathbf{w}|\mathcal{D})$ is intractable since it involves the normalizing factor $P(\mathcal{D})$, which in turn is obtained via marginalization of the likelihood through a high dimensional integral. Therefore, it is approximated through techniques such as variational inference \cite{hinton1993keeping,graves2011practical,blundell2015weight} or Markov Chain Monte Carlo methods \cite{Zhang2020Cyclical}. In variational inference, as adopted here, a tractable distribution $q(\mathbf{w}|\boldsymbol{\theta})$ called the ``variational posterior'' is considered, which has parameters $\boldsymbol{\theta}$. For instance, if the variational posterior is a Gaussian distribution the corresponding parameters are its mean and standard deviation, $\boldsymbol{\theta} = (\boldsymbol{\mu}_\theta,\boldsymbol{\sigma}_\theta)$. 
The optimal value of parameters $\boldsymbol{\theta}$ is obtained by minimizing the statistical dissimilarity between the true and variational posterior distributions. The  dissimilarity is measured through the KL divergence $ \text{KL}\left[ q(\mathbf{w}|\boldsymbol{\theta})\:||\: P(\mathbf{w}|\mathcal{D}) \right]$. This yields the following optimization problem:

\begin{equation} \label{eq:kl_min}
\begin{split}
  \boldsymbol{\theta}^* &= \arg \min_{\boldsymbol{\theta}} \text{KL}\left[ q(\mathbf{w}|\boldsymbol{\theta})\:||\: P(\mathbf{w}|\mathcal{D}) \right]\\
    &=  \arg \min_{\boldsymbol{\theta}} \int q(\mathbf{w}|\boldsymbol{\theta}) \log \left[\frac{q(\mathbf{w}|\boldsymbol{\theta})}{P(\mathbf{w})P(\mathcal{D}|\mathbf{w})} P(\mathcal{D})\right] d\mathbf{w}\,.   
\end{split}
\end{equation}
This leads to the following loss function for BNN that has to be minimized:

\begin{equation}\label{eq:elbo}
    \mathcal{F}_\text{KL}(\mathcal{D},\boldsymbol{\theta}) = \text{KL}\left[ q(\mathbf{w}|\boldsymbol{\theta})\:||\: P(\mathbf{w}) \right] - \mathbb{E}_{q(\mathbf{w}|\boldsymbol{\theta})} [\log P(\mathcal{D}|\mathbf{w})]\,.
\end{equation}
This loss function balances the simplicity of the prior and the complexity of the data through its first and second terms respectively, yielding regularization \cite{blundell2015weight,thiagarajan2021explanation}. 

Once the parameters $\boldsymbol{\theta}$ are learned, the BNNs can predict the charge density at any new input descriptor $\mathbf{x}$. In this work, the mean of the parameters ($\boldsymbol{\mu}_\theta$) are used to make point estimate predictions of the BNN.

 \subsection{Uncertainty quantification}
The variance in the output distribution $P(\rho|\mathbf{x},\mathcal{D})$ in Eq.~\eqref{eq:marg_rho} is the measure of uncertainty in the BNN's prediction. Samples from this output distribution can be drawn in three steps: In the first step, a $j^{th}$ sample of the set of parameters, $\widehat{\mathbf{w}}_{j=1,...,N_s}$, is drawn from the variational posterior $q(\mathbf{w}|\boldsymbol{\theta})$ which approximates the posterior distribution of parameters $P(\mathbf{w}|\mathcal{D})$. Here, $N_s$ is the number of samples drawn from the variational posterior of parameters. In the second step, the sampled parameters are used to perform inference of the BNN ($f_N$) to obtain the $j^{th}$ prediction $\widehat{\rho}_j = f_N^\mathbf{\widehat{w}_j}(\mathbf{x}) $. In the third step, the likelihood is assumed to be a Gaussian distribution: $P(\rho|\mathbf{x},\mathbf{\widehat{w}_j}) = \mathcal{N}(\widehat{\rho}_j,\sigma(\mathbf{x}))$,  whose mean is given by the BNN's prediction, $\widehat{\rho}_j$, and standard deviation by a heterogenous observation noise, $\sigma(\mathbf{x})$. A sample is drawn from this Gaussian distribution $\mathcal{N}(\widehat{\rho}_j,\sigma(\mathbf{x}))$ that approximates a sample from the distribution $P(\rho|\mathbf{x},\mathcal{D})$.  
The total variance of such samples can be expressed as: 

\begin{equation} \label{eq:var}
    \text{var}(\rho) = \sigma^2(\mathbf{x}) +\left[ \frac{1}{N_s} \sum_{j=1}^{N_s} \left(\widehat{\rho}_j\right)^2 - \left(\mathbb{E}(\widehat{\rho}_j)\right)^2 \right]\,.
\end{equation}
Here, $\mathbb{E}(\widehat{\rho}_j) = \frac{1}{N_s} \sum_{j=1}^{N_s}f_N^\mathbf{\widehat{w}_j}(\mathbf{x})$. The first term, $\sigma^2(\mathbf{x})$, in Eq.~\eqref{eq:var} is the aleatoric uncertainty that represents  the inherent noise in the data and is considered irreducible. The second term (in the square brackets) in Eq.~\eqref{eq:var} is the epistemic uncertainty, that quantifies the modeling  uncertainty.

In this work, the aleatoric uncertainty is learned via the BNN model along with the charge densities $\rho$. Therefore, for each input $\mathbf{x}$, the BNN learns two outputs: $f_N^\mathbf{w}(\mathbf{x})$ and $\sigma (\mathbf{x})$. For a Gaussian likelihood, the noise $\sigma$ is learned through the likelihood term of the loss function Eq.~\eqref{eq:elbo} following \cite{kendall2017uncertainties} as:
\begin{align}
 \log P(\mathcal{D}|\mathbf{w}) = \sum_{i=1}^{N_d} -\frac{1}{2} \log \sigma_i^2 - \frac{1}{2\sigma_i^2}(f_N^\mathbf{w}(\mathbf{x}_i)-\rho_i)^2\,.
\end{align}
Here, $N_d$ is the size of the training data set.
The aleatoric uncertainty,  $\sigma$, enables the loss to adapt to the data. The network learns to reduce the effect of erroneous labels by learning a higher value for $\sigma^2$, which makes the network more robust or less susceptible to noise. On the other hand, the model is penalized for predicting high uncertainties for all points through the $\log \sigma^2$ term. 

The epistemic uncertainty is computed by evaluating the second term of Eq.\eqref{eq:var}, via sampling $\mathbf{\widehat{w}_j}$ from the variational posterior. 
 
\subsection{Transfer Learning using multi-scale data}
Conventional DFT simulations for smaller systems are considerably cheaper than those for larger systems, as the computational cost scales cubically with the number of atoms present in the simulation cell.
However, the ML models cannot be trained using simulation data from small systems alone. This is because, smaller systems are far more constrained in the number of atomic configurations they can adopt, thus limiting their utility in simulating a wide variety of materials phenomena. Additionally, the electron density from simulations of smaller systems  differs from that of larger systems, due to the effects of periodic boundary conditions. 

To predict accurately across all length scales while reducing the cost of training data generation via DFT simulations, we use a transfer learning approach here. Transfer learning is a machine learning technique where a network, initially trained on a substantial amount of data, is later fine-tuned on a smaller dataset for a different task, with only the last few layers being updated while the earlier layers remain unaltered \cite{ zhuang2020comprehensive}. The initial layers (called ``frozen layers'') capture salient features of the inputs from the  large dataset, while the re-trained layers act as decision-makers and adapt to the new problem.

Transfer learning has been used in training neural network potentials, first on Density Functional Theory (DFT) data, and subsequently using datasets generated using more accurate, but expensive quantum chemistry models \cite{smith2019approaching}.  In contrast, in this work, transfer learning is employed to leverage the multi-scale aspects of the problem.
Specifically, the present transfer learning approach leverages the statistical  dissimilarity in data distributions between various systems and the largest system. This process is employed to systematically select the training data, ultimately reducing reliance on heuristics, as detailed in the supplemental material (see Fig. \ref{fig:histogram_rho_all}). 
This approach allows us to make electron density predictions across scales and system configurations, while significantly reducing the cost of training data generation.

In the case of aluminum, at first, we train the model using a large amount of data from DFT simulations of (smaller) $32$-atom systems. Subsequently, we freeze the initial one-third layers of the model and re-train the remaining layers of the model using a smaller amount ($40$\%) of data from simulations of (larger) $108$-atom systems. Further training using data from larger bulk systems was not performed, since the procedure described above already provides good accuracy (Figs.~\ref{fig:Al_SiGe_bar_plot_error},\ref{fig:transfer_learning_main}), which we attribute to the statistical similarity of the electron density of $108$ atom systems and those with more atoms (Fig.\ref{fig:histogram_rho_all} of the supplemental material). A similar transfer learning procedure is used for the SiGe model, where we initially train with data from $64$-atom systems and subsequently retrain using data from $216$-atom systems. Overall, due to the non-linear data generation cost using DFT simulations, the transfer learning  approach reduces training data generation time by over $50$\%.

\subsection{Postprocessing of ML predicted electron density}
One way to test the accuracy of the ML models is to compute quantities of interest (such as the total ground state energy, exchange-correlation energy, and Fermi level) using the predicted electron density, $\rho^{\text{ML}}$. Although information about the total charge in the system is included in the prediction, it is generally good practice to first re-scale the electron density before postprocessing \cite{alred2018machine, pathrudkar2022machine}, as follows:

\begin{align}
    \rho^{\text{scaled}}\left(\textbf{r}\right) = \rho^{\text{ML}}(\textbf{r})\frac{N_{\text{e}}}{\displaystyle\int_{\Omega}\rho^{\text{ML}}(\textbf{r})d\textbf{r}}\,.
    \label{eq:scaled_density}
\end{align}
Here, $\Omega$ is the periodic supercell used in the calculations, and $N_{\text{e}}$ is the number of electrons in the system. Using this scaled density, the Kohn-Sham Hamiltonian is set up within the SPARC code framework, which was also used for data generation via AIMD simulations \cite{xu2021sparc,xu2020m, ghosh2017sparc}. A single step of diagonalization is then performed, and the energy of the system is computed using the Harris-Foulkes formula \citep{harris1985simplified, foulkes1989tight}. The errors in predicting $\rho^{\text{ML}}(\textbf{r})$, and the ground state energy thus calculated, can be seen in Fig. \ref{fig:Al_SiGe_bar_plot_error}. More detailed error values can be found in Table \ref{tab:Al_error} and Table \ref{tab:SiGe_error} in the supplemental material.

\begin{acknowledgments}
This work was supported by grant DE-SC0023432 funded by the U.S. Department of Energy, Office of Science. This research used resources of the National Energy Research Scientific Computing Center, a DOE Office of Science User Facility supported by the Office of Science of the U.S. Department of Energy under Contract No.~DE-AC02-05CH11231, using NERSC awards BES-ERCAP0025205, BES-ERCAP0025168, and BES-ERCAP0028072.
SG and SP acknowledge Research Excellence Fund from MTU. ASB acknowledges startup support from the Samueli School Of Engineering at UCLA, as well as funding from UCLA’s Council on Research (COR) Faculty Research Grant. ASB also acknowledges support through a UCLA SoHub seed grant and a Faculty Career Development Award from UCLA's Office of Equity, Diversity, and Inclusion. The authors would like to thank UCLA's Institute for Digital Research and Education (IDRE), the Superior HPC facility at MTU, and the Applied Computing GPU cluster at MTU for making available some of the computing resources used in this work. ASB would like to acknowledge valuable discussions with Lin Lin (University of California, Berkeley) and Ellad Tadmor (University of Minnesota, Minneapolis). {SP would like to acknowledge the Doctoral Finishing Fellowship awarded by the Graduate School at Michigan Technological University.} The authors would like to acknowledge Nikhil Admal and Himanshu Joshi (University of Illinois Urbana-Champaign) for sharing some of the atomistic configurations with defects that were used in this work. 
\end{acknowledgments}

\subsection*{Author contributions}
SP developed the machine learning framework. PT worked on uncertainty quantification aspects of the machine learning model. SA performed DFT calculations and subsequent post-processing. ASB and SG were involved in conceptualization, methodological design, supervision, and securing funding/resources. All authors contributed to writing the manuscript.

\subsection*{Competing Interests}
\noindent The authors declare no competing interests.

\subsection*{Data Availability}\label{subsec:Data_availability}
\noindent Raw data were generated at Hoffman2 High-Performance Compute Cluster at UCLA's Institute for Digital Research and Education (IDRE) and National Energy Research Scientific Computing Center (NERSC). Derived data supporting the findings of this study are available from the corresponding author upon request.

\subsection*{Code Availability}\label{subsec:Code_availability}
Codes supporting the findings of this study are available from the corresponding author upon reasonable request.

\section*{Supplemental Material}
\begin{figure*}[htbp]
    \centering    \includegraphics[width=0.9\linewidth]{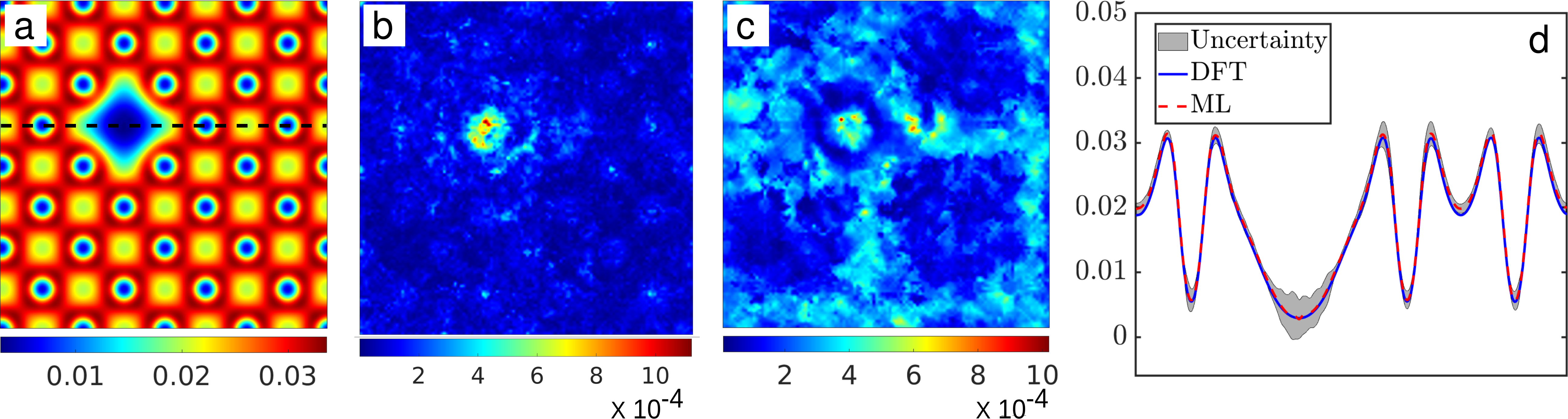}
    \caption{Uncertainty quantification for a 256 atom aluminum system with mono vacancy defect. From left: i) ML prediction of the electron density shown on the defect plane, ii) Epistemic uncertainty iii) Aleatoric uncertainty iv) Uncertainty shown on the black dotted line from the ML prediction slice. The uncertainty represents the bound $\pm 3\sigma$, where, $\sigma$ is the total uncertainty.}
    \label{fig:uq_defect}
\end{figure*}

\begin{figure*}[htbp]
    \centering
   \includegraphics[width=0.9\linewidth]{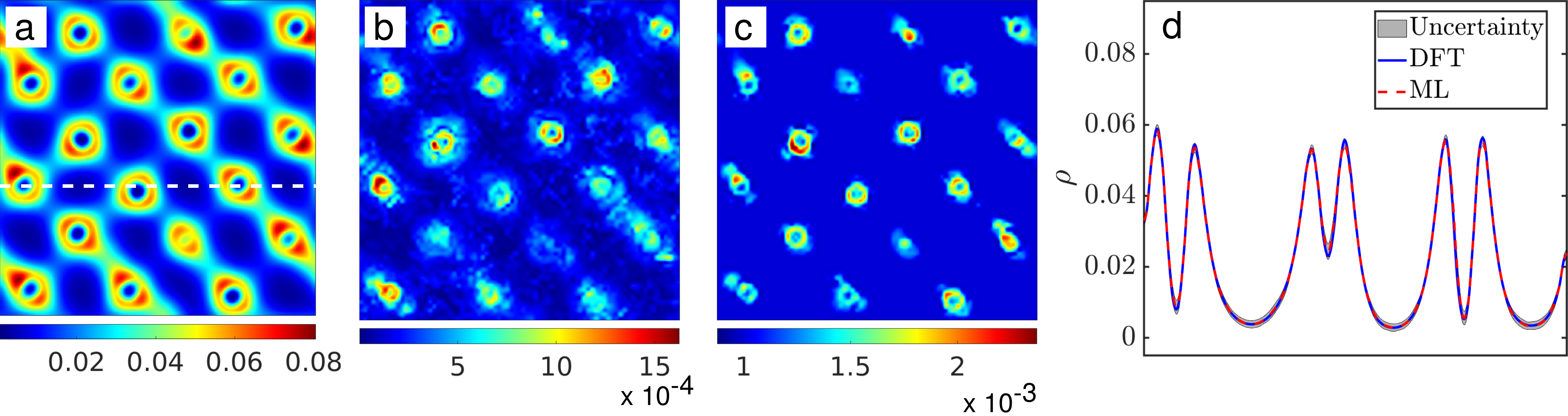}
    \caption{Uncertainty quantification Si$_{0.5}$Ge$_{0.5}$ system containing 216 atoms. (a) ML prediction of the electron density, (b) Epistemic Uncertainty (c) Aleatoric Uncertainty (d) Total Uncertainty shown along the dotted line from the ML prediction slice. The uncertainty represents the bound $\pm 3\sigma_{total}$, where, $\sigma_{total}$ is the total uncertainty.}
    \label{fig:SiGe5050}
\end{figure*}

\begin{figure*} [htbp]
    \centering
    \includegraphics[width=0.9\linewidth]{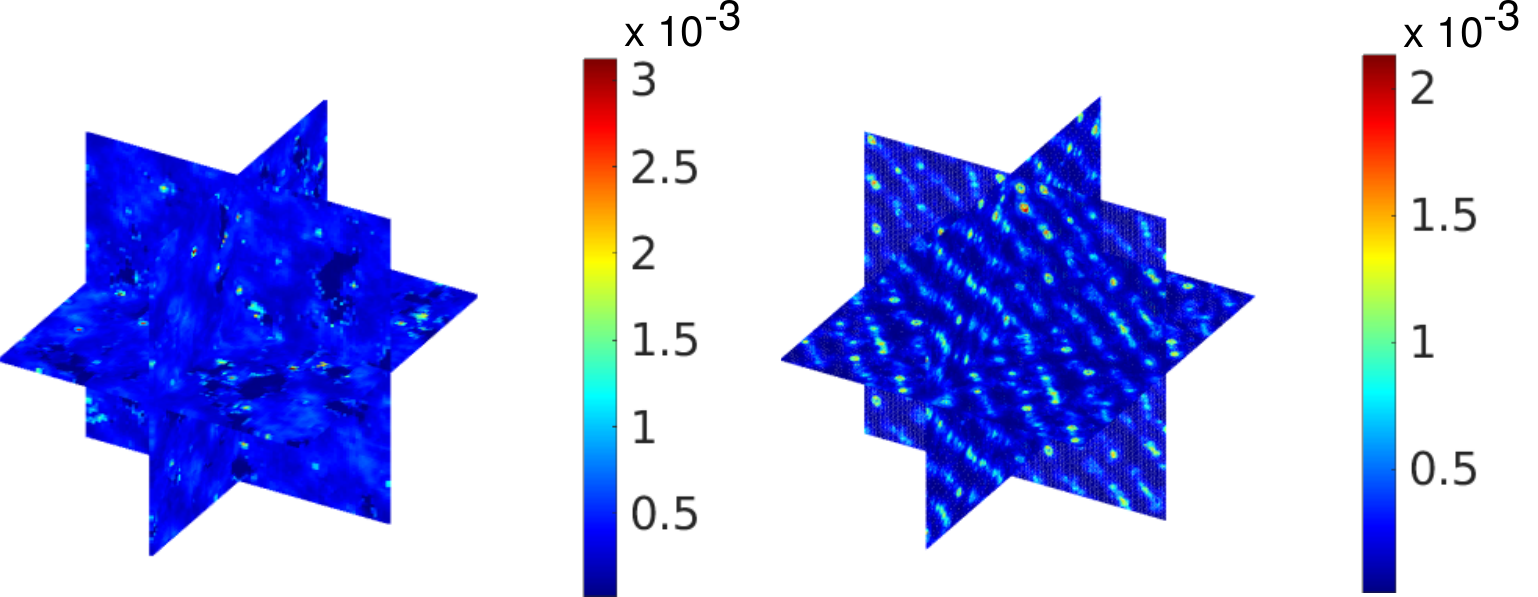}
    \caption{(Left) Total uncertainty for the Al system ($\sim 4.1$ million atoms) shown in Fig. \ref{fig:4M} of the main text. (Right) Total uncertainty for the SiGe system  ($\sim 1.4$ million atoms)  shown in Fig. \ref{fig:1M_SiGe} of the main text (right). }
    \label{fig:4Muq}
\end{figure*} 

\begin{figure} [htbp]
    \centering
    \includegraphics[width=0.9\linewidth]{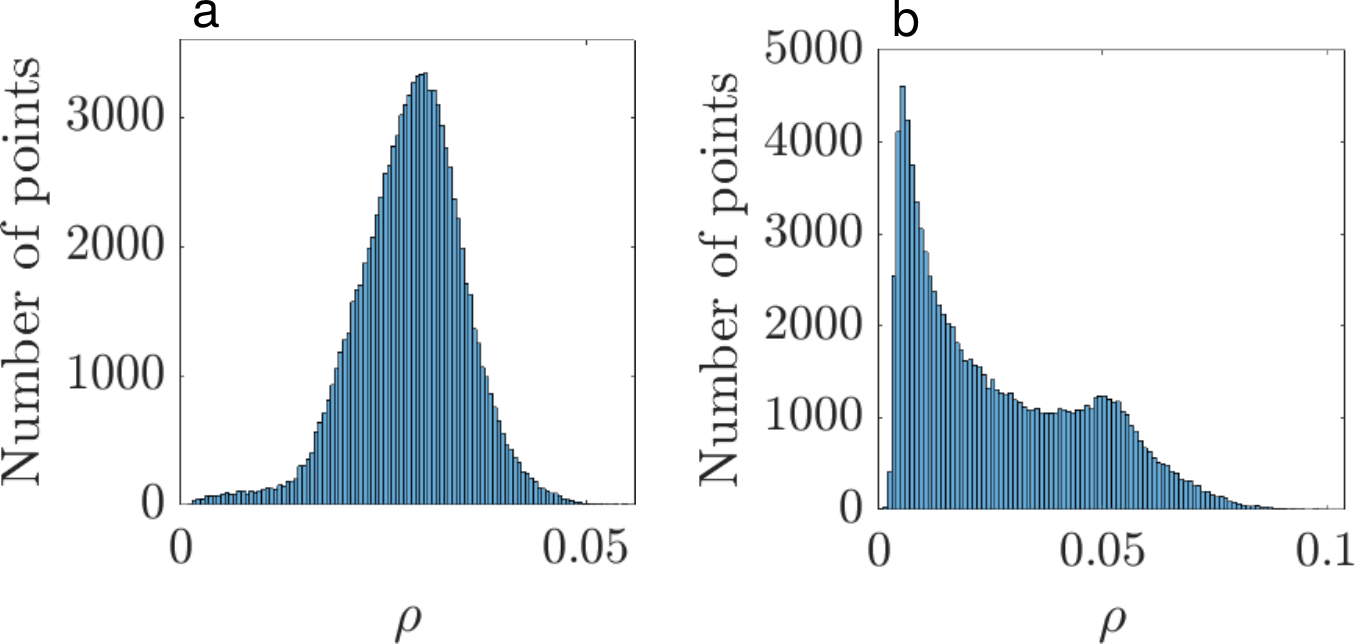}
    \caption{Histogram showing the distribution of charge density ($\rho$) for (a) aluminum and (b) SiGe.}
    \label{fig:histogram_rho}
\end{figure} 

\begin{figure*} [htbp]
    \centering
    \includegraphics[width=0.9\linewidth]{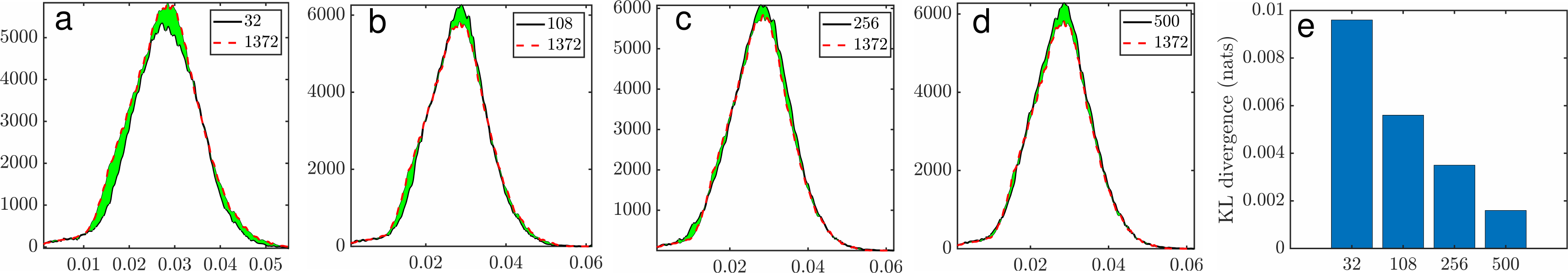}
    \caption{(a-d) Comparison of the histograms of electron density of aluminum for the largest system with that of smaller systems. The shaded green areas show the difference between the histograms. The largest aluminum system has $1372$ atoms, whereas the smaller systems have $32$, $108$, $256$, and $500$ atoms. 
    e) Kullback–Leibler (KL) divergence between the probability distributions corresponding to the  histograms in a-d and that of the largest system. The values of the KL divergence decreases with the increase in system size.}
    \label{fig:histogram_rho_all}
\end{figure*}

\section{Efficient generation of atomic neighborhood descriptors}\label{desc_sm}
The atomic neighborhood descriptors to encode the atomic neighborhood of the grid point are $||\textbf{r}_i - \textbf{R}_J||$ and    $\frac{(\textbf{r}_i - \textbf{R}_K) \cdot (\textbf{r}i - \textbf{R}_{S})}{||\textbf{r}_i - \textbf{R}_K|| \, || \textbf{r}i - \textbf{R}_{S} ||}$, as described in the section \ref{sec:Methods} of the main text. Our implementation of descriptor generation employs a tree data structure to reduce computational complexity and is outlined as a pseudocode in Algorithm \ref{desc_gen_code}.  

The descriptors described above satisfy the following conditions outlined in \cite{huo2022unified} and \cite{himanen2020dscribe}:
(i) invariance with respect to rotations and translations of the system 
(ii) invariance with respect to the permutation of atomic indices, i.e., the descriptors are independent of the enumeration of the atoms.
(iii) for a given atomic neighborhood, the descriptors are unique.
(iv) the descriptors encode the atomic neighborhood effectively while keeping the overall count low.
(v) the descriptors generation process is computationally inexpensive and uses standard linear algebra operations.

\begin{algorithm}
    \caption{Generation of Descriptors} \label{desc_gen_code}
    \begin{algorithmic}
        \State $M$ = Number of nearest neighbor atoms to compute distances
        \State $M_a$ = Number of nearest neighbor atoms to compute angles
        \State $k$ = Number of angles obtained for each $M_a$ atoms
        \State Build supercell by extending unit cell in all directions 
        \State KDTree = K-D tree for atoms in supercell
        \For{$\mathbf{g}$}  \Comment{$\mathbf{g}$: grid point}

        \State $D \gets$ distances to $M$ nearest atoms from $\mathbf{g}$ using K-D tree

        \For{$j = 1$ to $M_a$}
            \State $\mathbf{a}_i$ \Comment{coordinates of $i^{th}$ nearest atom from $g$ using K-D tree}
            \State $\mathbf{v}_1 \gets \mathbf{a}_i -\mathbf{g}$
            \For{$j = 1$ to $k$}
                \State $\mathbf{A}_j$ \Comment{coordinates of $j^{th}$ nearest atom from $\mathbf{a}_{i}$}
                \State $\mathbf{v}_2 \gets \mathbf{A}_j - \mathbf{g}$
                \State $\mathcal{A}_{ij} \gets \frac{\mathbf{v}_1 \cdot \mathbf{v}_2}{||\mathbf{v}_1||  \;||\mathbf{v}_2||}$  
            \EndFor
        \EndFor
        \State $\mathcal{A} \gets $ flatten($\mathcal{A}$)
        \State descriptors $\gets$ [$D, \mathcal{A}$]
        \EndFor
    \State Note: Inner two \textbf{for} loops are vectorized and Outermost \textbf{for} is parallelized in the implementation
    \end{algorithmic}
\end{algorithm}

Descriptors are obtained by implementing a parallelized version of Algorithm \ref{desc_gen_code}. In the case of SiGe systems, instead of explicitly encoding the species information, we follow \cite{chandrasekaran2019solving} and concatenate the descriptors obtained for Si and Ge, to form inputs to the neural network. To encode the relative placement of Si and Ge atoms with respect to each other, we also consider the cosine of angles between Si and Ge atoms formed at the grid point for the SiGe case.

\section{Computational Efficiency}
Computational time comparison between DFT calculation and ML prediction is given in tables  \ref{tab:Al_time} and \ref{tab:SiGe_time} for aluminum and SiGe, respectively. DFT calculations were performed using CPUs, whereas the ML predictions used a combination of GPU (inference step) and CPU (descriptor generation) resources.

{The primary contributor to ML prediction time is descriptor generation, constituting the majority of the computational effort and the remaining time is neural network inference (See Tables \ref{tab:Al_time} and \ref{tab:SiGe_time}). Given that neural network inference is well-suited for GPU execution and is commonly performed on GPUs, our assessment of parallelization performance focuses on descriptor generation time. In Figure \ref{fig:scaling_with_procs}, we present the parallelization performance of descriptor generation for the Aluminum system with 500 atoms. This parallelization was executed using the MATLAB's 'parfor' function on NERSC Perlmutter CPUs and we observe $66.6$\% strong scaling for $64$ processors.}

The DFT and ML calculations presented in this work were performed through a combination of resources, namely, desktop workstations, the Hoffman2 cluster at UCLA's Institute of Digital Research and Education (IDRE), the Applied Computing GPU cluster at MTU, and NERSC's supercomputer, Perlmutter. Every compute node of the Hoffman2 cluster has two $18$-core Intel Xeon Gold 6140 processors ($24.75$ MB L3 cache, clock speed of $2.3$ GHz), $192$ GB of RAM and local SSD storage. Every compute node on Perlmutter has a $64$-core AMD EPYC 7763 processor ($256$ MB L3 cache, clock speed of $2.45$ GHz), $512$ GB of RAM and local SSD storage. The GPU resources on Perlmutter consist of NVIDIA A100 Tensor Core GPUs. The GPU nodes used at UCLA and MTU consist of Tesla V100 GPUs. 

\underline{Large system generation:} {The million atom systems presented were generated by repeating one of the available test systems in all three directions and adding random perturbations in the atomic coordinates for each atom in the resulting system. This process ensures that the million-atom system is distinct from the smaller system employed in its generation and that the atomic neighborhoods generated within the million-atom system are not identical to those in the smaller system. Additionally, it is noteworthy that the systems replicated to achieve the million-atom configurations are entirely excluded from the training dataset (e.g. in the case of Aluminum, 1372 atom system was employed to generate the 4.1 million-atom system, while the training process utilized 32 and 108 atom systems. In the case of SiGe, a 512 atom system was used to generate the 1.4 million atom system). The perturbations used were sampled from a normal distribution with a zero mean and a 0.1 Bohr standard deviation. The choice of standard deviation was deliberate, aiming to prevent impractical distances between atoms and ensure realistic configurations.}

\underline{Large system calculations:} We present electron density calculation for  Al and SiGe systems, each with an excess of a million atoms, in Fig. \ref{fig:4M} and Fig. \ref{fig:1M_SiGe} of the main text, respectively. To predict the charge density while avoiding memory overload issues, we partition these multi-million atom systems into smaller systems, while retaining the atomic neighborhood information consistent with the larger original systems. In the case of aluminum, we break down the $4.1$M atom system into smaller units comprising $1372$ atoms and a grid consisting of $175^3$ points. Computation of descriptors for this $1372$-atom chunk takes approximately $34.72$ seconds on a desktop workstation system equipped with a $36$-core Intel(R) Xeon(R) Gold 5220 CPU @ 2.20GHz. Subsequently, the charge density prediction requires approximately $1.6$ seconds on an Nvidia V100 GPU.  Overall, the charge density prediction for the $4.1$M Al system takes around $30.72$ hours of wall time on combined CPU and GPU resources.  
 
Analogously, for SiGe, we partition the $1.4$M atom system into smaller systems composed of $1000$ atoms and a grid with dimensions of $132^3$ points. The computation of descriptors for this $1000$-atom SiGe chunk  requires $22.17$ seconds on the aforementioned desktop system. The subsequent charge density prediction takes approximately $1.1$ seconds. Overall, it takes around $6.8$ hours of wall time on combined CPU and GPU resources, to predict the electron density of the SiGe system with $1.4$M atoms.

Thus, the techniques described here make it possible to routinely predict the electronic structure of systems at unprecedented scales, while using only modest resources on standard desktop systems.  
\begin{table}[htbp] 
\begin{tabular}{|cc|c|c|c|c|}
\hline
\multicolumn{2}{|c|}{Number of Atoms}                                                                                       & 32    & 108    & 256    & 500    \\ \hline
\multicolumn{2}{|c|}{DFT Time (CPU)}                                                                                              & 466   & 11560  & 112894 & 245798 \\ \hline
\multicolumn{1}{|c|}{\multirow{4}{*}{\begin{tabular}[c]{@{}c@{}}ML \\ Time\end{tabular}}} & Descriptor Generation           & 43.25 & 151.52 & 367.54 & 739.58 \\ \cline{2-6} 
\multicolumn{1}{|c|}{}                                                                    & $\rho$ Prediction (CPU) & 2.76  & 9.67  & 23.46     & 47.20   \\ \cline{2-6} 
\multicolumn{1}{|c|}{}                                                                    & $\rho$ Prediction (GPU) & 0.60  & 0.64   & 0.75   & 0.99   \\ \cline{2-6} 
\multicolumn{1}{|c|}{}                                                                    & Total (With GPU)                & 43.85 & 152.16 & 368.29 & 740.57 \\ \hline
\multicolumn{2}{|c|}{DFT time / Total ML time}                                                                              & 10.63 & 75.97  & 306.53 & 331.90 \\ \hline
\end{tabular}
\caption{Comparison of DFT and ML wall times for prediction of electron density for an aluminum system. All times are in seconds. The DFT calculations were performed on Hoffman CPUs, ML descriptor generation was done on Hoffman CPUs, and the ML inference was performed on Tesla V100 GPUs. }
\label{tab:Al_time}
\end{table}
\begin{table}[htbp] 
\begin{tabular}{|cc|c|c|c|c|}
\hline
\multicolumn{2}{|c|}{Number of Atoms}                                                                                       & 64    & 216    & 512    & 1000   \\ \hline
\multicolumn{2}{|c|}{DFT Time}                                                                                              & 185   & 4774   & 51247  & 281766 \\ \hline
\multicolumn{1}{|c|}{\multirow{4}{*}{\begin{tabular}[c]{@{}c@{}}ML \\ Time\end{tabular}}} & Descriptor Generation           & 38.82 & 115.23 & 291.45 & 611.2  \\ \cline{2-6} 
\multicolumn{1}{|c|}{}                                                                    & $\rho$ Prediction (CPU) & 2.22  & 7.37   & 17.37  & 33.05  \\ \cline{2-6} 
\multicolumn{1}{|c|}{}                                                                    & $\rho$ Prediction (GPU) & 0.50  & 0.62   & 0.75   & 0.89   \\ \cline{2-6} 
\multicolumn{1}{|c|}{}                                                                    & Total (With GPU)                & 39.32 & 115.85 & 292.20 & 612.09 \\ \hline
\multicolumn{2}{|c|}{DFT time / Total ML time}                                                                              & 4.70  & 41.21  & 175.38 & 460.33 \\ \hline
\end{tabular}
\caption{Comparison of DFT and ML wall times for prediction of electron density for a \ce{SiGe} system. All times are in seconds. The DFT calculations were performed on Perlmutter CPUs, ML descriptor generation was done on Perlmutter CPUs and the ML inference was performed on Tesla V100 GPUs.}
\label{tab:SiGe_time}
\end{table}

\begin{figure*} [t!]
    \centering
    \includegraphics[width=0.4\linewidth]{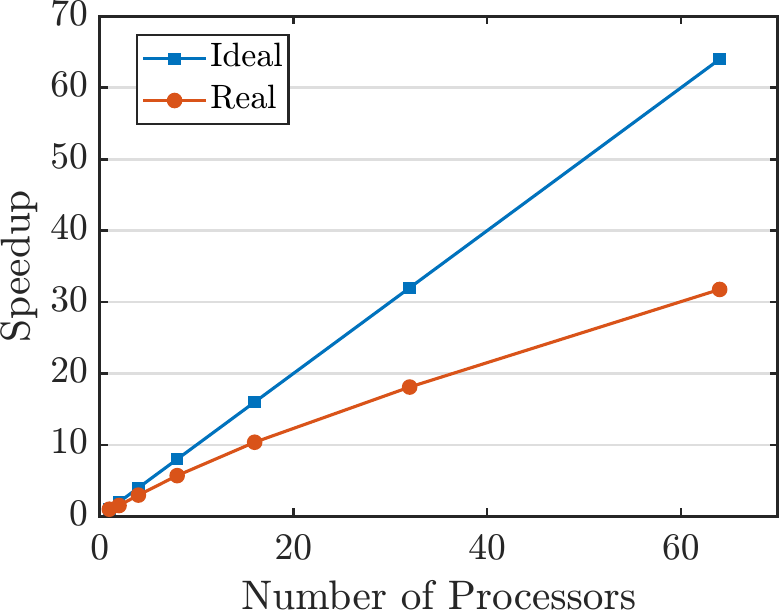}
    \caption{{Speedup of ML prediction time with respect to number of processors (strong parallel scaling). The plot is shown for a $500$ atom Aluminum system. Speedup is obtained with reference to 1 processor. The computation was performed on NERSC Perlmutter CPUs.}}
    \label{fig:scaling_with_procs}
\end{figure*} 

\section{Feature Convergence Analysis} 
Algorithm \ref{algo:1} and Algorithm \ref{algo:2} describe the process used to obtain the optimal number of descriptors. In algorithm  \ref{algo:1} only distances (set\,\RNum{1}) are considered as descriptors. The size of the set\,\RNum{1} (i.e. $M$) is selected for which the RMSE for the test dataset converges. 

\begin{algorithm}
    \caption{Optimal nearest neighbors} \label{algo:1}
    \begin{algorithmic}
    \State $M  = 0$  \Comment{Initialization} 
    \State $\epsilon_{0} = \epsilon_{-m} = \delta_1 = \delta_2 = $ A large  number \Comment{Initialization} 
    \State $\eta$ = tolerance in RMSE 
    \While{ $\delta_1 \geq \eta \; \& \; \delta_2 \geq \eta $ } 
        \State $M = M + m$ \Comment{Increase $M$ by $m \in \mathbb{Z^+}$}  
        \State $N_\text{set\,\RNum{1}} \gets M$ \Comment{$M$ nearest atoms}
        \State $N \gets N_\text{set\,\RNum{1}}$ \Comment{Only set\,\RNum{1} descriptors}
        \State Compute $N$ descriptors 
        \State Train $f_N$ \Comment{Train the BNN}
        \State $\epsilon_{M} \gets$ RMSE \Comment{Compute RMSE}
        \State $\delta_1 \gets \vert \epsilon_M - \epsilon_{M-m} \vert $ 
        \State $\delta_2 \gets \vert \epsilon_M - \epsilon_{M-2m} \vert \, $
    \EndWhile
    \State $M = M-2m$
    \end{algorithmic}
\end{algorithm}

As an illustration, for the aluminum systems, following algorithm \ref{algo:1} we use an increment of $m =10$. The algorithm  converges to $M = 60$ as seen in Fig.\,\ref{fig:feature_convergence} of the main text. Therefore, the set\,\RNum{1} consists of $60$ descriptors. Next, Set\,\RNum{2} descriptors consist of angles subtended at the grid point by a pair of atoms taken from the set of $M$ neighboring atoms in the set\,\RNum{1} determined by algorithm~\ref{algo:1}. Each pair of the neighboring atoms forms an angle at the grid point, yielding a total of $M(M-1)/2$ angles, which quickly becomes computationally intractable with increasing $M$. To alleviate this issue, we reduce the number of Set\,\RNum{2} descriptors by eliminating large angles, which are not expected to play a significant role. This amounts to choosing angles originating from $M_a < M$ atoms closest to the grid point, and the $k-$ nearest neighbors of each of these atoms. This yields a total of $M_a\times k$ angle descriptors. For various fixed values of $k$, we iteratively choose $M_a$ till the RMSE over the test dataset converges (Fig.\,\ref{fig:feature_convergence} of the main text).

\begin{algorithm}
    \caption{Optimal number of angles} \label{algo:2}
    \begin{algorithmic}
    \State $k  = 0 $ \Comment{Initialization}
    \State $\epsilon_{0} = \epsilon_{-m} = \delta_1 = \delta_2 = \delta_3 = $ A large number \Comment{Initialization}
    \State $\eta$ = tolerance in RMSE
    \While { $\delta_3 \geq \eta $ } 
        \State $k = k + 1$ 
        \State $M_a = 0$
        \While { $\delta_1 \geq \eta \; \& \; \delta_2 \geq \eta $ } 
            \State $M_a = M_a + m_a$ \Comment{ Increase $M_a$ by $m_a \in \mathbb{Z^+}$} 
            \State $N_\text{set\,\RNum{2}} \gets M_a\times k$ \Comment{$k$ neighbors of each of $M_a$ nearest atoms}
            \State $N \gets N_\text{set\,\RNum{1}} + N_\text{set\,\RNum{2}}$ \Comment{Number of total descriptors} 
            \State Compute $N$ descriptors 
            \State Train $f_N$ \Comment{Train the BNN}
            \State $\epsilon_{M_a} \gets$ RMSE \Comment{Compute RMSE}
            \State $\delta_1 \gets \vert \epsilon_{M_a} - \epsilon_{M_a-m_a} \vert $ 
            \State $\delta_2 \gets \vert \epsilon_{M_a} - \epsilon_{M_a-2m_a} \vert \, $
        \EndWhile
        \State $M_a = M_a-2m_a$
        \State $\epsilon'_k \gets \epsilon_{M_a}$ 
        \State $\delta_3 \gets \vert \epsilon'_{k} - \epsilon'_{k-1} \vert \, $
    \EndWhile
    \State $k = k-1$
    \end{algorithmic}
\end{algorithm}

Following algorithm \ref{algo:2} we use an increment of $m=5$. Fig.\,\ref{fig:feature_convergence} of the main text shows the convergence plot for angles for $k=2, 3$, and $4$. For $M=60$, the RMSE value is the minimum for   $k=3$. The RMSE value for   $k=3$ converges at $M_a = 15$, which results in a total of $M_a\times k = 45$ angles. Therefore, set\,\RNum{2} consists of $45$ descriptors. To summarize, following the present feature selection strategy, the total number of descriptors used for the aluminum model is $N = N_{\text{set\,\RNum{1}}} + N_{\text{set\,\RNum{2}}} = 105$.

We found that including scalar triple products and scalar quadruple products in the descriptor, in addition to the dot products, did not improve the accuracy of the ML model. To interpret why this is the case, we observe that the (normalized) scalar triple product can be interpreted in terms of the corner solid angle (polar sine function) of the parallelepiped generated by three vectors starting at the given grid point and ending at three atoms chosen in the neighborhood of the grid point. However, this quantity can also be calculated through the dot products between these vectors and is, therefore, already incorporated in the second set of descriptors. Therefore, the scalar triple product does not furnish any additional information. 
Similar arguments can be made for quadruple and higher products. 

\section{Details on Uncertainty Quantification}
We provide additional results on uncertainty quantification (UQ) in this section. One of the key advantages of the inbuilt UQ capabilities of the present ML model is that it allows us to assess the model's generalizability. To illustrate this, we consider systems with defects and varying alloy compositions. The uncertainty estimates of a model trained without any defect data in training are shown in Fig.~\ref{fig:uq_defect_train} of the main text. The model is more confident in its prediction of defects even if a small amount (single snapshot) of defect data is added in training. This is evident by comparing Fig.~\ref{fig:uq_defect} and main text Fig.~\ref{fig:uq_defect_train}. This result is in agreement with the fact that unavailability or insufficient training data could yield high epistemic uncertainties at locations where such incompleteness of data exists. In addition to high uncertainty, the error at the defect location increases when  data from systems with defects are not used in training. This implies a positive correspondence between error and uncertainty in the Bayesian neural network model. A similar effect of higher uncertainty for unknown compositions is observed for the SiGe systems. Since the model is trained only with data from SiGe systems with 50-50 composition, the uncertainties quantified for this composition shown in Fig.~\ref{fig:SiGe5050}  is less in comparison to the prediction for 60-40 composition (Fig.~\ref{fig:uq_SiGe_4060} of the main text). However, the uncertainty for the 60-40 composition is not significantly higher than the 50-50 composition, demonstrating the  generalization capability of the ML model.  

 {
    In the following, we investigate the correlation between error and epistemic uncertainty. The epistemic uncertainty is chosen since it captures the uncertainty due to modeling error. We found positive correlations between the uncertainty and the error for configurations that were not present in the training and therefore exhibit higher errors. Examples include vacancies in Aluminum and alloy compositions away from the training data, as shown in Fig.~\ref{fig:uq_err_corr}. 
    We have also observed that for systems similar to training data, the errors as well as uncertainties are quite low, and do not exhibit strong correlations.  This indicates that for systems predicted with high uncertainties, uncertainty values may be used to identify regions with high error.}

\begin{figure} [htbp]
    \centering
    \subfigure[{256 atom Aluminum with single vacancy defect}]{\includegraphics[width=0.4\linewidth]{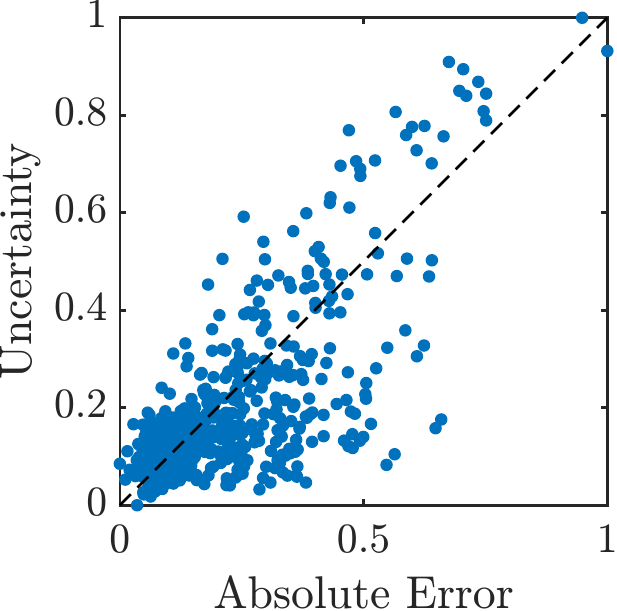}}
    \subfigure[{108 atom Aluminum with double vacancy defect}]{\includegraphics[width=0.4\linewidth]{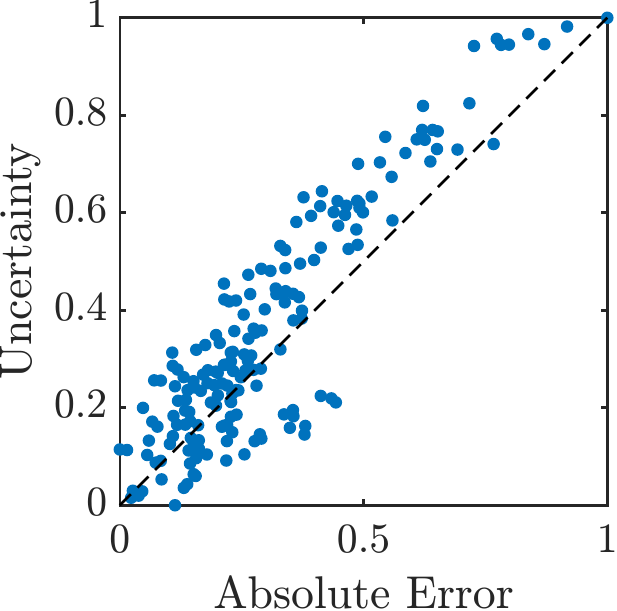}}
    \subfigure[{216 atom Si$_{0.4}$Ge$_{0.6}$}]{\includegraphics[width=0.4\linewidth]{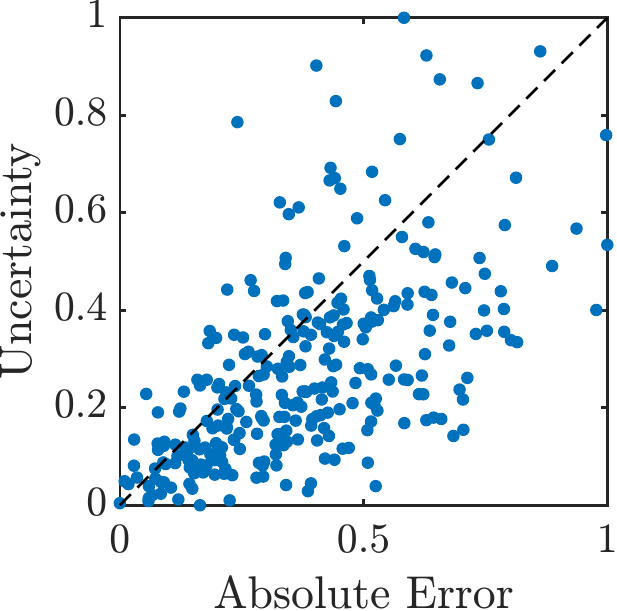}}
        \caption{{Correlation between epistemic uncertainty and error. All three cases show a positive correlation with $R = 0.75, 0.90, 0.59$, respectively. The uncertainty values and absolute error values are normalized using the min-max method. Each data point in the plots corresponds to uncertainty and error values are averaged over the neighborhood that is used to compute descriptors for the data point.}}
    \label{fig:uq_err_corr}
\end{figure} 

Results of uncertainty quantification $\approx$ 4.1 million atom aluminum system and $\approx$ 1.4 million atom SiGe system are shown in Fig.~ \ref{fig:4Muq}. With an increase in system size, we extrapolate farther away from the system size included in the training data. Despite this, the total uncertainty of millions of atom systems is similar to that of smaller systems. This implies that the model can predict  systems with millions of atoms with the same level of confidence as smaller systems, which in turn assures  the accuracy of the predictions. 
Looking ahead, we plan to further enhance the credibility of million-atom predictions by validating against results obtained from upcoming and state-of-the-art techniques involving Density Functional Theory (DFT) computations at a large scale \cite{SURYANARAYANA2018288, das2022dft, gavini2023roadmap, das2023large}.

We found that the ML model is less confident in predicting charge densities near the nucleus in comparison to the away from the nucleus for various systems, which is reflected in the high values of uncertainties at those locations. We attribute this to fewer grid points close to the nuclei, and the availability of more data away from them. This imbalance in the data is evident from the histograms for the distribution of charge densities shown in Fig.~\ref{fig:histogram_rho}, where grid points with low values of the electron density --- as is the case with points very close to the nuclei --- are seen to be very few. 

\section{Details on the advantages of transfer learning}
\begin{figure*} [t!]
    \centering
    \subfigure[Al system]{\includegraphics[width=0.4\linewidth]{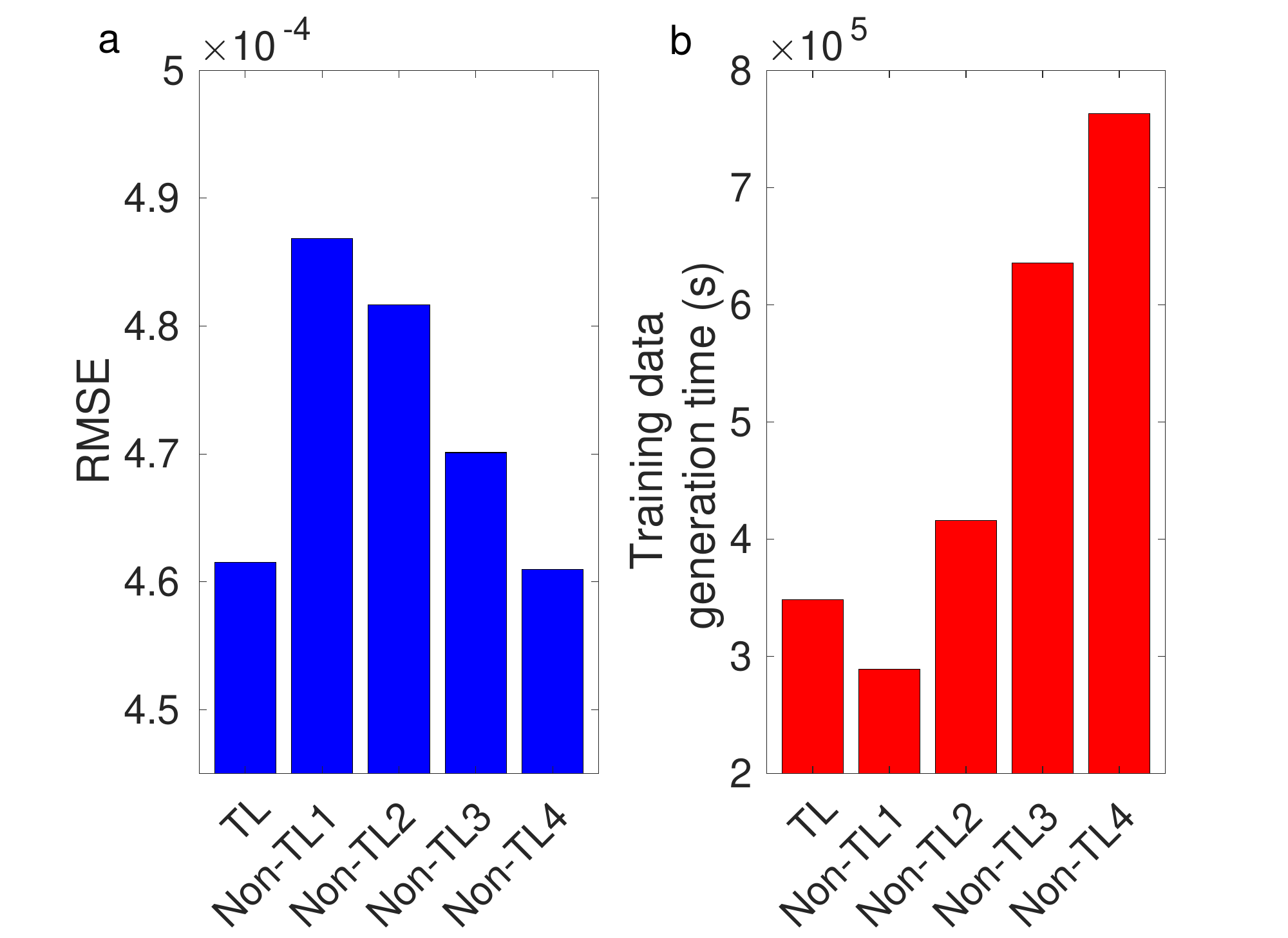}}
    \subfigure[SiGe System]{\includegraphics[width=0.4\linewidth]{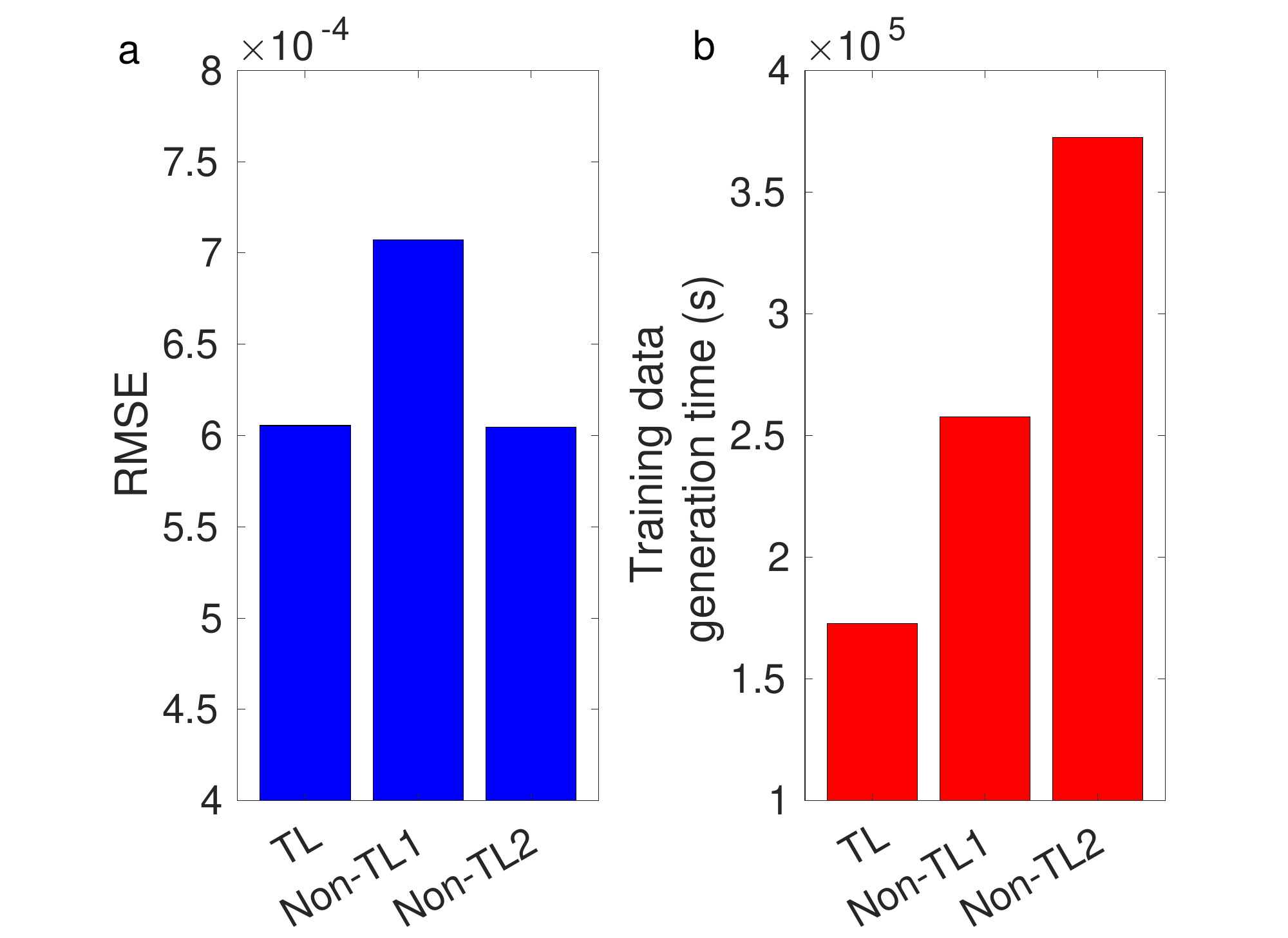}}
    \caption{Comparison of (a) error and (b) training data generation time between models with and without transfer learning.}
    \label{fig:al_tl_supp}
\end{figure*} 
\begin{figure*} [t!]
    \centering    
    \subfigure[]{\includegraphics[width=0.75\linewidth]{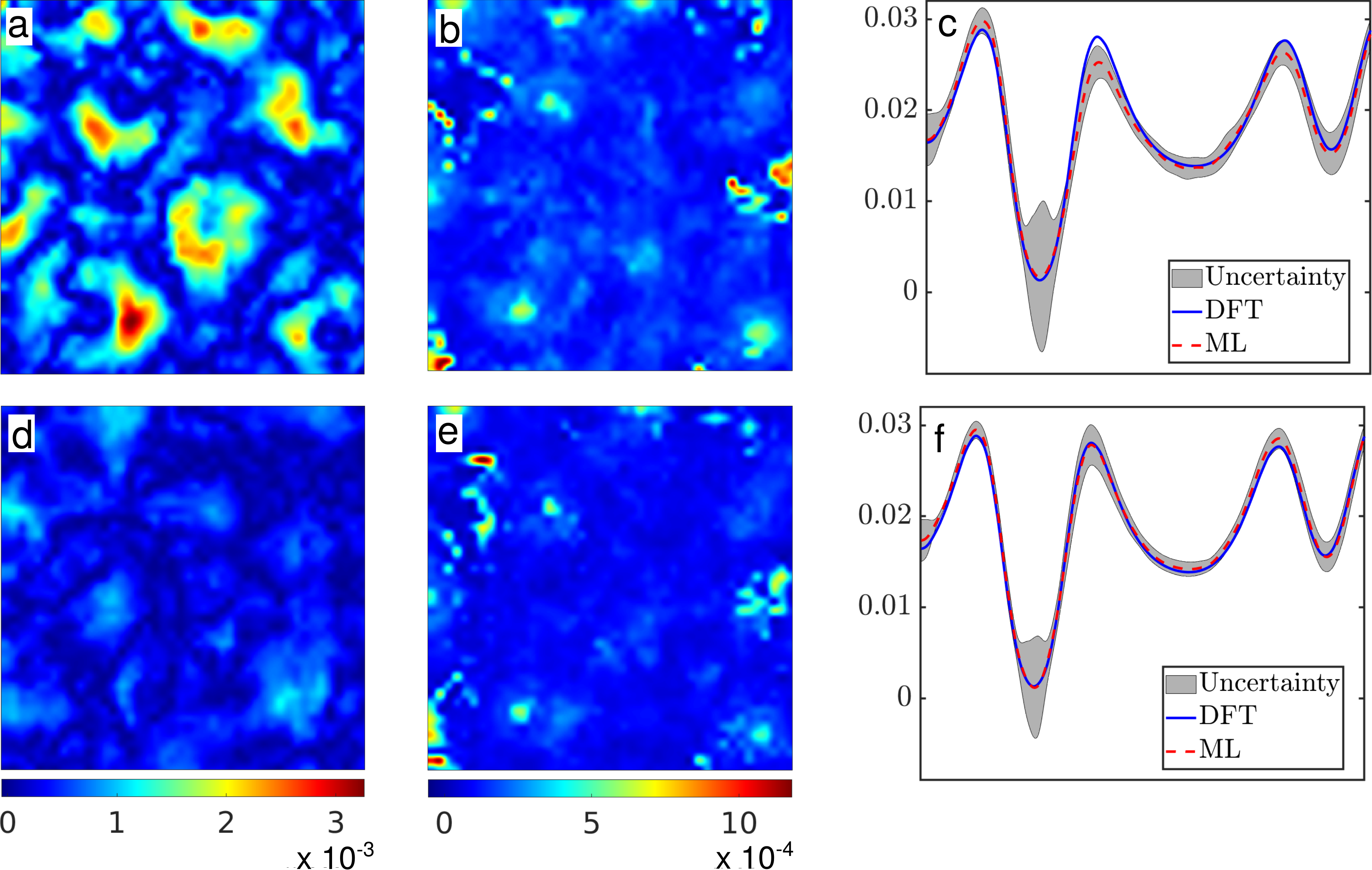}} 
    \subfigure[]{\includegraphics[width=0.35\linewidth]{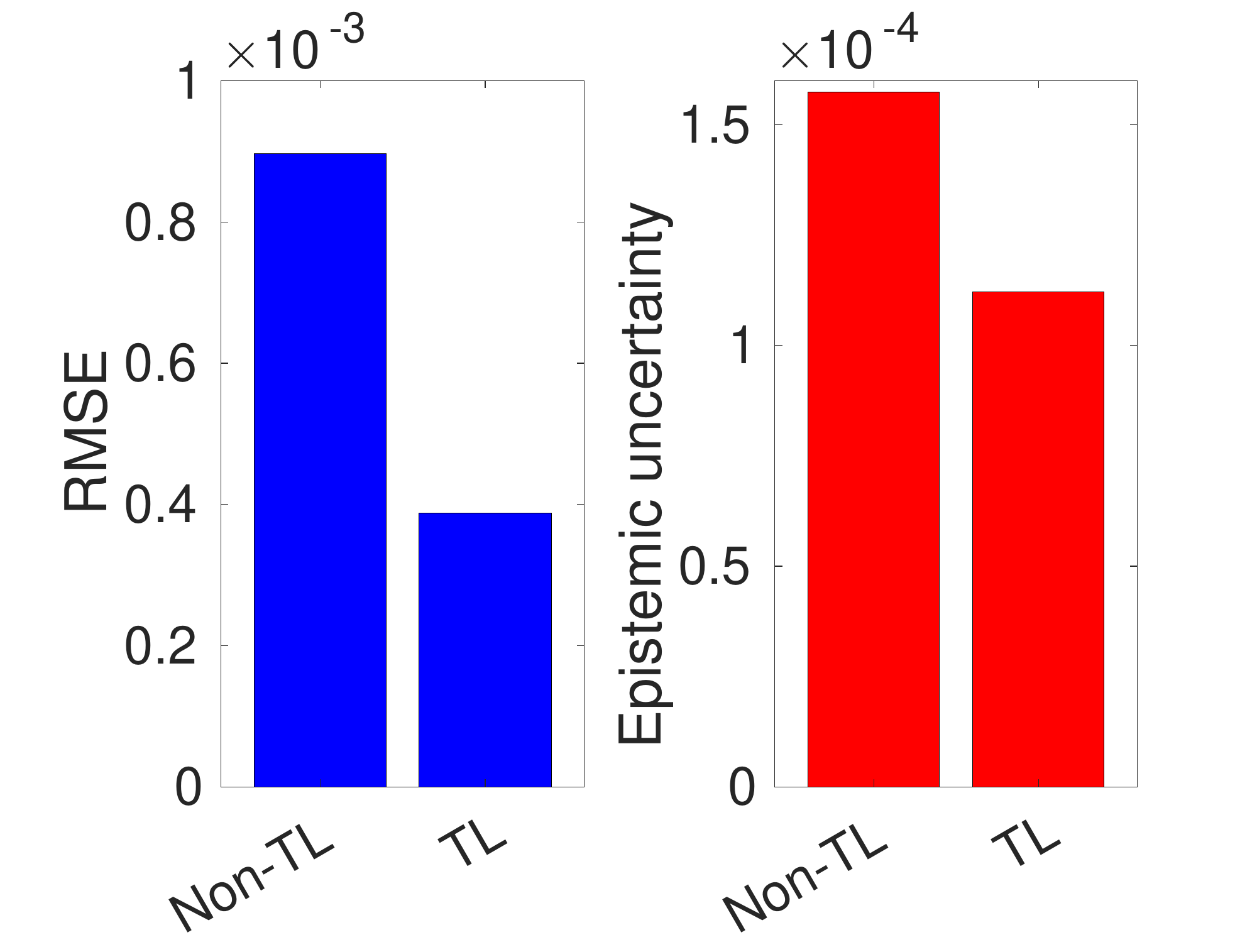}}
    \caption{(i) Decrease in error and uncertainty for a larger system (1372 atom) with transfer learning. Comparison is shown between predictions by a non-TL model trained using data only from the 32-atom system i(a-c) and a TL model trained by transfer learning using additional data from the 108-atom system i(d-f). The slice considered is shown in Fig. \ref{fig:1372uq}(a) of the main text. i(a and d) Error in ML prediction, i(b and e) Epistemic uncertainty, i(c and f) Total uncertainty along a line, as shown in Fig. \ref{fig:1372uq}(a) of the main text. Color bars are the same for i(a) and (c), and i(b) and (d). (ii) Bar plot showing a decrease in RMSE error and epistemic uncertainty. ii(a) The decrease in RMSE error is 56\% and ii(b) the decrease in the mean epistemic uncertainty is 29\%.}
    \label{fig:uq_defect_decrease}
\end{figure*} 
As demonstrated in prior research \cite{zepeda2021deep} and in this work, employing data from larger systems for training enhances the accuracy of machine learning models. However, the following question persists: what is the appropriate largest sizes of the training system to achieve a sufficiently accurate machine learning model that works across scales? To answer this question, we propose the following approach.

To ensure accurate predictions for bulk systems (comprising thousands or more atoms), it is imperative that our model be trained on data that statistically resembles such systems. Small-scale systems with only a few tens of atoms may not adequately represent the bulk limit, primarily due to the periodicity constraints inherent in simulations. This calls for training the model using larger systems. To determine appropriate training system sizes that adequately represents bulk systems, we employ the Kullback-Leibler (KL) divergence \cite{kullback1951information}. We consider the largest available system as the most faithful representation of bulk systems and use it to determine the largest size of the training systems. For the case of  Aluminum, a system consisting of $1372$ atoms can be reliably calculated using KS-DFT and is chosen as the reference. We compare the electron density distributions from various available systems against this reference system.  The KL divergence values then guide us in selecting the largest training system  needed to train a model that can accurately predict even at large scales (relevant to the reference system). Specifically, the largest training systems chosen by us contain $108$ atoms, as these systems are found to be sufficiently statistically similar to the $1372$-atom reference system (as illustrated in Fig.~\ref{fig:histogram_rho_all}). 
This meticulous selection process guarantees that our machine learning model is accurate at large scales while providing a judicious stopping point to our transfer learning scheme by determining the largest system needed for training. Thus, we present an approach that answers the question of selecting training system size and reduces the reliance on ad hoc heuristics for doing so. 

The transfer learning approach \cite{weiss2016survey} significantly reduces the root-mean-square error of a test dataset while costing much less computation for the training data generation. To depict this, a comparison of the transfer learned model with various non-transfer learned models is shown  in Fig.~\ref{fig:al_tl_supp}. 

We found that transfer learning helps to reduce the error and uncertainty in prediction for larger systems. By adding data from the $108$-atom aluminum systems in training, during the transfer learning approach, we significantly reduce the error (by 56\%) and uncertainty (by 29\%) of the predictions for a $1372$-atom test system in comparison to a non-TL model trained using data only from the $32$-atom systems, as shown in Fig.~\ref{fig:uq_defect_decrease}.

\section{Details on Bayesian Neural Network} \label{sec:DetailsBNN}

\textbf{Architecture:} We use a  Densenet \cite{huang2017densely} type architecture with three Dense blocks for the Bayesian Neural networks in this work. Each Dense block is composed of three hidden layers with $250$ nodes per layer and a GELU activation function \cite{hendrycks2016gaussian}. The skip connections in the Densenet-type architecture are weighted by a trainable coefficient. These skip connections have multiple advantages. Firstly, they prevent gradients from diminishing significantly during backpropagation. Further, they facilitate improved feature propagation by allowing each layer to directly access the feature generated by previous layers. Finally, these skip connections promote feature reuse, thereby substantially reducing the number of parameters. Such skip connections have been used for electron density predictions in the literature \cite{zepeda2021deep}.

Due to the stochastic weights of Bayesian neural networks, each weight is represented by its mean and standard deviation. Thus, the number of parameters in a Bayesian neural network is twice as compared to a deterministic network with the same architecture. In addition, the output of the Bayesian Neural networks used in this work has two neurons, one for predicting the charge density ($\rho$) and the other for predicting the aleatoric uncertainty ($\sigma$).

\textbf{Training Details:} {The parameters of the BNNs for the 32-atom Al system and 64-atom SiGe systems were initialized randomly with values drawn from the Gaussian distribution. The mean of the parameters were initialized with values drawn from  $\mathcal{N}(0,0.1)$. The standard deviations were parameterized as $\sigma = \log(1+\exp(\tau))$ so that $\sigma$ is always non-negative. The parameter $\tau$ was initialized with values drawn from $\mathcal{N}(-3,0.1)$. The priors for all the network parameters were assumed to be Gaussians: $\mathcal{N}(0,0.1)$. With these initializations and prior assumptions the initial models (i.e. model for 32-atom Al system and 64-atom SiGe system) were trained using standard back-propagation for BNNs.} The Adam optimizer \citep{kingma2014adam} was used for training and the learning rate was set to $10^{-3}$ for all the networks used in this work. {In the case of transfer learning, we freeze both the mean and standard deviation of the initial one-third layers of the model and re-train the mean and standard deviations of the remaining layers of the model. The prior assumptions, initialization of the learnable parameters, and their learning procedures remained the same as described above for the 32-atom Al and 64-atom SiGe systems.} The training time for the \ce{Al} and \ce{SiGe} systems are presented in Table~\ref{tab:train_time}. All the Bayesian Neural networks are trained on NVIDIA A100 Tensor Core GPUs 

The amount of data used in training for the two systems is as follows: 
\begin{itemize}
    \item Al: 127 snaps from 32 atom data and in addition 25 snaps from 108 atom data. The 108 atom data has $90 \times 90 \times 90$ grid points, while the 32 atom system has $60 \times 60 \times 60$ grid points.
    \item  SiGe: 160 snaps of 64 atom data and in addition 30 snaps of 216 atom data. The 64 atom system has $53 \times 53 \times 53$ grid points, while the 216 atom system has $79 \times 79 \times 79$ grid points. 
\end{itemize}

\begin{table}[htbp]
    \centering
    \begin{tabular}{|c|c|c|c|c|}
        \hline
         \multirow{2}{*}{System} & \multirow{2}{*}{Size} & \multirow{2}{*}{Epochs} & \multicolumn{2}{c|}{Training wall time (s)}  \\ \cline{4-5} 
          &&& Per epoch & Total \\ \hline
         \multirow{2}{*}{Al} & 32 &20 &906 & \multirow{2}{*}{31060} \\
         &108&20&647& \\ \hline
         \multirow{2}{*}{SiGe} & 64 &20 &651 & \multirow{2}{*}{18030} \\
         &216&10&501 & \\ \hline
    \end{tabular}
    \caption{GPU Training times for the BNNs. The training was performed on the NVIDIA Tesla A100 GPU. }
    \label{tab:train_time}
\end{table}

\textbf{Validation and Testing Details:} 20\% of the  data from the systems used for training is used as validation data. Testing is performed on {snapshots not used for training and validation,} and systems that are larger than those used for generating the training data in order to determine the accuracy in electron density prediction.


\section{Postprocessing results}
In tables \ref{tab:Al_error} and \ref{tab:SiGe_error} we compare the errors in the electron densities and the ground state energies for various \ce{Al} and \ce{SiGe} systems. We see errors well below the millihartree per atom range for total energies, even in the presence of defects and some degree of compositional variations --- these systems being quite far from the ones used to generate the training data. The average L$^1$ norm per electron between ML  and DFT electron densities for the largest available aluminum system (containing $1372$ atoms --- this is the largest aluminum system for which the DFT calculations could be carried out reliably within computational resource constraints), is $1.14\times10^{-2}$. In the case of SiGe, where the largest available system consists of $1728$ atoms, the average L$^1$ norm per electron is $8.25\times10^{-3}$. We observe that the errors for these largest systems are somewhat smaller than the typical errors associated with the systems listed in Tables \ref{tab:Al_error} and  \ref{tab:SiGe_error}, contradictory to what is anticipated. This can be attributed to the fact that the available AIMD trajectories for larger systems are typically not long enough (due to computational constraints) to induce significant variations in atomic configurations with respect to the equilibrium configuration, unlike the longer AIMD trajectories available for smaller systems. Consequently, the largest systems tested here are more amenable to accurate prediction, resulting in lower errors. 

{The time for the calculation of the total energy and forces from ML-predicted densities via postprocessing involves computation of the electrostatic, exchange correlation and band-energy terms, and uses a single diagonalization step to compute wave-function dependent quantities. Therefore, its computational time is similar to that of a single self-consistent field (SCF) step in a regular DFT calculation, provided the same eigensolver is used. For reference, using the MATLAB version of the SPARC code \cite{xu2020m} on a single CPU core, the postprocessing time is about 174 seconds for a 32 atom aluminum system while it is about 1600 seconds for 108 atoms. This also includes the time for computation of the Hellmann-Feynman forces. We would also like to mention here that this postprocessing step  can be significantly sped up by the ML prediction of other relevant quantities, such as the band energy and electrostatic fields \cite{shi2021machine}. As for the atomic forces, i.e., energy derivatives with respect to atomic coordinates,  automatic differentiation of the underlying neural networks can be employed to speed up calculations. All of these constitute ongoing and future work.}

\begin{table*}[htbp]
\resizebox{0.8\textwidth}{!}{%
\begin{tabular}{|c|c|c|c|c|c|}
\hline
\multirow{3}{*}{Case} & Accuracy of & Ground-state  & Exch.~Corr. & Fermi  & Max error in \\
 & electron density & energy & energy & level & eigenvalue \\
 & (L$^1$ norm per electron) &  (Ha/atom) & (Ha/atom) & (Ha) & (Ha)\\\hline 
Entire test data set & $2.62\times10^{-2}$ & $2.33\times10^{-4}$ & $4.36\times10^{-4}$ & $4.61\times10^{-4}$ & $4.58\times10^{-3}$ \\ \hline
\ce{Al} ($32$ atoms) & $2.27\times10^{-2}$ & $1.30\times10^{-4}$ & $1.07\times10^{-3}$ & $9.80\times10^{-4}$ & $4.10\times10^{-3}$ \\ \hline
\ce{Al} ($108$ atoms) & $1.67\times10^{-2}$ & $9.33\times10^{-5}$ & $9.82\times10^{-5}$ & $1.13\times10^{-4}$ & $1.87\times10^{-3}$ \\ \hline
\ce{Al} ($256$ atoms) & $3.93\times10^{-2}$ & $5.60\times10^{-4}$ & $4.18\times10^{-4}$ & $2.03\times10^{-4}$ & $6.67\times10^{-3}$ \\ \hline
\ce{Al} ($500$ atoms) & $3.96\times10^{-2}$ & $4.11\times10^{-4}$ & $2.41\times10^{-4}$ & $5.04\times10^{-4}$ & $8.52\times10^{-3}$ \\ \hline
\ce{Al} vacancy defects & $1.92\times10^{-2}$ & $9.80\times10^{-5}$ & $1.42\times10^{-4}$ & $2.98\times10^{-4}$ & $3.85\times10^{-3}$ \\ \hline
Strain imposed \ce{Al} & $2.54\times10^{-2}$ & $1.75\times10^{-4}$ & $8.91\times10^{-4}$ & $6.64\times10^{-4}$ & $3.11\times10^{-3}$ \\ \hline
\end{tabular}
}
\caption{Accuracy of the ML predicted electron density in terms of the L$^1$ norm per electron, calculated as $\frac{1}{N_\text{e}}\times\displaystyle\int_{\Omega}\left|\rho^{\text{scaled}}(\textbf{r})-\rho^{\text{DFT}}(\textbf{r})\right|d\textbf{r}$, for various test cases for an FCC aluminum bulk system ($N_{\text{e}}$ is the number of electrons in the system). Also shown in the table are errors in the different energies as computed from $\rho^{\text{scaled}}$. The test data set for post-processing was chosen such that it covered examples from all system sizes, configurations, and temperatures. For calculating the relevant energies, $\rho^{\text{scaled}}$ was used as the initial guess for the electron density, and a single Hamiltonian diagonalization step was performed. Energies were then computed.}
\label{tab:Al_error}
\end{table*}

\begin{table*}[htbp]
\resizebox{0.8\textwidth}{!}{%
\begin{tabular}{|c|c|c|c|c|c|}
\hline
\multirow{3}{*}{Case} & {Accuracy of} & Ground-state  & Exch.~Corr. & Fermi  & Max error in \\
 & electron density & energy & energy & level & eigenvalue \\
 & (L$^1$ norm per electron) &  (Ha/atom) & (Ha/atom) & (Ha) & (Ha)\\\hline 
Entire test data set & $1.93\times10^{-2}$ & $1.47\times10^{-4}$ & $9.34\times10^{-4}$ & $1.43\times10^{-3}$ & $7.29\times10^{-3}$ \\ \hline
\ce{Si_{0.5}Ge_{0.5}} ($64$ atoms) & $1.51\times10^{-2}$ & $8.08\times10^{-5}$  & $1.40\times10^{-3}$  & $8.71\times10^{-4}$ & $5.07\times10^{-3}$ \\ \hline
\ce{Si_{0.5}Ge_{0.5}} ($216$ atoms) & $1.90\times10^{-2}$ & $1.18\times10^{-4}$ & $2.50\times10^{-4}$ & $3.08\times10^{-4}$ & $4.99\times10^{-3}$ \\ \hline
\ce{Si_{0.5}Ge_{0.5}} ($512$ atoms) & $2.50\times10^{-2}$ & $2.57\times10^{-4}$ & $3.70\times10^{-4}$ & $1.32\times10^{-3}$ & $1.27\times10^{-2}$ \\ \hline
\ce{Si_{0.5}Ge_{0.5}} vacancy defects & $1.70\times10^{-2}$ & $9.68\times10^{-5}$ & $2.36\times10^{-4}$ & $2.82\times10^{-3}$ & $6.85\times10^{-3}$ \\ \hline
\ce{Si_{x}Ge_{1-x}} \footnotesize{($x\neq 0.5$)} & $2.39\times10^{-2}$ & $2.54\times10^{-4}$ & $2.41\times10^{-3}$ & $1.25\times10^{-3}$ & $9.36\times10^{-3}$ \\ \hline
\end{tabular}
}
\caption{Accuracy of the ML predicted electron density in terms of L$^1$ norm per electron, calculated as $\frac{1}{N_\text{e}}\times\displaystyle\int_{\Omega}\left|\rho^{\text{scaled}}(\textbf{r})-\rho^{\text{DFT}}(\textbf{r})\right|d\textbf{r}$, for various test cases for \ce{Si_{0.5}Ge_{0.5}} ($N_{\text{e}}$ is the number of electrons in the system). Also shown in the table are errors in the different energies as computed from $\rho^{\text{scaled}}$. The test data set for post-processing was chosen such that it covered examples from all system sizes and temperatures.  For calculating the relevant energies, $\rho^{\text{scaled}}$ was used as the initial guess for the electron density, and a single Hamiltonian diagonalization step was performed. Energies were then computed. For \ce{Si_{x}Ge_{1-x}}, we used $x=0.40,0.45,0.55,0.60$.
}
\label{tab:SiGe_error}
\end{table*}

\section{Calculation of the bulk modulus for aluminum}
We show a comparison between some material properties calculated using the electron density predicted by the ML model, and as obtained through DFT calculations. Specifically, we compute the optimum lattice parameter and the bulk modulus for aluminum --- these corresponding to the first and second derivatives of the post-processed energy curves (Fig.~\ref{fig:bulk_modulus} of the main text), respectively. A summary of our results can be found in Table \ref{tab:bulk_modulus}. It can be seen that bulk modulus differs by only about $1\%$, while the lattice parameters are predicted with even higher accuracy. Notably, the predicted lattice parameter and the bulk modulus are very close to experimental values \cite{raju2002high}, and the deviation from experiments is expected to decrease upon using larger supercells to simulate the bulk, a trend also seen in  Table \ref{tab:bulk_modulus}. This is consistent with the overall results shown in the main manuscript and further reinforces the predictive power of our model for non-ideal systems.

\section{Comparison with models based on other descriptors}

{In the main text, we have presented errors achieved in electron density prediction by our model. The results indicate that our approach is generally as accurate as (and in some cases outperforms) previous work \cite{zepeda2021deep, chandrasekaran2019solving}. To further compare it with existing similar approaches, we compare it with electron density predictions made via the well known SNAP descriptors \cite{thompson2015spectral, ellis2021accelerating}. Specifically, we have compared the relative L1 error (as defined in \cite{zepeda2021deep}) on 29 test snapshots using the dataset of an Aluminum system with 32 atoms.  We used the same training dataset and employed a neural network for both the descriptors. Both the descriptors yield nearly identical L1 errors (although the distribution of errors is different as shown in Fig. \ref{fig:spd_vs_snap}). At the same time, the calculation of the scalar product descriptors employed here exhibits computational efficiency, requiring about 50\% less time than generation of the SNAP descriptors. To ensure a fair and accurate comparison of descriptor computation time, the computations for both descriptors were performed on a single-core CPU. We utilized the data of Be 128 atoms provided by \cite{fiedler2023predicting} and the SNAP code provided by \cite{MalaGit, ellis2021accelerating}, for comparing descriptor calculation time.}

   \begin{figure}[htbp]
        \centering
        \subfigure[Error Histogram] {\includegraphics[width=0.45\linewidth]{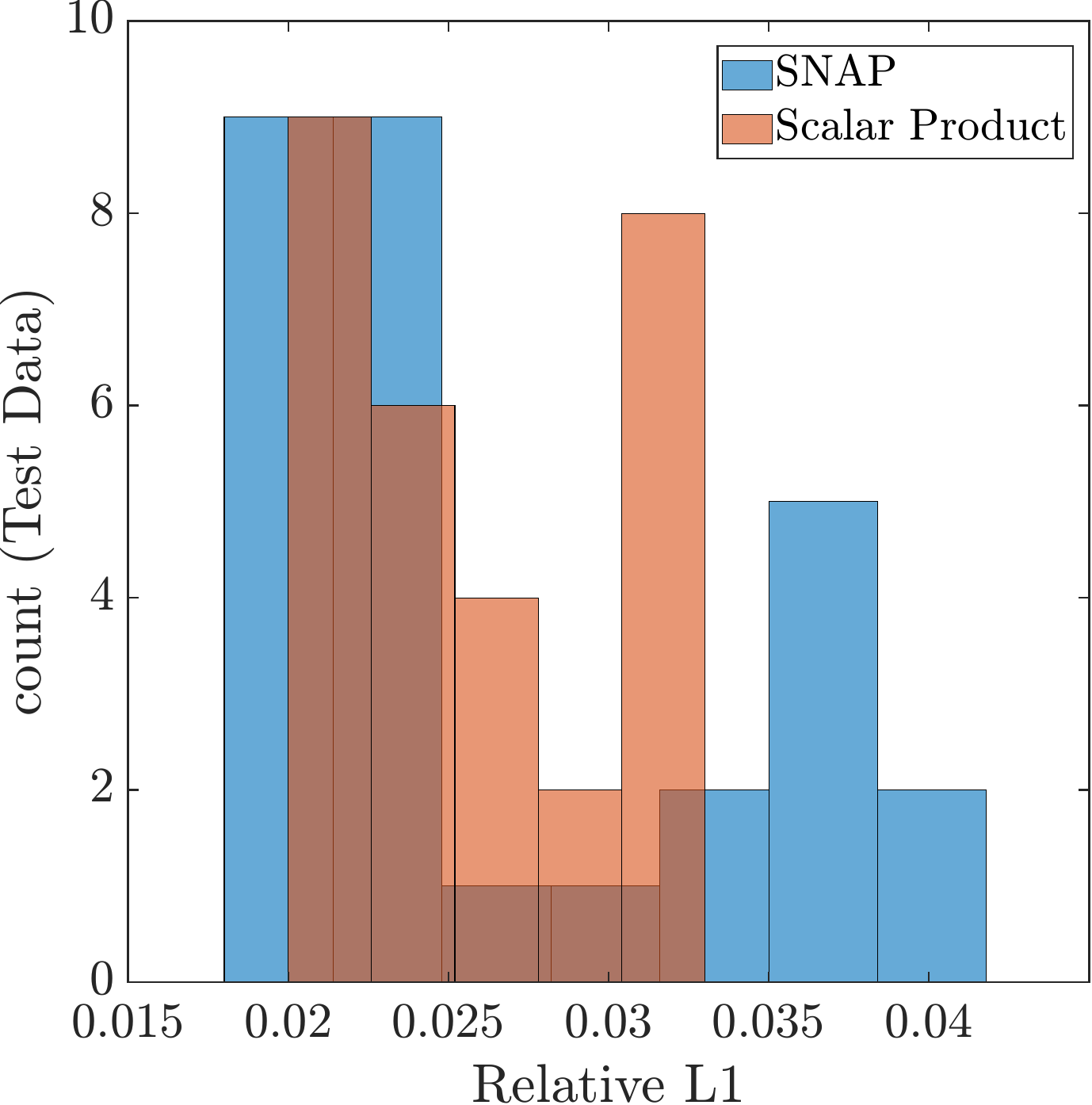}}
        \subfigure[Time comparison] {\includegraphics[width=0.45\linewidth]{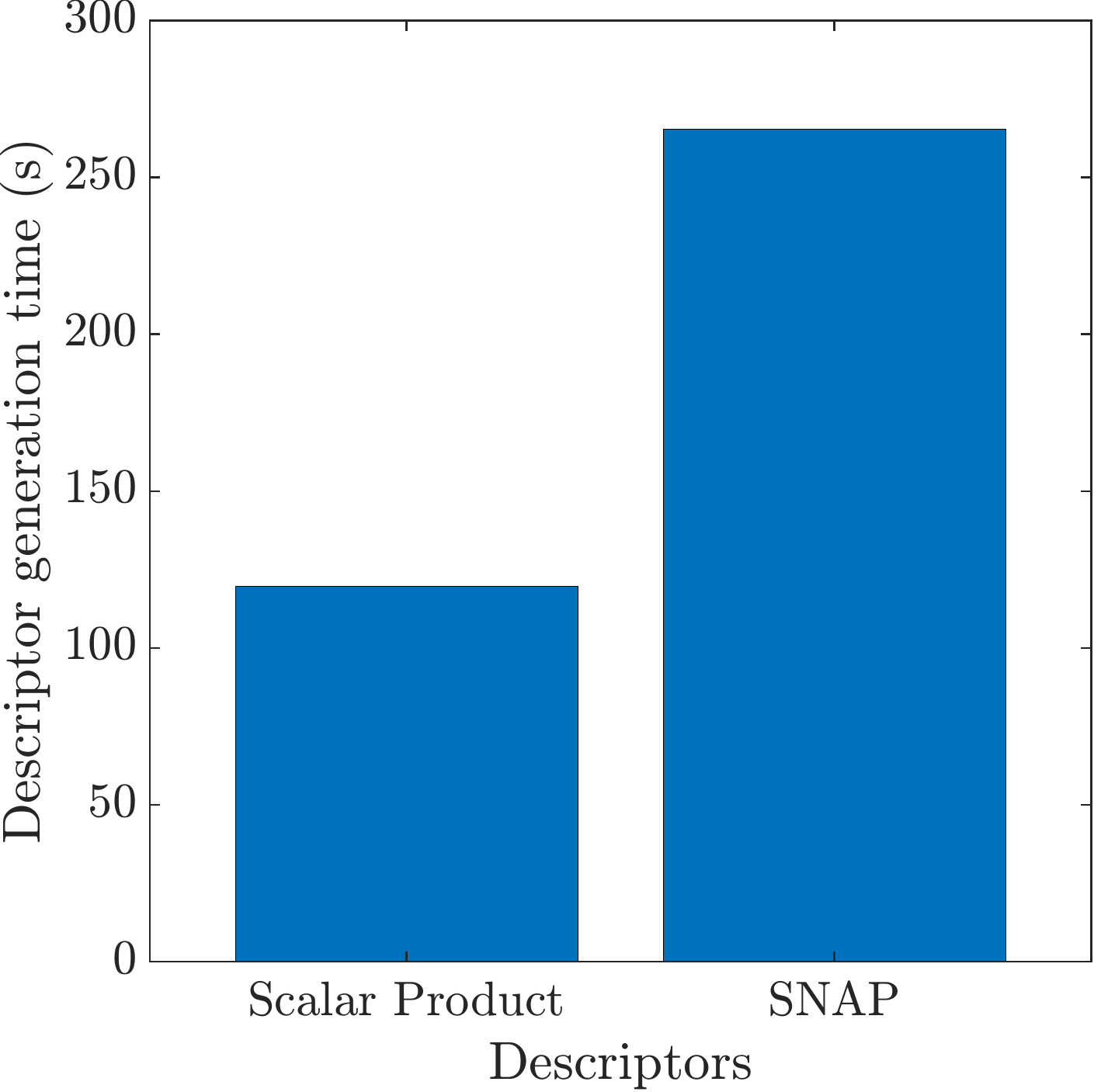}}
        \caption{Comparison with SNAP descriptors}
        \label{fig:spd_vs_snap}
    \end{figure}

\section{Equivariance of the model}

{In this section we show numerically that our model is equivariant, i.e., the predicted electron density is invariant with respect to overall rotation, translation, and permutation of atomic indices of the underlying material system. As mentioned in \cite{koker2023higher}, equivariance can be achieved by designing invariant features and predicting the electron density as a scalar valued variable. Since our model is based on these strategies, our machine learning model is expected to be equivariant, theoretically. We substantiate this claim numerically in Figure \ref{fig:showing_equi}.} 

\begin{figure*}[htbp]
    \centering
    \includegraphics[width=0.7\linewidth]{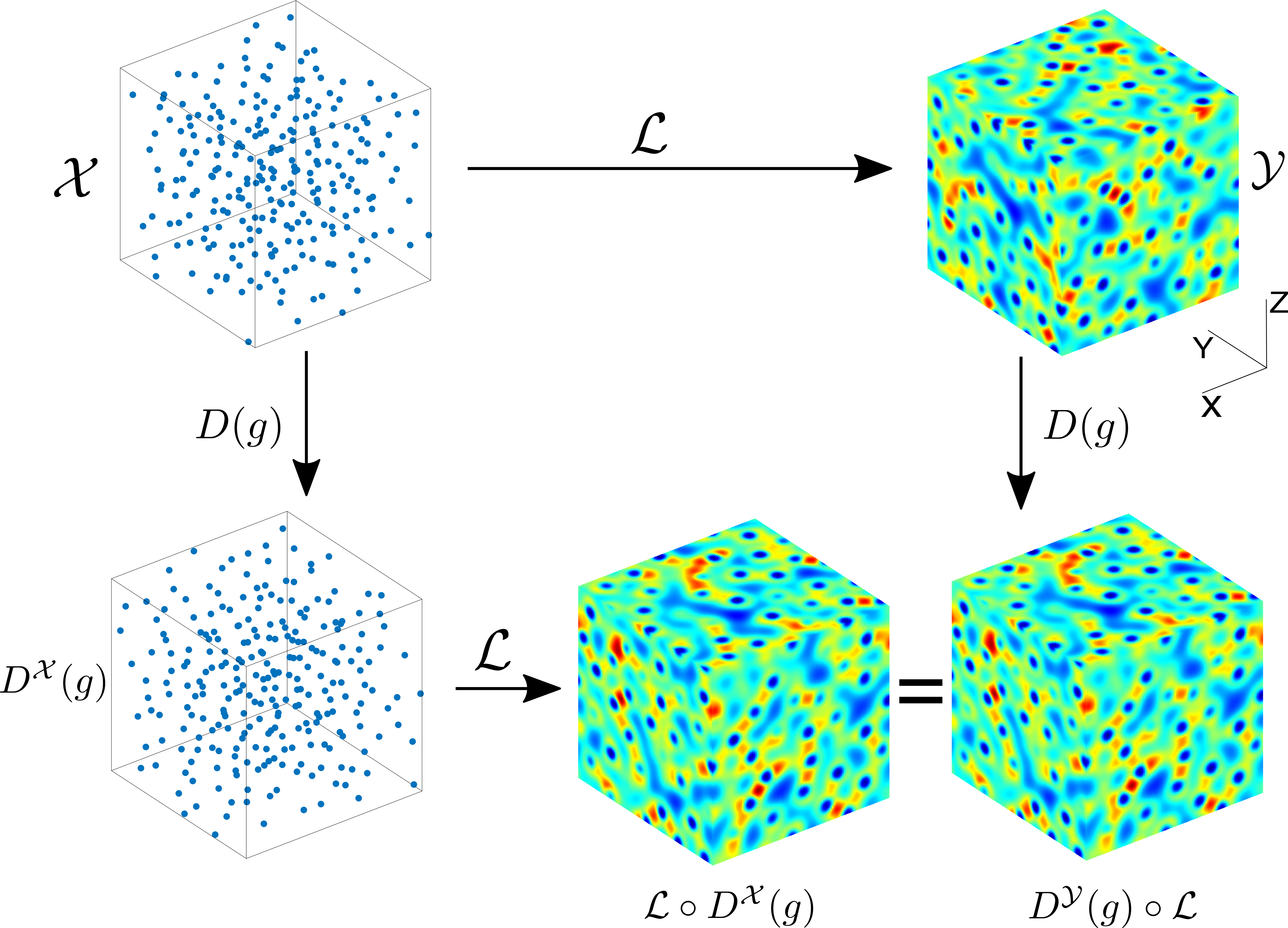}
    \caption{{Schematic showing preservation of equivariance in our model. $\mathcal{X}$ is 256 atom Aluminum system at high temperature (chosen such that there are no obvious intrinsic rotational symmetries of the system). $\mathcal{Y}$ is the corresponding electron density. $D(g)$ corresponds to the rotation of $\frac{\pi}{2}$ around the Y-axis. $\mathcal{L}$ is the composite map from the system to the electron density. We observe numerically that, $\vert \vert \mathcal{L} \circ D^{\mathcal{X}}(g) -D^{\mathcal{Y}}(g) \circ \mathcal{L} \vert \vert _{\infty} \approx 10^{-10}$. $\vert \vert \mathcal{L} \circ D^{\mathcal{X}}(g) -D^{\mathcal{Y}}(g) \circ \mathcal{L} \vert \vert _{\infty}$ is not exactly zero because of roundoff errors in billions of floating point operations involved in descriptor calculations and forward propagation through neural networks. Thus, $\mathcal{L} \circ D^{\mathcal{X}}(g) = D^{\mathcal{Y}}(g) \circ \mathcal{L}$, and hence equivariance is preserved.}  }
    \label{fig:showing_equi}
\end{figure*}

\begin{table*}[htbp]
\resizebox{0.55\textwidth}{!}{%
\begin{tabular}{|c|c|c|c|c|}
\hline
Material property & $2\times2\times2$ supercell & $3\times3\times3$ supercell \\ \hline
Lattice parameter (Bohr) & $7.4294$ ($7.4281$) & $7.5208$ ($7.5188$) \\ \hline
Bulk modulus (GPa) & $92.2774$ ($92.7708$) & $75.7977$ ($76.3893$) \\ \hline
\end{tabular}
}
\caption{A comparison between the calculated lattice parameter and the bulk modulus for aluminum using $\rho^{\text{ML}}$ and $\rho^{\text{DFT}}$ (DFT values in parentheses). We observe that the predicted lattice parameter closely matches the value given by DFT calculations. The ``true'' optimized lattice parameter for \ce{Al}, using a fine k-space mesh, is found to be $7.5098$ Bohr while experimental values are about $7.6$ Bohr \citep{cooper1962precise}). The ML predicted value of the bulk modulus matches the DFT value very closely, which itself is very close to the experimental value of approximately 76 GPa \cite{raju2002high}, at room temperature. 
}
\label{tab:bulk_modulus}
\end{table*}

\newpage
\clearpage
\bibliography{main}

\end{document}